\documentclass[11pt]{article}
\usepackage[top=1in, bottom=1in, left=1in, right=1in]{geometry}
\usepackage{amsmath}
\usepackage{mathrsfs}
\usepackage{epsfig}
\usepackage{graphicx}
\usepackage{authblk}
\usepackage{cite}
\usepackage{url}
\usepackage{hyperref}
\usepackage{amsmath}
\usepackage{amsbsy}
\usepackage{amscd}
\usepackage{amsopn}
\usepackage{amstext}
\usepackage{amsxtra}
\usepackage{color}
\usepackage{lscape}
\newcommand{\bb}[1]{{#1}}

\begin{document}

\title{Temporal profiles of avalanches on  networks  }

\author{James P Gleeson}
\affil{MACSI, Department of Mathematics and Statistics, University of Limerick, Ireland; james.gleeson@ul.ie}
\author{Rick Durrett}
\affil{Department of Mathematics, Duke University, Durham, NC, USA}

\date{31 May 2017} 

\maketitle

\begin{abstract}
An avalanche or cascade occurs when one event causes one or more subsequent events, which in turn may cause further events in a chain reaction. Avalanching dynamics are studied in many disciplines, with a recent focus on average avalanche shapes, i.e., the temporal profiles of avalanches of fixed duration. At the critical point of the dynamics the rescaled average avalanche shapes for different durations collapse onto a single universal curve. We apply Markov branching process theory to derive an equation governing the average avalanche shape for cascade dynamics on networks. Analysis of the equation at criticality demonstrates that nonsymmetric average avalanche shapes (as observed in some experiments) occur for certain combinations of dynamics and network topology. We give examples using numerical simulations of models for information spreading, neural dynamics, and behaviour adoption {and we propose simple experimental tests to quantify whether cascading systems are in the critical state.}
\end{abstract}

\begin{figure}
\centering
\epsfig{figure=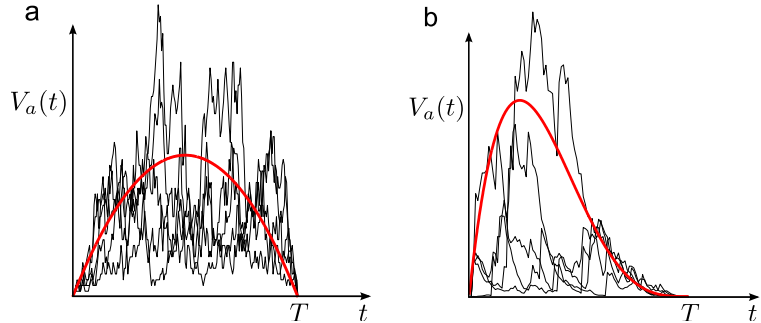,width=10 cm} 
\caption{{\bf $\mathbf{\left|\right.}$ Examples of average avalanche shapes.} In each panel, the black curves show five examples of individual avalanches that all have duration $T$.  The average avalanche shape for the duration $T$ (red curve) is found by averaging the temporal profiles of all such avalanches. Typically, the average avalanche shape is symmetric (e.g., parabolic) as in panel (a), but nonsymmetric avalanche shapes like panel (b) have also been observed (e.g., Fig.~S4 of \cite{Friedman12}).}\label{figschematic}
\end{figure}

The dynamics of avalanches or cascades are studied in many disciplines. Examples include the spreading of disease (or information) from human to human \cite{Pinto11,BorgeHolthoefer13}, avalanches of neuron firings in the brain \cite{Friedman12}, and the ``crackling noise'' exhibited by earthquakes and magnetic materials \cite{Sethna01}. Of particular interest are cases with dynamics poised at a critical point, where universal scalings of avalanches are observed. The most commonly studied feature of such systems is the distribution of avalanche sizes, which has a power-law scaling at the critical point. The observation of heavy-tailed distributions of avalanche sizes has therefore been used to indicate whether a system is critical. However, power-law distributions can also arise from mechanisms other than criticality \cite{Stumpf12,Newman05}, so recently attention has focussed more upon the temporal aspects of avalanches, which  also exhibit universal characteristics at criticality.

The average avalanche shape is determined by averaging the temporal profiles of all avalanches that have a fixed duration $T$. At criticality, the average avalanche shape is a universal function of the rescaled time $t/T$, meaning that the average avalanche shapes for different durations can be rescaled to collapse onto a single curve \cite{Sethna01}. This feature has recently been used as a sensitive test for criticality in a range of dynamics, from the Barkhausen effect in ferromagnetic materials \cite{Papanikolaou11} to neural avalanches \cite{Friedman12,Beggs12} and electroencephalography (EEG) recordings from hypoxic neonatal cortex \cite{Roberts14}. While average avalanche shapes are typically symmetric (e.g., parabolic) functions of time, nonsymmetric (left-skewed)  avalanche shapes  have also been observed in experiments. For example, early  observations of nonsymmetric avalanche shapes in experiments on Barkhausen noise \cite{Sethna01} raised doubts about whether the theoretical model used in \cite{Mehta02,Kuntz00} was in the correct universality class. Although this discrepancy between theory and experiment was later resolved by a more detailed theory for avalanche propagation \cite{Zapperi05,Colaiori08},  several instances of nonsymmetric avalanche shapes (e.g., the neural avalanches in  \cite{Friedman12}) still lack explanation. Despite some progress in modelling avalanche profiles using random walks \cite{Baldassarri03,Roberts14} and self-organized criticality models \cite{Laurson05,Rybarsch14,Massobrio15,Hesse14}, the factors that cause nonsymmetric average avalanche shapes remain poorly understood.

The characteristics of avalanches that occur on networks  depend on both the network connectivity and the node-to-node dynamics of the cascade \cite{Larremore12}. Cascading models have been applied, for example, to power-grid blackouts \cite{Dobson12}, epidemic outbreaks \cite{Pastor-Satorras15}, and to the propagation of memes (pieces of digital information) through online social networks \cite{Cheng14,Goel15}. The  distribution of avalanche sizes at criticality is known to depend non-trivially on the degree distribution of the underlying network \cite{Goh03}, but the time-dependence of cascades has not been studied from this perspective.

 In this paper we focus  on the temporal profile of cascades, i.e., the average avalanche shape, and how it is affected by the network degree distribution. Using a mathematical derivation of the average avalanche shape for Markovian dynamics \bb{(in both critical and noncritical cases)} we demonstrate that---as in other universality-breaking examples \cite{Radicchi15}---networks with heavy-tailed degree distributions can give rise to qualitatively different results from those found on networks with finite-variance degrees. However, the dynamics of the avalanching process are also important: we show that in fact it is the interaction between the dynamics and the network topology that determines whether average avalanche shapes are symmetric or not.

\section*{Results}

\subsection*{Average avalanche shapes}
To define the {average avalanche shape} we consider the set $S_T$ of all avalanches that are of duration $T$ (meaning that the avalanche has terminated at a time $T$ after its first event, with no further events occurring at any time $t>T$). Each avalanche $a$ in the set $S_T$ is described by a function $V_a(t)$, which is the number of events that occur at time $t$ in that avalanche (so $V_a(t)=0$ for $t>T$). The avalanche shape for the duration $T$ is defined as the average of the functions $V_a(t)$, taken over all the avalanches $a$ in set $S_T$, see Fig.~\ref{figschematic}.

\subsection*{Dynamics and networks}\label{sec:networks}

We consider  networks that are defined by their degree distributions, but are otherwise maximally random (``configuration-model'' networks \cite{Newmanbook}). For undirected networks, the degree distribution $p_k$ is the probability that a randomly-chosen node has degree (number of neighbours) equal to $k$. For directed networks, the joint degree distribution $p_{j k}$ is the probability that a random node has in-degree $j$ and out-degree $k$. We denote the mean degree by $z$ (so $z=\sum_k k p_k$ for undirected networks and $z=\sum_{j,k} k p_{j k} = \sum_{j,k} j p_{j k}$ for directed networks). Such configuration-model networks are locally tree-like, which facilitates the use of the branching-process approximations that we employ.  We assume that the networks consist of a single connected component  {(a strongly connected component in the case of a directed network \cite{Newmanbook})} and that they are large enough to permit us to use infinite-size approximations.

Our focus is on discrete-state dynamics, where each node of the network can be in one of a set of discrete states at each moment in time; transitions between states may occur continuously in time, or only at discrete time steps. Cascades occur when nodes successively switch to one specific state, which we will call the ``active'' state; we will generically refer to all other states as ``inactive''. Once a node is activated (i.e.,  once it transitions to the active state), it affects its neighbouring nodes by increasing the probability that they will also become activated at a later time. We focus on unidirectional dynamics, meaning that in the case where a node activates some of its neighbouring nodes and then subsequently becomes inactive, the neighbouring nodes cannot directly re-activate it. One important class of such dynamics includes cases where an activated node cannot subsequently return to the inactive state, and so cannot be re-activated (this class is called ``monotonic dynamics'' in \cite{PorterGleeson16}). Another class takes place on tree-like directed networks, which have negligible numbers of loops, so that activation of node $i$ can affect its out-neighbours, but there exists no path for the out-neighbours to subsequently  affect the state of node $i$ (even if node $i$ returns to the inactive state).
Each cascade is assumed to be  initiated by a randomly-chosen single node, called the ``seed'' node, which is activated at the beginning of the process while all other nodes are inactive; subsequent to the activation of the seed node, the cascade of activation of nodes proceeds according to the rules of the model under consideration.

One example of such dynamics is given by threshold models. In a threshold model, each node $i$ of an undirected network possesses a positive threshold $R_i$  that is assigned randomly from a distribution. When an inactive node $i$ of degree $k_i$ is chosen for updating it considers the number $m_i$ of its neighbours who are active, and makes a decision according to the rules of the specific model. In the Watts threshold model \cite{Watts02}, for example, node $i$ becomes active if the fraction $m_i/k_i$ is greater than, or equal to, the node's threshold $R_i$, i.e., the node activates if the fraction of its neighbours who are active is sufficently large, otherwise the node remains inactive. An alternative type of threshold model is that of Centola and Macy  \cite{Centola07}, wherein node $i$ activates if the total number of active neighbours (rather than the fraction of such neighbours) is large enough: $m_i\ge R_i$ (see also the discussion of threshold models and their relation to coordination games in \cite{PorterGleeson16}).

Cascades also occur in the neuronal dynamics model of \cite{Friedman12}, where each node in a weighted, directed network represents a neuron. The weight $\phi_{i j}$ on the edge connecting node $i$ to node $j$ is chosen at random from a uniform distribution on the interval $(0,\phi_\text{max})$, where $\phi_\text{max}$ is a tunable parameter \cite{Deville08,Deville10}. (In \cite{Friedman12} the values of $\phi_{i j}$ are instead inferred from experiments on neural networks). Using the same model parameters as \cite{Friedman12}, the neurons are modelled by binary-state elements as a very simple approximation of integrate-and-fire dynamics: whenever neuron $i$ fires (becomes active), it causes neuron $j$ to become active (in the next discrete time step) with probability $\phi_{i j}$. After a neuron fires, it is returned to the inactive state in the next time step. In \cite{Friedman12}, exogenous input noise is added to the system to ensure continuous neural activity. Since we focus purely on the avalanche dynamics, we instead randomly select a node to be the seed node of the cascade and activate it in the first time step, and record the ensuing avalanche of activations.

A final example of cascade dynamics is given by the model of \cite{GleesonPRL14} for meme propagation on a directed social network (like Twitter). In this model, each node (of $N$) in an directed network represents a user of the social network. The out-degree $k_i$ of a node $i$ is the number of its ``followers'' in the network: these are the users that receive the ``tweets'' (or distinct pieces of digital information, generically called ``memes'') sent by node $i$. Each user also retains a memory of the last meme received from the nodes it follows via a ``screen'' that is overwritten when a new meme is received. (More realistic models that incorporate longer memory are described in \cite{Weng12,Gleeson16}). In each time step (with $\Delta t=1/N$) one node is chosen at random and with probability $\mu$ the node ``innovates'' by creating a new meme, placing it on its screen, and tweeting this meme to all its followers (where the new meme overwrites any existing memes on their screens). Alternatively (with probability $1-\mu$), the chosen node ``retweets'' the meme that is currently on its screen. A newly-innovated meme can therefore experience an avalanche of popularity as it is retweeted multiple times before it eventually is forgotten by all users in the network, at which time the avalanche terminates. The analyses of \cite{GleesonPRL14} and \cite{Gleeson16} show that the avalanche dynamics of the memes are critical in the limit $\mu \to 0$ and subcritical for $\mu>0$. 

{Other examples of unidirectional dynamics to which our results apply include the zero-temperature random-field Ising model \cite{Sethna01} and susceptible-infected-recovered (SIR) disease-spread models  with fixed recovery times \cite{Pastor-Satorras15}, such as the Independent Cascade model for information diffusion \cite{Kempe03}.}

\subsection*{The offspring distribution}
The central quantity in our branching process analysis is called the {offspring distribution}, denoted by $q_k$ ($k=0,1,2,\ldots$). Roughly speaking, this distribution gives the likelihood that if a cascade of activation reaches a node (i.e., if one of the node's neighbours becomes active), that the node will activate and expose $k$ other neighbouring nodes to potential activation (see Supplementary Note~1 for details of the branching process approximation). For an undirected network, {the offspring distribution $q_k$ can usefully be expressed in terms of a simple degree-dependent quantity $\hat{q}_k$ that is defined as}
\begin{equation}
\quad {\hat q_k}^\text{ (undir)}= \frac{k+1}{z} p_{k+1} v_{k+1}, \label{qk}
\end{equation}
noting that the probability of reaching a node of degree $k+1$ by travelling along a random edge is $\frac{k+1}{z} p_{k+1}$, and if this edge spreads the activation to the node it has $k$ remaining inactive neighbours.
The quantity $v_k$ is the probability that a node $i$ of degree $k$ is {vulnerable} \cite{Watts02}, meaning that the activation of a single neighbouring node (at time $t_1$) will lead to the activation of node $i$ at some time $t>t_1$, assuming that no other neighbour of node $i$ becomes active by time $t$. {Note that the sum $r=\sum_{k=0}^\infty \hat{q}_k \le 1$ is the probability that a node reached by travelling along a random edge is vulnerable. The relationship between $\hat{q}_k$ and the offspring distribution $q_k$ is given by equation~(\ref{fPGF}) below, but our main qualitative results depend only upon the large-$k$ scaling of $\hat{q}_k$ (see Supplementary Note 1).}

For a directed network, {the definition of $\hat q_k$} is
\begin{equation}
\quad { \hat q_k }^\text{ (dir)}= \sum_{j} \frac{j}{z} p_{j k} v_{j k}, \label{qjk}
\end{equation}
where the factor $\frac{j}{z} p_{j k}$ represents the probability of reaching a node of in-degree $j$ and out-degree $k$ by travelling along a random edge of the network. The vulnerability
 $v_{j k}$ is the probability that the activation of a single in-neighbour (at time $t_1$) of a node $i$ (of in-degree $j$ and out-degree $k$)  will lead to the activation of node $i$ at some time $t>t_1$, assuming no other in-neighbour of node $i$ becomes active by time $t$. {The probability $r$ is defined as for the undirected case, and the relationship between $\hat{q}_k$ and the offspring distribution is again given by equation~(\ref{fPGF}) below.}


The { branching number} defines whether the dynamical process on a given network is subcritical, critical, or supercritical. The branching number is the mean of the offspring distribution, i.e., the expected number of ``children'' per ``parent'', and it can be expressed as
\begin{equation}
\xi = \sum_k k\, q_k{=\sum_k k\, \hat{q}_k}, \label{eq3}
\end{equation}
{see Supplementary Note 1. The}
 value $\xi=1$ is the critical value, separating the subcritical case ($\xi<1$) from the supercritical case ($\xi>1$). In the critical case, power-law distributions of avalanche sizes are observed \cite{GleesonPRL14,Gleeson16} but in this paper we focus on the temporal profiles of the avalanches.

\subsection*{Average avalanche shape}\label{sec:shape}
The detailed derivation of our results from the theory of Markov branching processes is given in Supplementary Note~2. For models with continuous-time updating a particularly simple result is found: the average avalanche shape $A(t)$ for avalanches of duration $T$ can be expressed as
\begin{equation}
A(t) = \frac{Q(T-t)\left[ f'\left(Q(T)\right) - f'\left( Q(T-t)\right)\right]}{f\left( Q(T-t)\right) - Q(T-t)} \label{At},
\end{equation}
where $f(s)$ is the generating function for the offspring distribution,
\begin{equation}
f(s) = \sum_k q_k s^k { = \frac{1}{r}\sum_k \hat{q}_k \left(1-r+r s \right)^k}\label{fPGF}
\end{equation}
and $Q(t)$ is the fraction of avalanches that are extinct by time $t$, which is given by the solution of the ordinary differential equation
\begin{equation}
\frac{d Q}{d t} = f(Q) -Q, \,\text{ with } Q(0)=0.\label{qeqn}
\end{equation}
Thus, the average avalanche shape for a given offspring distribution {$q_k$} and duration $T$ can be calculated by solving only one ordinary differential equation, equation~(\ref{qeqn}), and then using the solution function $Q(t)$ in the explicit formula of equation~(\ref{At}). We also show (in Supplementary Note~5) that the average avalanche shape for discrete-time updating can be found in a similar fashion, but the resulting expression is less amenable to analysis than the continuous-time case. However, qualitatively similar results are found for both continuous-time and discrete-time updating (see Supplementary Fig.~2), so we focus mainly on the solution given by equations~(\ref{At}) and (\ref{qeqn}).

\begin{figure}
\centering
\epsfig{figure=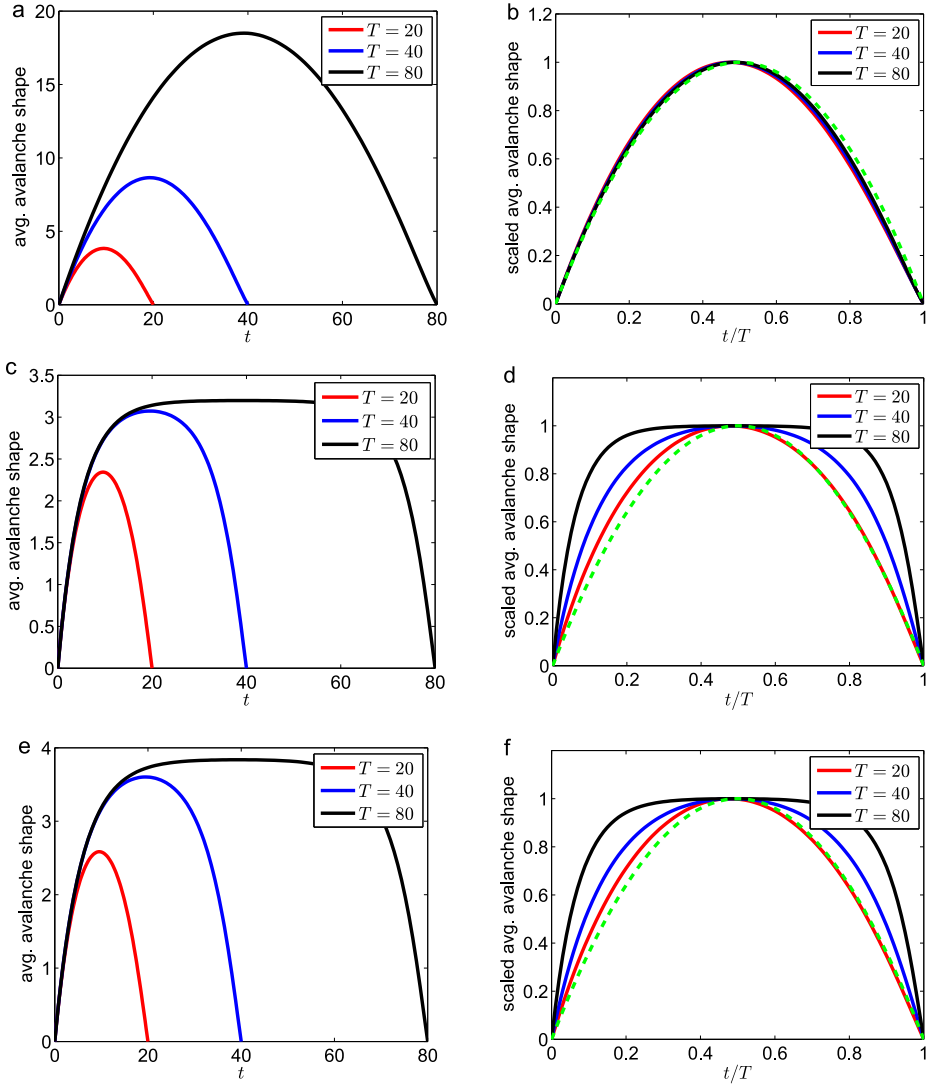,width=13.0 cm}
\caption{ {\bf $\mathbf{\left|\right.}$ Symmetric average avalanche shapes.} (Left column: panels (a), (c) and (e)): Average avalanche shapes  from equation~(\ref{At}), for Poisson offspring distribution $q_k$, with mean $\xi$. Each panel in the right column (panels (b), (d) and (e)) shows the same functions as in the panel to its left, but plotted (as, for example, in \cite{Papanikolaou11}) in rescaled time $t/T$ and rescaled vertically to have a maximum value of 1. Panels (a) and (b): critical case with $\xi=1$; panels (c) and (d): subcritical case,  $\xi=0.8$; panels (e) and (f): supercritical case, $\xi=1.2$.
The dashed green curve in panels (b), (d), and (f) is the parabola $4 \frac{t}{T}\left(1-\frac{t}{T}\right)$. }\label{fig1}
\end{figure}

In Figure~\ref{fig1} we show the avalanche shapes that are given by equation~(\ref{At}) in the case where the offspring distribution $q_k$ is a Poisson distribution with mean $\xi$. In the critical case ($\xi=1$) we see from Figs.~\ref{fig1}(a) and \ref{fig1}(b) that a rescaling of time and of avalanche height causes the avalanche shapes for different durations to collapse onto a single symmetric curve: this scaling collapse (although not the shape of the curve) is predicted by the universality arguments of \cite{Sethna01}.\bb{ Note that a case where equation~(\ref{qeqn}) is exactly solvable (binary fission) is examined in Supplementary Note~3 and is shown to give symmetric avalanche shapes, which are parabolic at the critical point. For subcritical ($\xi<1$) avalanches the profiles are symmetric but non-parabolic (Fig.~\ref{fig1}(c)) and do not collapse onto a universal curve (Fig.~\ref{fig1}(d)). For supercritical ($\xi >1$) avalanches (where we only consider those avalanches which terminate at a finite time $T$: a non-zero fraction of avalanches also exist that never terminate), very similar symmetric shapes are observed to the subcritical case (Figs.~\ref{fig1}(e) and \ref{fig1}(f)).}

Next, we consider the case where the offspring distribution has a power-law tail:
\begin{equation}
q_k \sim C\, k^{-\gamma} \quad\text{ as } k \to \infty, \label{qkasy}
\end{equation}
with exponent $\gamma$ in the range $2<\gamma<3$ so that the second moment of the offspring distribution is infinite. (Here, and throughout the paper, we use $C$ to denote a constant prefactor in an asymptotic scaling relation). Such cases are of practical interest because scale-free degree distributions are well-known \cite{DurrettRandomGraphDynamics,Newmanbook,Radicchi15} to strongly affect dynamics on networks, and (as we discuss below) a scale-free degree distribution can, for certain dynamics, lead to offspring distributions such as equation~(\ref{qkasy}). Note that the second moment of the offspring distribution is related to the second derivative of the generating function of equation~(\ref{fPGF}) evaluated at $s=1$, so this case corresponds to $f''(1)=\infty$.

\begin{figure}
\centering
\epsfig{figure=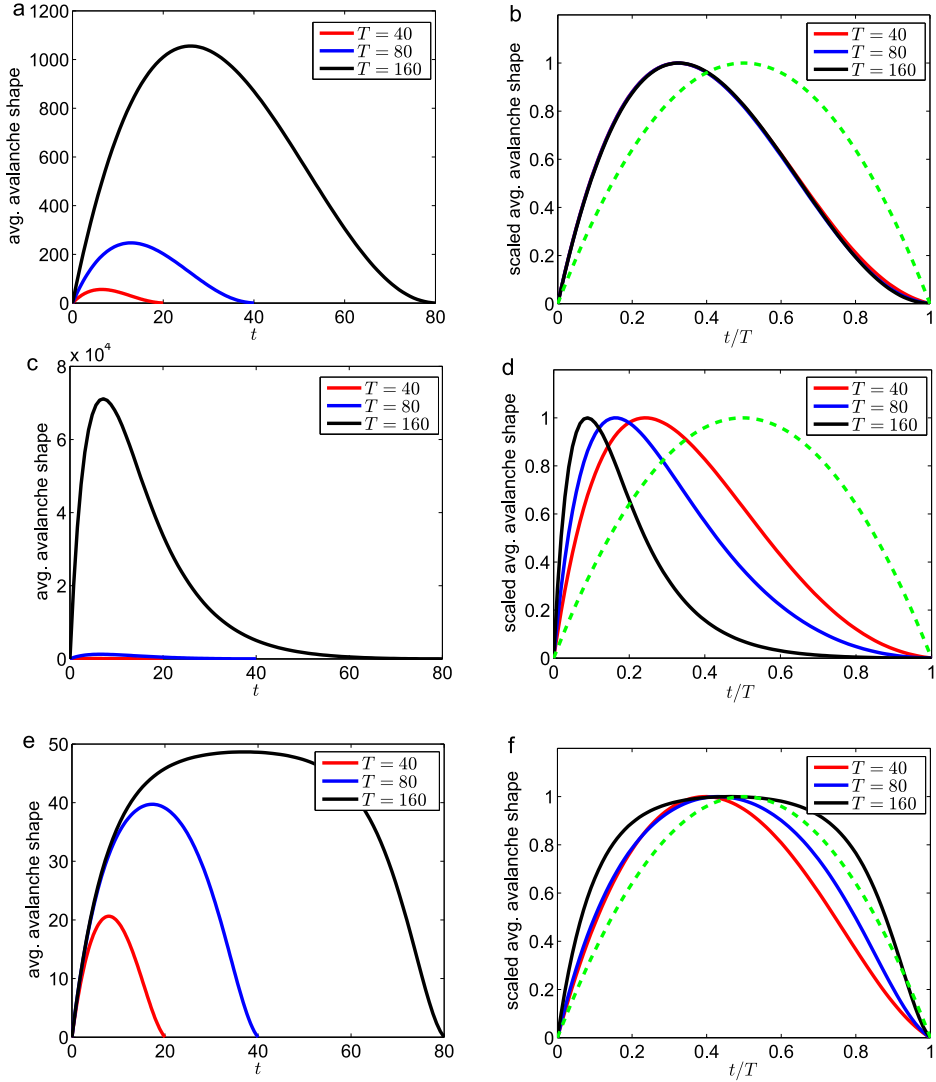,width=13.0 cm}
\caption{\bb{  {\bf $\mathbf{\left|\right.}$ Nonsymmetric average avalanche shapes.}  Left column: Average avalanche shapes  from equation~(\ref{At}), for power-law offspring distribution: $q_k = C\, k^{-\gamma}$ for $k\ge 1$, with exponent $\gamma=2.5$ and constant $C$ chosen to give the branching number $\xi$.  Each panel in the right column (panels (b), (d) and (e)) shows the same function as in the panel to its left, but rescaled as in Fig.~\ref{fig1}. Panels (a) and (b): critical case with $\xi=1$; panels (c) and (d): subcritical case,  $\xi=0.9$; panels (e) and (f): supercritical case, $\xi=1.15$; the dashed green curve is the same parabola as in Fig.~\ref{fig1}.} }\label{fig2}
\end{figure}

Using offspring distributions of the form (\ref{qkasy}) in equations~(\ref{At}) and (\ref{qeqn}) gives the avalanche shapes \bb{  shown in Figure~\ref{fig2}.
Clearly, the avalanche shapes---both critical and noncritical cases---are nonsymmetric, with a leftward skew. (This contrasts with the right-skewed avalanche shapes found from random-walk models with long memory \cite{Baldassarri03}).  A detailed asymptotic analysis of the governing equation (Supplementary Note~6) enables us to conclude that, in the critical case as $T\to\infty$ (and $T-t\to\infty$), the average avalanche shape scales as}
\begin{equation}
A(t) \sim \left\{ \begin{array}{cl}
                    C \frac{t}{T}\left(T-t\right) & \text{ if } f''(1)\text{ is finite,}\\ \\
                    C \frac{t}{T}\left(T-t\right)^\frac{1}{\gamma-2} & \text{ if } q_k \propto k^{-\gamma}\text{ as }k\to \infty, \text{ with }2<\gamma<3,
                    \end{array}
                    \right. \label{Aasy}
\end{equation}
where the constant prefactor $C$ is independent of $T$.
Note that parabolic avalanche shapes (with peak at $t=T/2$) are seen in the large-$T$ limit whenever the offspring distribution $q_k$ has finite second moment, but the shape is nonsymmetric (with peak at $t=(\gamma-2)T/(\gamma-1)<T/2$) for power-law distributions with $\gamma$ between 2 and 3.

\begin{figure}
\centering
\epsfig{figure=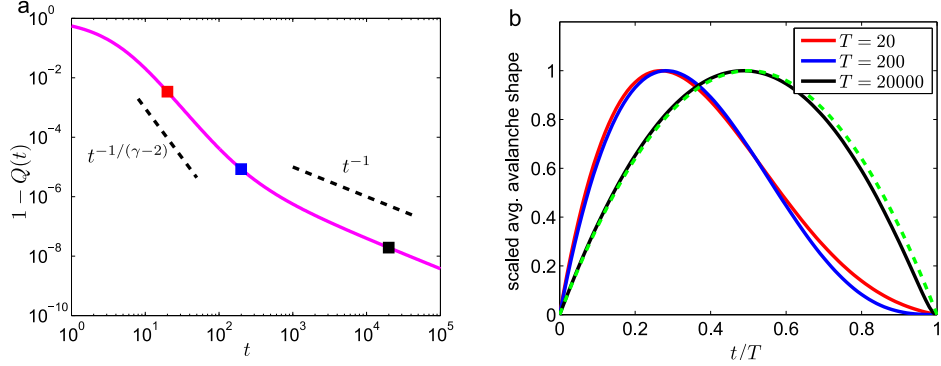,width=13.0 cm}
\caption{ {\bf $\mathbf{\left|\right.}$ Effects of truncated power-law offspring distribution.} Survival function $1-Q(t)$ (panel (a)) and rescaled avalanche shape (panel (b)) for offspring distributions with truncated power-law form: $q_k\propto k^{-\gamma} e^{-k/\kappa}$ with $\gamma=2.3$, $\kappa=10^6$. The constant of proportionality is determined by the criticality condition $\xi=1$. The coloured squares in panel (a) mark the durations $T=20$, $T=200$, $T=2\times 10^4$ of the avalanches whose average shapes are plotted in panel (b). Note the $T=20$ and $T=200$ cases have the nonsymmetric profiles typical of power-law $q_k$, but the $T=2\times 10^4$ case closely matches the parabolic profile expected for offspring distributions with finite second moment; \bb{the dashed green curve is the same parabola as in Fig.~\ref{fig1}.}}\label{figtrunc}
\end{figure}

True power-law tails are never seen in real networks, due to finite-size effects.
If the offspring distribution instead has a truncated power-law form, with an exponential cutoff for $k\gg \kappa$:
\begin{equation}
q_k \sim C\, k^{-\gamma} e^{-\frac{k}{\kappa}} \quad\text{ as } k \to \infty, \label{trunc}
\end{equation}
and with $2<\gamma<3$, then the avalanche shapes for all durations $T$ do not collapse onto a single curve, even at criticality. As the asymptotic analysis of Supplementary Note~6 reveals, the shape for large $T$ is determined by the survival function $1-Q(t)$: this is the fraction of avalanches that remain alive at a time $t$ after they begin. For the offspring distribution of equation~(\ref{trunc}), the survival function $1-Q(t)$ scales as $t^{-\frac{1}{\gamma-2}}$ for early times, but as $t^{-1}$ for later times; the crossover from one regime to the other is determined by the exponential cutoff $\kappa$ in the offspring distribution (see Fig.~\ref{figtrunc}(a)). Therefore, it is possible to observe nonsymmetric shapes for avalanches with relatively short durations $T$, but the longer-duration avalanches (the $T\to \infty$ limit) revert to the parabolic shape typical of offspring distributions with finite second moment, see Fig.~\ref{figtrunc}(b).

\subsection*{Other characteristic temporal shapes}\label{sec:other}
The  Markov branching process approach that we use to derive the average avalanche shape in   equation~(\ref{At}) can also be applied to calculate other temporal characteristics of avalanches. In Supplementary Note~2, for example, we  derive a formula for the variance of the avalanche shape (i.e., the variance of the set of functions $\{V_a(t)\,|\,a\in S_T\}$, see Methods). The overall shape of the standard deviation is found to be similar to the average avalanche shape: The asymptotic result for the critical case (in the limit $T\to\infty$ and $T-t\to\infty$, see Supplementary Note~6) is that the coefficient of variation at time $t$ for avalanches of duration $T$ is
\begin{equation}
\text{CV}(t) = \frac{\sqrt{\text{variance}}}{A(t)} \sim \left\{ \begin{array}{cl}
                     \frac{1}{\sqrt{2}} & \text{ if } f''(1)\text{ is finite,}\\ \\
                    \sqrt{\frac{3-\gamma+(\gamma-2)\frac{t}{T}}{(\gamma-1)\frac{t}{T}}} & \text{ if } q_k \propto k^{-\gamma}\text{ as }k\to \infty, \text{ with }2<\gamma<3.
                    \end{array}
                    \right. \label{e10}
\end{equation}
Note that the coefficient of variation is constant (independent of $t$ and $T$) for the case where parabolic avalanche shapes occur, so the standard deviation also has a parabolic profile. However, in the power-law case with $\gamma<3$, the shape of the standard deviation is even more skewed than that of the average avalanche shape profile: the coefficient of variation  limits to a constant as $t\to T$, but it diverges  like  $1/\sqrt{t}$ as $t\to 0$.

The universality of the average avalanche shape has been used in analyzing experimental data; specifically, the collapse of shapes for avalanches of different durations can help identify whether the dynamics is critical or not \cite{Sethna01,Friedman12,Roberts14,Papanikolaou11}.
However, one drawback of the average avalanche shape is that it requires an accurate assessment of the time $T$ at which each avalanche terminates. Pinpointing such termination times can be difficult in empirical data, especially for avalanches of information on social networks, many of which exhibit very long lifespans \cite{Lerman12,Lerman16,Gleeson16}. Another characteristic temporal shape that we can calculate analytically is the average shape of all avalanches that have not terminated by a given observation time $T$: such a characteristic may prove easier to calculate for empirical data than the standard avalanche shape. In Supplementary Note~4 we show that the average non-terminating avalanche shapes at various observation times $T$ collapse onto a single curve when the dynamics are critical. The universal curve is again found to have a parabolic form (but with peak at $t=T$) if $f''(1)$ is finite, and to have a skewed (non-parabolic) shape if the offspring distribution is power-law. \bb{Moreover, the calculation of the average non-terminating avalanche shape requires less data than that of the average avalanche shape (see Supplementary Note 7), which may make it a useful diagnostic tool in experimental studies.}

As we demonstrate with our numerical simulations below, an even simpler temporal profile can provide a very sensitive measure of whether an avalanching system is critical or not. The average number of events observed at a time $t$ after an avalanche begins is given by the average of $V_a(t)$ over the entire set of avalanches, regardless of the duration of the avalanche. (Note avalanches that terminated at a time $T$ with $T<t$ contribute zero to the measure at time $t$). In Supplementary Note~4 we  show that the average number of such events at time $t$ is an exponentially decaying function of $t$ if the dynamics is subcritical, an exponentially growing function of $t$ for supercritical dynamics, and is a constant (independent of $t$) for the critical case. \bb{This temporal characteristic is particularly useful in understanding the numerical simulation results below, and it can equally well be applied to experimental data.}


\subsection*{Numerical simulations}\label{sec:simulations}

In this section we report on numerical simulations of unidirectional dynamics on networks, to assess the applicability of the branching process theory developed above. We  run numerical simulations of  example dynamics on synthetic (configuration-model) and real-world networks, recording average avalanche shapes (and other temporal characteristics) for comparison with our theory. The branching process paradigm is only an approximation for dynamics on networks, as its assumptions are invalidated by the existence of loops in the network and by the finite size of the networks \cite{Newmanbook}. We demonstrate that while such features indeed impact upon the agreement with theoretical results, the main qualitative feature of critical dynamics that we have identified---the appearance of nonsymmetric avalanches shapes when the offspring distributions defined by equations~(\ref{qk}) or (\ref{qjk}) have sufficiently heavy tails---is indeed observable in numerical simulations of cascade dynamics on  networks.

\begin{figure}
\centering
\epsfig{figure=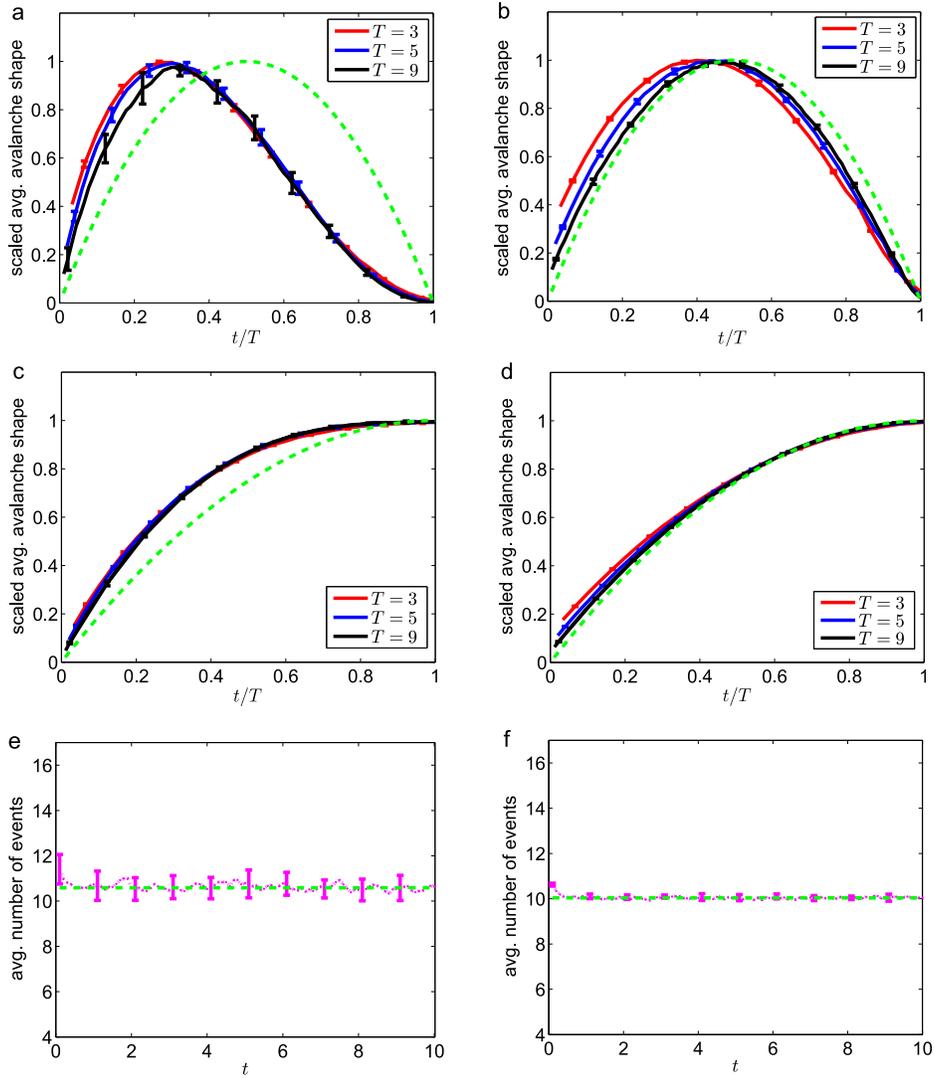,width=13.0 cm}
\caption{ {\bf $\mathbf{\left|\right.}$ Twitter cascades model.} Numerical simulation results for critical meme-popularity avalanches within the model of \cite{GleesonPRL14}. The panels in the left column show results from a network with a power-law out-degree distribution, with exponent $\alpha=2.5$. The results in the right column are from a network where every node has exactly $z=10$ out-neighbours.
Panels (a) and (b): rescaled average avalanche shapes. Panels (c) and (d): rescaled average non-terminating avalanche shapes. Panels (e) and (f): average number of tweets of the meme per unit time.Here, and in Figs.~\ref{figneuro}--\ref{figSNAP}, the green dashed curve in panels (a) and (b) is the parabola described in Fig.~\ref{fig1}, while that in panels (c) and (d) is the half-parabola $\frac{t}{T}\left(2-\frac{t}{T}\right)$. In panels (e) and (f), the green dashed line is the constant value representing the late-time average value of the number of events (the average is taken over the second half of the time range, i.e., $t$ values between 5 and 10). See Methods for definition of error bars.}\label{figCIC}
\end{figure}

Figure~\ref{figCIC} shows results from simulations of the continuous-time (Markovian) meme propagation model of \cite{GleesonPRL14}. Memes are tweeted from user to user according to the rules of the model; the popularity of each meme is tracked in the simulations, and the number of events in the  avalanche profile of a meme is  the number of times it is tweeted within a time interval.
 We record the average avalanche shape determined by all memes whose avalanches terminate at time $T$ (using a bin of duration $0.5$, so $T=9$, for example, includes all avalanches that terminate at times in the range $[8.5,9]$).
Two network structures are compared (see Methods): the panels in the left column of Fig.~\ref{figCIC} (panels (a), (c) and (e)) present results for a network with scale-free out-degree distribution $p_k \propto  k^{-\alpha}$  with $\alpha=2.5$, while the panels in the right column are for a network where every node has exactly $z=10$ followers (note the mean degree of the two networks are approximately equal). In both cases the in-degree distribution is Poisson, and in-degrees and out-degrees of nodes are independent. We set the innovation parameter $\mu$ to zero, so we expect the dynamics to be critical (from equations~(\ref{qjk}), (\ref{eq3}), and (\ref{vjkCIC})). 

According to our theory, the rescaled average avalanche shape curves for different (and sufficiently large) avalanche durations $T$ should collapse onto a single curve. We see good agreement with this prediction in Fig.~\ref{figCIC}(a): note the distinctively nonsymmetric shape of the collapsed curve. Although the average avalanche shapes for the $z$-regular out-degree network in Fig.~\ref{figCIC}(b) are not fully converged by $T=9$, they are evidently approaching the parabolic profile expected for the case where the offspring distribution has finite second moment.
Panels (c) and (d) contrast the average non-terminating avalanche shapes that are found on the two networks. On the power-law network the shape is clearly non-parabolic (panel (c)), while the match to the asymptotic expression found in Supplementary Note 6 is very good for the case of finite $f''(1)$ (panel (d)). \bb{On both types of network the error bars (see Methods for definition) are smaller than in panels (a) and (b), reflecting the fact that the set of avalanches that have not terminated by time $T$ is typically much larger than the set of avalanches that terminate exactly at $T$, so the average shape is better estimated using the larger data set.}
Panels (e) and (f) show the average number of events (tweets) per unit time. The criticality of the process on both networks is reflected in the fact that this measure remains constant over time.

\begin{figure}
\centering
\epsfig{figure=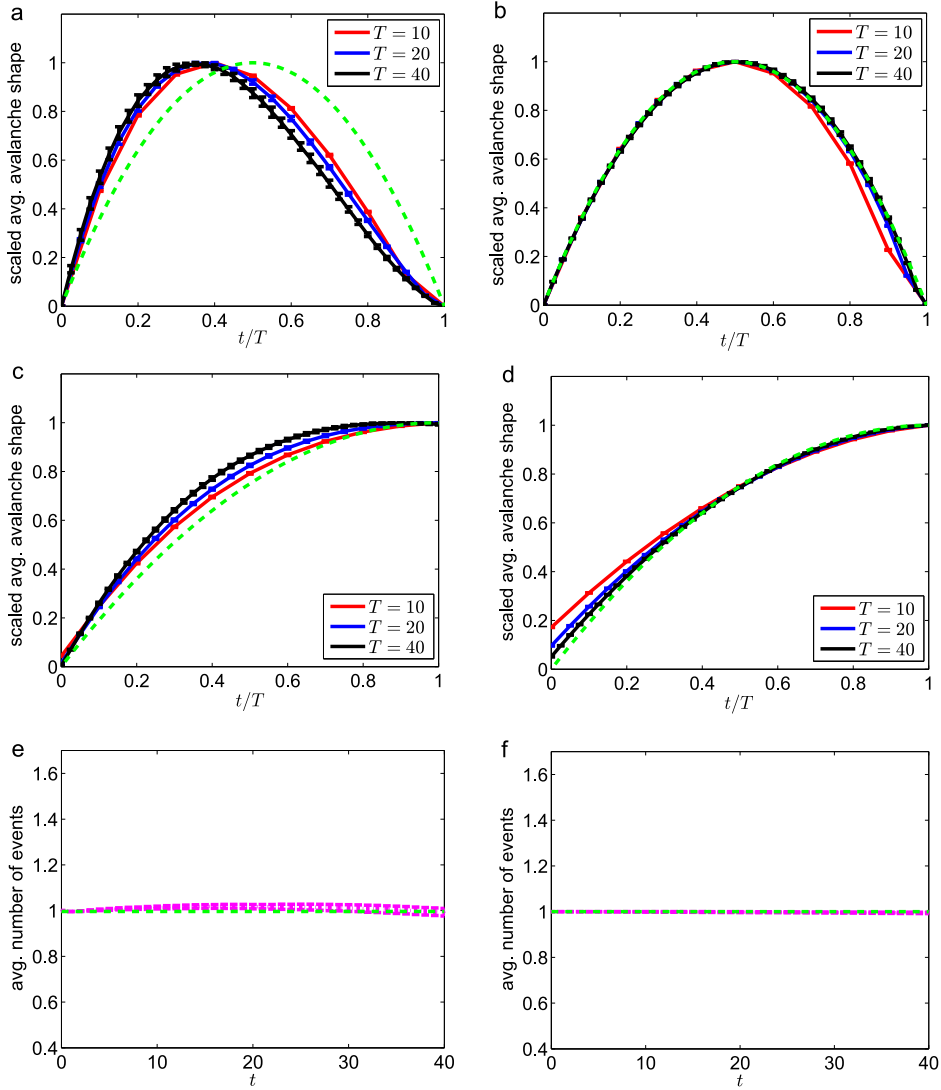,width=13.0 cm}
\caption{ {\bf $\mathbf{\left|\right.}$ Neuronal avalanches model. }Results of numerical simulation of the neuronal dynamics model of \cite{Friedman12} at criticality.
 As in Fig.~\ref{figCIC}, the results in the left column are for a network with power-law out-degree distribution ($\alpha=2.5$), while those in the right column are for a network with $z$-regular out-degrees.
 Panels (a) and (b): rescaled average avalanche shapes. Panels (c) and (d): rescaled average non-terminating avalanche shapes. Panels (e) and (f): average number of firing neurons per discrete time step.The green dashed line in panels (e) and (f) shows the constant value 1 that is expected for the average number of events in this critical discrete-time branching process. See Methods for definition of error bars.
}\label{figneuro}
\end{figure}

Figure~\ref{figneuro} shows results obtained from simulations of the neuronal dynamics model of \cite{Friedman12}, where the parameter $\phi_\text{max}$ is tuned so as to poise the dynamics near to criticality. As in Fig.~\ref{figCIC}, the left column of results is for a network with power-law out-degree distribution of exponent $\alpha=2.5$, while those in  the right column are for a $z$-regular network. All neurons are synchronously updated in each discrete time step and we record the number of activated neurons at each step as the ``events'' of the avalanche; the avalanche terminates when no new neurons are activated.
Although the details of this discrete-time case differ from the continuous-time case of Fig.~\ref{figCIC}, the results again qualitatively agree with theory. As expected (see Methods), we see nonsymmetric average avalanche shapes and non-parabolic average non-terminating avalanche shapes in panels (a) and (c), while the corresponding results on the  finite-second-moment network (panels (b) and (d)) agree closely with the parabolic asymptotic shapes of equations~(\ref{Aasy}) and Supplementary Note 6.

\begin{figure}
\centering
\epsfig{figure=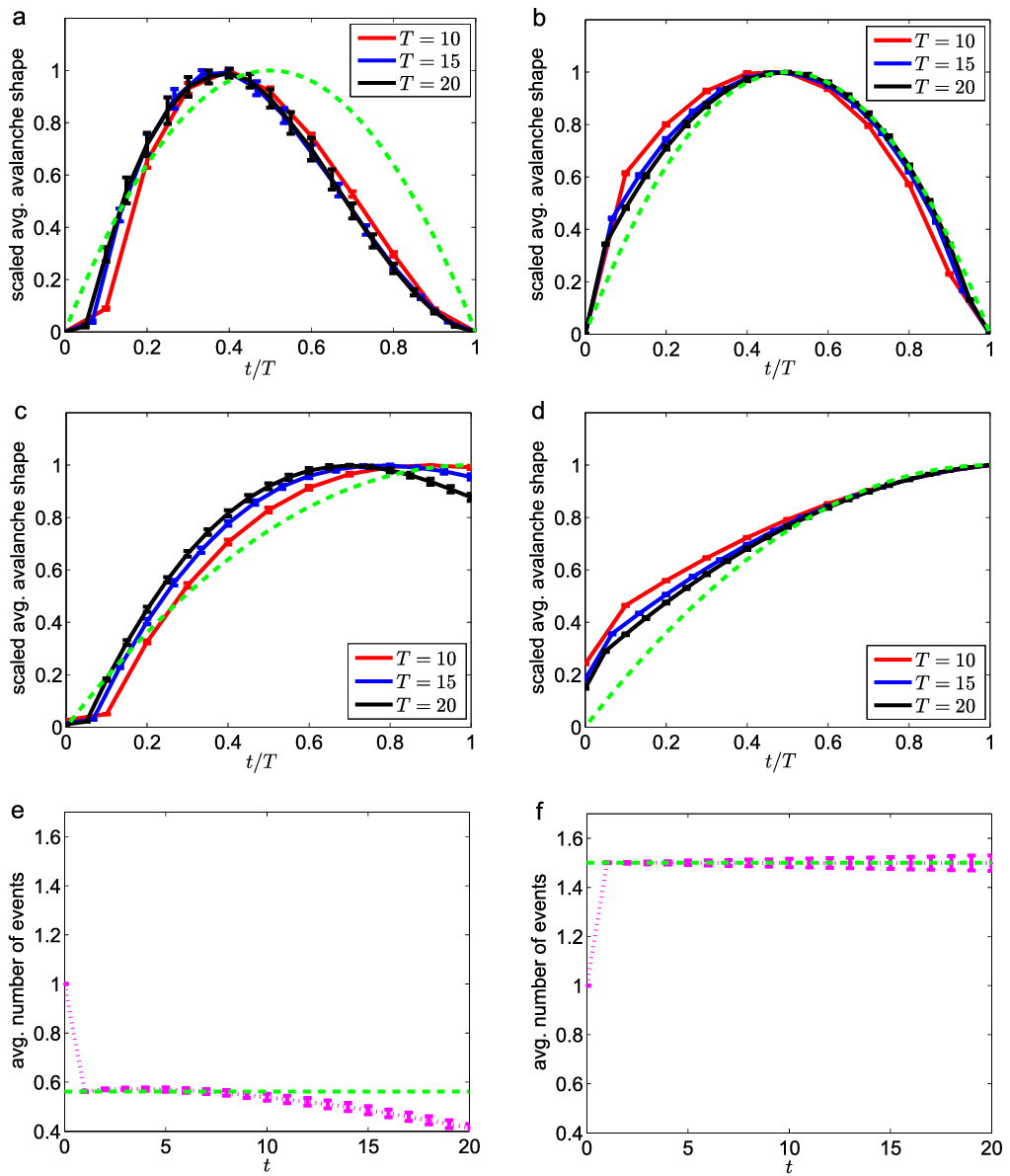,width=13.0 cm}
\caption{ {\bf $\mathbf{\left|\right.}$ Behaviour-adoption model. }Results of numerical simulation of a Centola-Macy threshold model  \cite{Centola07} at criticality.
Left column results are for a scale-free network ($\alpha=3.3$); right column results are for a $z$-regular random network.
Panels (a) and (b): rescaled average avalanche shapes. Panels (c) and (d): rescaled average non-terminating avalanche shapes. Panels (e) and (f): average number of newly-activated nodes per discrete time step. The green dashed line in panels (e) and (f) shows the constant value $z^2/\left<k^2-k\right>$ expected for the average number of events at criticality (see Methods). See Methods for definition of error bars.
 }\label{figthreshold}
\end{figure}

In Figure~\ref{figthreshold} we consider threshold dynamics on an undirected network. Specifically, we show results for a Centola-Macy threshold model, with a uniform distribution of thresholds on the interval $(0,\theta_\text{max})$, where the parameter $\theta_\text{max}$ is tuned to place the dynamics close to criticality.  The updating is synchronous, i.e.,  all nodes are updated at each discrete time step, and the number of avalanche events at each time step is the number of nodes that are newly activated.
For the Centola-Macy dynamics we expect (see Methods) nonsymmetric average avalanche profiles when the power-law exponent $\alpha$ of the network degree distribution lies between 3 and 4; note we use $\alpha=3.3$ in the left column of Fig.~\ref{figthreshold}. It is noteworthy that the network degree distribution has  finite variance in this case: it is the interaction between the network topology and the Centola-Macy dynamics that leads to nonsymmetric avalanche shapes.
(In contrast, for the Watts threshold model with uniformly distributed thresholds, nonsymmetric profiles appear only for networks with degree exponents $\alpha$ in the range $2<\alpha<3$, i.e., for degree distributions with infinite variance).



 In this case, we can clearly see one of the limitations of the branching process approximation: the seed node for each cascade is chosen uniformly at random from all the nodes, but subsequently activated nodes in the cascade are reached with a probability proportional to their degree, as in equation~(\ref{qk}). Therefore the branching process picture does not correctly capture the first step of the cascade and this discrepancy can be seen in the early-time shape of the avalanches, particularly in panels (a) and (b) of Fig.~\ref{figthreshold}. The theory could be extended to deal with this issue (as in \cite{Iribarren11} for example), but the effects on the average profiles diminish as longer-duration avalanches are considered.

A more serious limitation of the theory's accuracy is presented by Fig.~\ref{figthreshold}(e), where we see the deviation of the average number of events away from the constant value that indicates critical dynamics. In fact, the finite size of the network and the heavy-tailed degree distribution mean that the activated nodes are quickly (within about 10 time steps) replacing inactive nodes throughout a significant fraction of the network, so that some of the ``new branches'' emanating from an activated node are in fact connected to previously-activated nodes, contrary to the branching process assumption of independence. As a result, the spreading efficiency decreases over time, and the branching process---despite initially being at criticality---becomes subcritical. Nevertheless, as Fig.~\ref{figthreshold}(a) shows, the average avalanche shapes still collapse quite well, and clearly are nonsymmetric; the average shape of non-terminating avalanches (in panel (c)) is a more sensitive indicator of the loss of criticality due to the finite size of the network.


\begin{figure}
\centering
\epsfig{figure=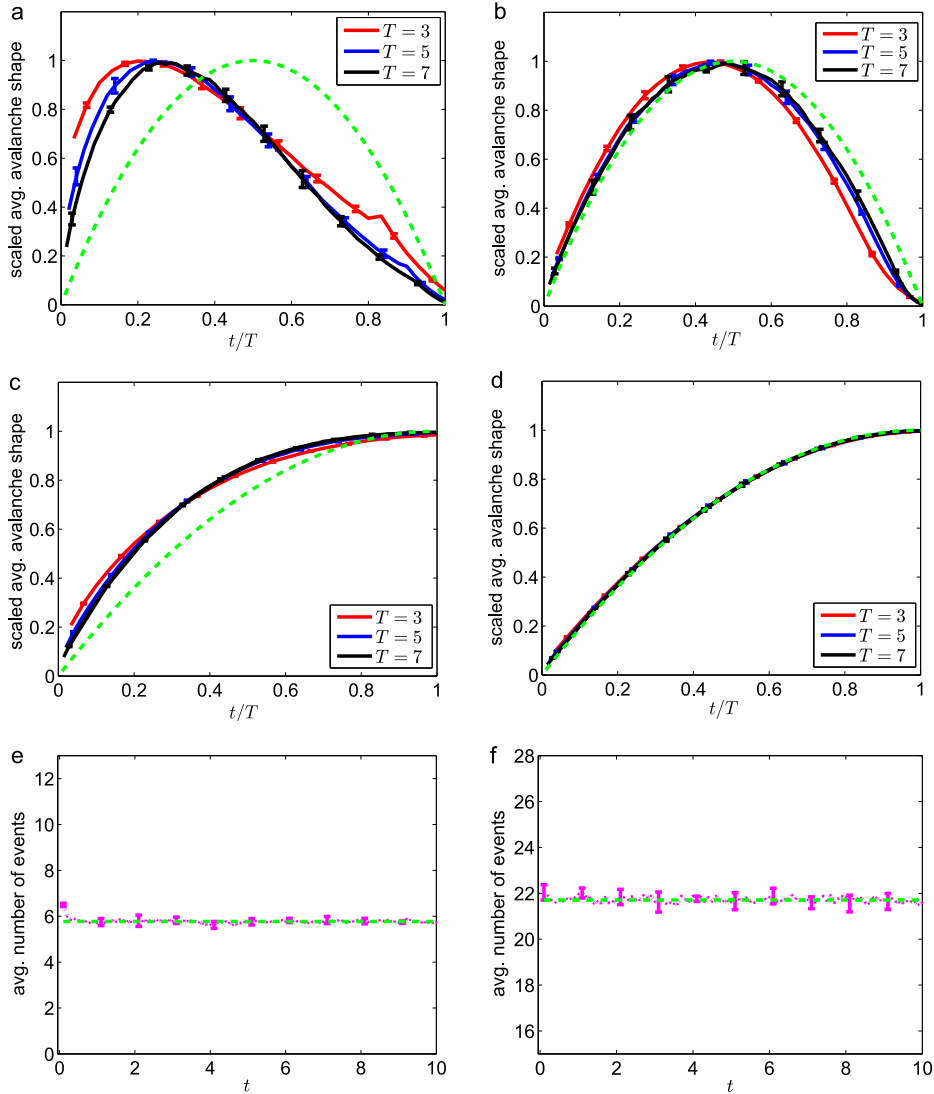,width=13.0 cm}
\caption{ {\bf $\mathbf{\left|\right.}$ Cascades on empirical Twitter network.} Numerical simulation results for critical meme-popularity avalanches \cite{GleesonPRL14}. The results in the left column are for the Twitter network of \cite{SNAPdata,SNAPref}, while those in the right column are for a directed Erd\H{o}s-R\'{e}nyi network with the same mean degree as the Twitter network (see Supplementary Note 8).
Although the empirical network is not tree-like, the qualitative predictions of our theory still hold: compare to the corresponding panels of Fig.~\ref{figCIC}. The green dashed curves are defined as in Fig.~\ref{figCIC}. See Methods for definition of error bars.}\label{figSNAP}
\end{figure}

All networks used in the simulations above were configuration-model networks. However, real-world networks are known to differ significantly from configuration-model networks with the same degree distribution \cite{Newmanbook}. For example, degree-degree correlations and closed triangles of nodes (``clustering'') are common in many social networks, but are relatively rare in the corresponding configuration-model networks. In our final example, we therefore use  the sample of the Twitter network that is made available for download from \cite{SNAPdata} by the authors of \cite{SNAPref}. Unlike a configuration-model directed network, this dataset contains a high proportion of reciprocated links (i.e., instances where node $i$ follows node $j$ and node $j$ also follows node $i$); in fact $48\%$ of all links in the network are reciprocated. This feature means that the network is not tree-like, and the branching process assumptions are not strictly true. Nevertheless, Fig.~\ref{figSNAP} shows that results from numerical simulations of the critical meme propagation model of \cite{GleesonPRL14} are in very good qualitative agreement with the predictions of our theory: a good collapse of the average avalanche shapes for different durations $T$ is found (Fig.~\ref{figSNAP}(a)), with a clear left skew that is consistent with the heavy-tailed distribution of number of Twitter followers found in empirical data \cite{Kwak10,Gleeson16}. \bb{By randomly rewiring the original network \cite{Karsai11}, we confirm that the left skew is due to the degree distribution of the network, and not to any meso- or macro-scopic structure of the network, see Supplementary Note 8.} As in other examples where tree-based theory is more accurate than expected on real-world networks \cite{Melnik11}, this result demonstrates that the predictions of the theory are quite robust to violations of the assumptions used in the mathematical derivation.

\bb{Another important assumption of the mathematical derivation is the Markovian nature of the dynamics. In Supplementary Note 9 we generalize the meme propagation model to include non-Markovian dynamics \cite{Iribarren09,Iribarren11}, so that the inter-event times between successive tweets of a user need not be exponentially distributed. Although we are limited to simulation results in this case, the qualitative predictions of the Markovian theory presented here again appear to be robust even to quite strongly non-Markovian dynamics.}

\section*{Discussion}

 In this paper we have examined the link between Markov branching processes and unidirectional cascade dynamics on networks, with a focus on the temporal profile of avalanches. Our main result is equation~(\ref{At}), which gives the average avalanche shape for avalanches \bb{(both critical and noncritical)} of duration $T$, and requires only the solution of a single ordinary differential equation (equation~(\ref{qeqn})). The input to this equation is the offspring distribution of the branching process, which is determined from the network structure and the dynamical system of interest by equation~(\ref{qk}) or equation~(\ref{qjk}). In our analysis of the avalanche shape given by equation~(\ref{At}), we have demonstrated that nonsymmetric avalanche shapes can arise at criticality when the offspring distribution has a power-law tail. Using numerical simulations of threshold models, neuronal dynamics, and online information-sharing, we show that nonsymmetric shapes can occur for common models running on networks with power-law degree distributions. However, it is important to note that a heavy-tailed degree distribution is not sufficient to guarantee nonsymmetric avalanche shapes: It is the interplay between the cascade dynamics and the network topology (as can be seen from the formulas in  equations~(\ref{gammaundir}) and (\ref{gammadir}) for the exponent of the offspring distribution) that determines the symmetry of the average avalanche shape.  The  results of numerical simulations verify the qualitative predictions of the branching process theory, despite finite-size effects and other violations of the assumptions of the theory. \bb{Further simulation studies (Supplementary Note 9) indicate that the qualitative results remain valid even for non-Markovian dynamics, although further theoretical investigation is clearly required to verify the regimes of validity.}

\bb{ In addition, our theoretical approach enables us to identify other characteristic temporal shape functions (e.g., the average non-terminating avalanche shape, see Supplementary Note 4) that may prove useful when experimentalists seek to identify critical behaviour from the temporal signatures in a data set with a limited number of avalanche time series.}
 Given the relevance of branching process descriptions to cascades in a range of fields (e.g., neuroscience \cite{Shew11}, social networks \cite{Gleeson16}, crackling noise \cite{Sethna01}, etc.), it is hoped that these insights may find many applications. We anticipate several possible directions for extensions of the methodology introduced here; notably, removing the Markovian assumption to apply a similar analysis for non-Markovian cascades \cite{Iribarren11,Gleeson16}, and extending the theory to multilayer networks \cite{Kivela14,Boccaletti14}.

\vspace{1cm}


\section*{Methods}

\subsection*{Vulnerabilities}
Here we give examples of how the vulnerabilities $v_k$ and $v_{j k}$ are calculated for the models introduced in the main text. Inserting these $v_k$ and $v_{j k}$  functions into equations~(\ref{qk}) and (\ref{qjk}), respectively, defines the offspring distribution for each of the models.

 The vulnerability $v_{k}$ for a threshold model on an undirected network  is  the probability that a node $i$, of degree $k$, activates when exactly one of its neighbours is active, i.e., when $m_i=1$. According to the rules of the Watts threshold model, for example, node $i$ will become active if its threshold $R_i$ is less than, or equal to, $1/k$, which is the fraction of its neighbours that are active when $m_i=1$. The probability that the node's threshold is less than this value is given by \begin{equation}
    v_{k}^\text{(Watts)}=F\left(\frac{1}{k}\right),\label{vWatts}
    \end{equation}
    where $F$  is the cumulative distribution function of the thresholds. In the Centola-Macy threshold mode, a node with one active neighbour will activate if its threshold is less than or equal to 1, and so the vulnerability in this case is
    \begin{equation}
     v_{k}^\text{(C-M)} = F(1).\label{VCM}
    \end{equation}

 In the neuronal dynamics model of \cite{Friedman12} a node is vulnerable if it becomes active when one of its in-neighbours fires. According to the rules of the model, this occurs with a probability that depends on the edge between the two nodes, but is chosen from a uniform distribution on $(0,\phi_\text{max})$, independently of the degrees of the nodes. The probability that a random node with in- and out-degrees $j$ and $k$ is vulnerable is therefore the average of this uniform distribution, i.e., $v_{j k} = \phi_\text{max}/2$.

 {In the model of \cite{GleesonPRL14} for meme propagation on a directed social network we focus on a chosen meme and assume that this meme has been tweeted by an in-neighbour of node $i$ (i.e., by one of the $j$ users followed by user $i$). The probability that node $i$ will subsequently retweet this meme (before user $i$'s memory is overwritten by other tweets it receives from the $j$ users it follows) is given, for $j \gg 1$, by
    \begin{equation}
    v_{j k} \approx \frac{1-\mu}{j},\label{vjkCIC}
    \end{equation}
    where $\mu$ is the innovation probability and $k$ is the number of followers of node $i$, see Sec.~IVA of \cite{Gleeson16}.}

\subsection*{Power-law offspring distributions from network dynamics} \label{sec:powerlaw}
In Supplementary Note 6 we show that nonsymmetric average avalanche shapes occur within Markovian branching processes when the offspring distribution $q_k$ has a power-law tail with exponent $\gamma$ between 2 and 3. Here we examine how such offspring distributions might arise from unidirectional dynamics on undirected and directed networks, using the relationships given by equations~(\ref{qk}) and (\ref{qjk}).

In the case of an undirected network, we suppose that the vulnerability $v_k$ depends on the node degree $k$ as
\begin{equation}
v_k \sim C\, k^{-\nu} \text{ as } k\to \infty. \label{nudef}
\end{equation}
Then, if the degree distribution of the network has a power-law tail:
\begin{equation}
p_k \sim C\, k^{-\alpha} \text{ as } k\to \infty, \label{pk1}
\end{equation}
the large-$k$ asymptotics of {$\hat q_k$} {(and hence of the offspring distribution $q_k$, see Supplementary Note 1) are} given by equation~(\ref{qk}) as
\begin{equation}
q_k \sim C \,k^{1-\alpha-\nu}
\end{equation}
and so we write
\begin{equation}
\gamma^\text{(undir)} = -1 + \alpha+\nu \label{gammaundir}
\end{equation}
for the power-law exponent of the offspring distribution. Note that in general the value of $\gamma$ will be different from the power-law exponent $\alpha$ of the network's degree distribution.

The case of a directed network is complicated by the existence of the joint distribution $p_{j k}$ of in-degree $j$ and out-degree $k$. If we assume the simplest case of nodes having independent in- and out-degree, then the joint distribution factorizes: $p_{j k} = p_j^\text{in}\, p_k$, and the large-$k$ behaviour of the vulnerability can be specified by the weighted sum over in-degrees as
\begin{equation}
\sum_j \frac{j}{z}\, p_j^\text{in}\, v_{j k } \sim C\, k^{-\nu}.\label{nujk}
\end{equation}
Assuming a power-law out-degree distribution, as in equation~(\ref{pk1}), equation~(\ref{qjk}) {and the equivalence of the power-law exponents of $q_k$ and $\hat{q}_k$} yields
\begin{equation}
q_k \sim C \, k^{-\alpha-\nu} \text{ as } k\to \infty,
\end{equation}
so that the power-law exponent of the offspring distribution is
\begin{equation}
\gamma^\text{(dir)} = \alpha+\nu. \label{gammadir}
\end{equation}

\subsection*{Details of numerical simulations}
Each of Figs.~\ref{figCIC} through \ref{figthreshold}   consists of two columns of panels. The left-hand column (i.e., panels (a), (c) and (e)) show results for a network with a power-law degree (or out-degree) distribution $p_k\propto k^{-\alpha}$ for $k \ge k_\text{min}$ (with $p_k=0$ for $k<k_\text{min}$). The right-hand column (panels (b), (d) and (f)) are for random regular networks, where every node has the same (out-)degree $z$.
\bb{ In each experiment, $n_A$ individual avalanches are simulated to calculate the average avalanche shape and other measures (see Supplementary Note 7 for a study of how the value of $n_A$ impacts upon the results shown).}  For each avalanche the seed node is changed and the order of node updates (for Figs.~\ref{figCIC} and \ref{figSNAP}) or the dynamical parameters (the edge weights $\phi_{i j}$ for Fig.~\ref{figneuro} or the node thresholds $R_i$ for Fig.~\ref{figthreshold}) are randomized. In order to quantify the robustness of the results, the complete experiment that results in the average avalanche shape is repeated a total of $n_R$ times and the error bars in the Figures denote the standard deviation of the measures over the set of replica experiments.

Figure \ref{figCIC} shows results for the meme propagation model of \cite{GleesonPRL14} at criticality (the innovation parameter $\mu$ is zero), using $n_A=10^6$ avalanches and $n_R=12$ replicas, on networks with $N=10^5$ nodes. The directed network with power-law out-degree distribution (panels (a), (c) and (e)) has exponent $\alpha=2.5$ and minimum out-degree of $k_{\text{min}}=4$, giving a mean degree of $z=10.6$. The panels in the right column are for a directed network where every node has exactly $k=10$ followers. In each case the followers are chosen uniformly at random from the set of all nodes, so in-degrees and out-degrees are independent.
Using the vulnerability from equation~(\ref{vjkCIC}) in equation~(\ref{nujk}) gives a $\nu$ value of $0$ for this model, so
equation~(\ref{gammadir}) predicts  nonsymmetric avalanche shapes for values of $\alpha$, the tail exponent of the network's out-degree distribution, between 2 and 3; note we use $\alpha=2.5$ in the left column of Fig.~\ref{figCIC}.

The neuronal dynamics model of \cite{Friedman12} is used for the simulations of Fig.~\ref{figneuro}, with the parameter $\phi_\text{max}$ set to its critical value of $2/z$.  Here the vulnerability $v_{j k}=\phi_\text{max}/2$ is independent of node degree $k$, and so $\nu=0$ in equation~(\ref{nujk}), giving $\gamma=\alpha=2.5$ from equation~(\ref{gammadir}) for the power-law network. We use  $n_A=10^7$ avalanches and $n_R=24$ replicas, on the same directed networks as used in Fig.~\ref{figCIC}.

Figure~\ref{figthreshold} is for the Centola-Macy threshold model with thresholds uniformly-distributed between 0 and $\theta_\text{max}={\left< k^2-k\right>}/{z}$. Using  $n_A=10^6$ avalanches and $n_R=24$ replicas, we run simulations on scale-free undirected networks with $\alpha=3.3$, $k_\text{min}=2$, $z=2.9$ and $N=10^6$ (left column panels) and on random $z$-regular graphs with $z=3$ and $N=10^5$ (right column panels). The vulnerability $v_k$ of equation~(\ref{VCM}) is a constant in this case, so $\nu=0$ in equation~(\ref{nudef}). As a result, equation~(\ref{gammaundir}) gives $\gamma=-1+\alpha$ as the exponent of the offspring distribution, and so nonsymmetric avalanche shapes are expected for power-law networks with exponent $\alpha$ between 3 and 4 (as in the left column of Fig.~\ref{figthreshold}).
However, note that if we instead consider the Watts threshold model with uniformly distributed thresholds, equation~(\ref{vWatts}) gives $\nu=1$, hence $\gamma=\alpha$, meaning that nonsymmetric avalanche shapes occur only for networks with degree exponents $\alpha$ in the range $2<\alpha<3$. The differing conditions for the two threshold models provide a good example of how the node-to-node dynamics and the network topology interact in a nontrivial fashion to generate nonsymmetric average avalanche shapes.

\bb{The green dashed lines in panels (e) and (f) of Fig.~~\ref{figthreshold} show the expected number of ``children'' events triggered by a seed node that is chosen uniformly at random (not with probability proportional to its degree, as in equation (\ref{qk})). The seed node has an average of $z$ neighbours, each of which activates with probability $F(1)=1/\theta_\text{max}$, giving an expected number of events (after the seeding event at $t=0$) equal to $z^2/\left< k^2-k\right>$. }

In Figure~\ref{figSNAP} we use the same meme propagation model as in Fig.~\ref{figCIC} (with $\mu=0$), running $n_A=1.4\times 10^6$ avalanches in $n_R=6$ replicas. The network substrate is the sampled Twitter network of \cite{SNAPdata,SNAPref}, which has mean degree $z=21.75$ and $N=81,306$ nodes.

\subsection*{Code availability} Matlab/Octave simulation codes for the examples used in this paper are available from  \\ \url{http://www.ul.ie/gleesonj/avalanches}.

\subsection*{Data availability} Network data for the examples used in this paper are available from  \\ \url{http://www.ul.ie/gleesonj/avalanches}.

\section*{Acknowledgements}
Helpful discussions with  Gareth Baxter, Sergey Dorogovtsev, Ali Faqeeh, Peter Fennell, Tom Hurd, Kristina Lerman, Kevin O'Sullivan, Mason Porter, and Tim Rogers  are gratefully acknowledged.
This work was supported by Science Foundation Ireland (grant numbers 11/PI/1026, 09/SRC/E1780, 15/SPP/E3125 and 16/IA/4470). We acknowledge the SFI/HEA Irish Centre for High-End Computing (ICHEC) for the provision of computational facilities.

%


\renewcommand{\figurename}{Supplementary Figure}
\setcounter{figure}{0}
\renewcommand{\theequation}{S\arabic{equation}}


\clearpage
\section*{Supplementary Note 1: Linking branching processes to discrete-state dynamics on networks} \label{app:BP}

In this  Supplementary Note we demonstrate how branching processes can be used to describe unidirectional dynamics on configuration-model networks.
\begin{figure}
\centering
\epsfig{figure=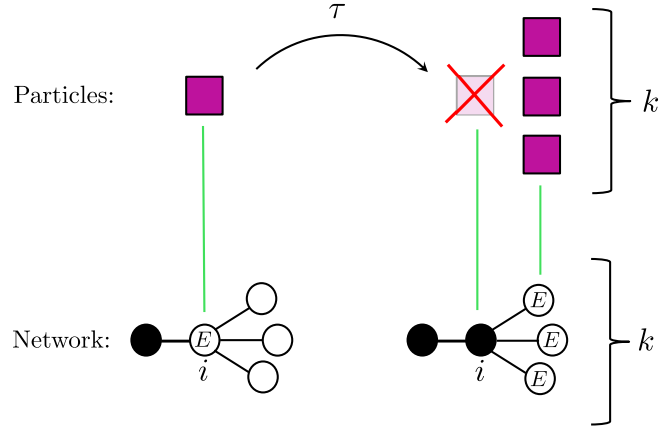,width=9.5 cm}
\caption{{\bf $\mathbf{\left|\right.}$ Linking branching processes to cascades on networks.}{ The green lines demonstrate the analogy between the particles of a continuous-time branching process (top), and the exposed vulnerable (EV) nodes of a cascade process on an undirected network (those nodes marked ``E'' in the bottom diagrams), as described in Supplementary Note~1. Black nodes are active and white nodes are inactive; note that only vulnerable nodes are shown here (so the actual degree of node $i$ is at least $k+1$). } }\label{fig:BP}
\end{figure}
The paradigm of continuous-time branching processes describes {particles} who each live for a lifetime $\tau$, where the value of $\tau$ is drawn from a prescribed distribution. In the Markov case, the distribution of $\tau$ values is exponential (and we choose the timescales so that the mean of the distribution is 1); in the Galton-Watson process time is discrete, so every particle has a lifetime of exactly 1 time unit. (The framework we describe here can also be used to link non-Markovian dynamics with the appropriate (Bellman-Harris) branching processes \cite{AthreyaNeybook}; however, in this paper we restrict our attention to the Markovian continuous-time, and discrete-time, cases).
At the end of its lifetime, the particle dies; simultaneously, a number  $k$ of identical ``children'' particles appear, where $k$ is a random integer from a distribution $q_k$ (the {offspring distribution}). Each avalanche of the branching process is initiated at time 0 by a single particle, and terminates at time $T$ when there are no particles remaining alive; the temporal evolution of the cascade process (i.e., the avalanche shape)  can be described in terms of the number of particles alive at time $t$, for times $t$ with $0<t<T$.

Supplementary Figure~\ref{fig:BP} shows how unidirectional dynamics on a network can be described in terms of a branching process.
\bb{ First, we define a node as ``exposed'' if it is currently inactive but one of its neighbouring nodes is active. We identify a particle of the branching process with an exposed vulnerable (EV) node of the network, i.e., a node that is currently inactive, but is linked to an active node, and that will eventually become active itself. An EV node (such as the node $i$ in Supplementary Figure~\ref{fig:BP}) remains in the EV state for a time $\tau$ until it becomes active. Since the node $i$ is then no longer an EV node, it can be considered to be a ``dead'' particle of the branching process. However, by becoming active, node $i$ has also simultaneously exposed all of its inactive neighbours, and if $k$ of these are vulnerable (where $k$ is less than or equal to the degree of node $i$)  then $k$ new EV nodes (particles) have been created. The number $k$ of these ``children particles'' depends on the degree of node $i$ and on the probability $r=\sum \hat{q}_k$ that a node reached along a random (in-) edge is vulnerable. An EV node has $k'$ inactive neighbours with probability $\hat{q}_{k'}/r$, and since each inactive neighbour is (independently) vulnerable with probability $r$, the offspring distribution (the probability that a particle generates $k$ children particles (EV nodes) when it dies) is given by
\begin{equation}
q_k =\frac{1}{r} \sum_{k'=k}^\infty \hat{q}_{k'}  \left( \begin{array}{c}
                                k' \\
                                k
                                \end{array} \right)r^k \left(1-r\right)^{k'-k} . \label{qksum}
\end{equation}
The corresponding generating function is given by equation~(5) of the main text.

By differentiating equation~(5) of the main text, it is straightforward to verify the simple relations
\begin{equation}
f'(1) = \sum_k k \,q_k = \sum_k k \,\hat{q}_k
\end{equation}
and
\begin{equation}
f''(1) = \sum_k k (k-1) q_k = r \sum_k k (k-1) \hat{q}_k.
\end{equation}
In particular, $f''(1)$ is infinite if $\hat{q}_k \sim C \,k^{-\gamma}$ as $k\to \infty$ for values of the power-law exponent $\gamma$ between 2 and 3. By considering the asymptotic expansion of $f(x)$ near $x=1$, as in Supplementary Equation~(\ref{PL}) of Supplementary Note 6, it can be seen that if $\hat{q}_k \sim C\, k^{-\gamma}$ then the offspring distribution $q_k$ also has a power-law tail, with the same exponent $\gamma$. }



\section*{Supplementary Note 2: Derivation of average avalanche shape}\label{app:derivation}

In our theoretical approach, we model the avalanche dynamics using a continuous-time  branching process \cite{AthreyaNeybook}. Each ``particle'' survives for a random lifetime $\tau$ (drawn from the exponential distribution with mean 1) and at the end of its lifetime, is replaced by $k$ ``children'', where the number of offspring $k$ is $0,1,2,\ldots$ with probability $q_k$. The mean number $\xi$ of children per parent is given by equation~(3) of the main text;  the critical case $\xi =1$ is of particular interest, but our derivation is also valid for other cases.

Following the notation of Chapter III of \cite{AthreyaNeybook}, we define $Z(t)$ as a one-dimensional continuous time Markov branching process representing the number of alive particles at time $t$. The (time-homogeneous) transition probabilities are defined as
\begin{equation}
P_{1 j}(t) = \text{Prob}\left\{ Z(\tau+t)=j | Z(\tau)=1\right\}
\end{equation}
and the corresponding generating function is
\begin{equation}
F(s,t) = \sum_k P_{1 k}(t) s^k.
\end{equation}
Note that the probability of extinction by time $t$ is
\begin{equation}
Q(t)\equiv \text{Prob}\left\{ Z(t)=0 | Z(0)=1\right\} = F(0,t).
\end{equation}

We consider an avalanche as a process starting at time $t=0$ with a single particle  ($Z(0)=1$), with lifetimes of particles that are exponentially distributed with parameter $1$. Upon death, a particle leaves $k$ offspring with probability $q_k$, $k=0,1,2,\ldots$; the  generating function for the offspring distribution is defined in equation~(5) of the main text,
with $\xi=f'(1)$.

For a fixed duration $T>0$, we define the {avalanche path probability} $\pi_n(t)$ to be the probability that $n$ particles are alive at time $t$ ($0\le t\le T$), conditioned on there being exactly one particle alive at time 0 and exactly one particle alive at time $T$:
\begin{equation}
\pi_n(t)=\text{Prob}\left\{ Z(t)=n | Z(0)=1 \text{ and } Z(T)=1\right\}. \label{pin}
\end{equation}
Because of the Markov property, $\pi_n(t)$ is proportional to the product of the probabilities of having a path that (i) has $n$ particles alive at time $t$ and (ii) goes from $n$ particles at time $t$ to one particle at time $T$, so
\begin{equation}
\pi_n(t) \propto P_{1 n} (t) P_{n 1}(T-t).
\end{equation}
We can write
\begin{equation}
P_{n 1}(T-t) = n \left[ P_{1 0}(T-t)\right]^{n-1} P_{1 1}(T-t)
\end{equation}
to reflect the fact that the $n$ particles active at time $t$ can be reduced to a single particle at time $T$ only if $n-1$ of them have no descendants alive at time $T$, with the remaining one (which can be chosen in $n$ ways) having one descendant at time $T$. The correctly normalized avalanche path probability distribution is therefore
\begin{equation}
\pi_n(t) = \frac{ P_{1 n}(t) \,n \left[ P_{1 0}(T-t)\right]^{n-1} P_{1 1}(T-t)}{\sum_n P_{1 n}(t)\, n \left[ P_{1 0}(T-t)\right]^{n-1} P_{1 1}(T-t)}. \label{1}
\end{equation}
The denominator of Supplementary Equation (\ref{1}) can be written as
\begin{equation}
F'\left(P_{1 0}(T-t),t\right) P_{1 1}(T-t),
\end{equation}
where, for brevity,  the prime denotes differentiation with respect to the generating function variable $s$ (i.e., the first argument of $F(s,t)$). Noting that $P_{1 0}(t)=F(0,t)=Q(t)$ and $P_{1 1}(t) = F'(0,t)$, the denominator of  Supplementary Equation (\ref{1}) can be further simplified to
\begin{equation}
F'\left(F(0,T-t),t\right) F'(0,T-t) = F'(0,T),
\end{equation}
where we have used the Markov property \cite{AthreyaNeybook}
\begin{equation}
F(s,T)=F\left(F(s,T-t),t\right).\label{Markov}
\end{equation}
Thus, the avalanche path probability distribution is
\begin{equation}
\pi_n(t) = \frac{ P_{1 n}(t) \,n \left[ P_{1 0}(T-t)\right]^{n-1} P_{1 1}(T-t)}{F'(0,T)},
\end{equation}
and the generating function for this distribution is
\begin{align}
G(s,t) & = \sum_n \pi_n(t) s^n \nonumber \\
&= \frac{1}{F'(0,T)} \sum_n P_{1 n}(t) \, n \left[ P_{1 0}(T-t)\right]^{n-1} P_{1 1}(T-t) s^n\nonumber \\
&= \frac{ s F'\left( s P_{1 0}(T-t),t\right) P_{1 1}(T-t)}{F'(0,T)}. \label{e25}
\end{align}

The average avalanche shape is determined by the expected number of particles alive at time $t$, where the expectation is over the set of avalanche paths conditioned as in Supplementary Equation~(\ref{pin}). We calculate this using the generating function $G(s,t)$ as follows:
\begin{align}
G'(1,t)& = \frac{1}{F'(0,T)} \left[ F'\left(P_{1 0}(T-t),t\right) P_{1 1}(T-t) + F''\left(P_{1 0}(T-t),t\right) P_{1 0}(T-t) P_{1 1}(T-t)\right] \nonumber\\
&= 1+\frac{F(0,T-t) F'(0,T-t) F''\left(F(0,T-t),t\right)}{F'(0,T)} \label{B}.
\end{align}

Since we are considering Markov branching processes, we can use the Kolmogorov forward and backward equations \cite{AthreyaNeybook} for the generating function $F(s,t)$ :
\begin{align}
\frac{\partial }{\partial t} F(s,t) &= \left( f(s)-s\right) F'(s,t) \nonumber\\
\frac{\partial }{\partial t} F(s,t) &= f\left(F(s,t)\right) - F(s,t) \label{KB},
\end{align}
and by eliminating $\partial F/\partial t$ from this pair of equations we write $F'(s,t)$ in terms of $F(s,t)$:
\begin{equation}
F'(s,t) = \frac{f\left(F(s,t)\right)-F(s,t)}{f(s)-s}\quad \text{ for } s\neq 1. \label{2}
\end{equation}
Differentiating with respect to $s$ and substituting $F'$ using Supplementary Equation (\ref{2}) yields
\begin{equation}
F''(s,t) = \frac{\left[f\left(F(s,t)\right)-F(s,t)\right]\left[ f'\left(F(s,t)\right) - f'(s)\right]}{\left[ f(s)-s\right]^2}, \label{e29}
\end{equation}
and evaluating at $s=F(0,T-t)$ and using Supplementary Equation (\ref{Markov}), we have
\begin{equation}
F''\left(F(0,T-t),t\right) = \frac{\left[f\left(F(0,T)\right)-F(0,T)\right]\left[ f'\left(F(0,T)\right) - f'\left(F(0,T-t)\right)\right]}{\left[ f\left(F(0,T-t)\right)-F(0,T-t)\right]^2}. \label{3}
\end{equation}
Using Supplementary Equations (\ref{2}) and (\ref{3}) in Supplementary Equation (\ref{B}) gives
\begin{align}
G'(1,t) & = 1 + \frac{F(0,T-t)\left[ f'\left(F(0,T)\right) - f'\left( F(0,T-t)\right)\right]}{f\left( F(0,T-t)\right) - F(0,T-t)} \nonumber\\
& = 1 + \frac{Q(T-t)\left[ f'\left(Q(T)\right) - f'\left( Q(T-t)\right)\right]}{f\left( Q(T-t)\right) - Q(T-t)} \label{4},
\end{align}
where $Q(t)=F(0,t)$ solves the ordinary differential equation obtained from the second of Supplementary Equations (\ref{KB}) at $s=0$:
\begin{equation}
\frac{d Q}{d t} = f(Q) -Q, \quad\text{ with } Q(0)=0.\label{5}
\end{equation}
We have therefore reduced the problem of calculating the average avalanche shape to the solution of a single ordinary differential equation. Given an offspring distribution and a duration $T$, we can easily solve Supplementary Equation~(\ref{5}) using standard numerical methods for $t=0$ to $T$, then substitute $Q(t)$ into Supplementary Equation~(\ref{4}) to obtain the average avalanche shape
\begin{equation}
A(t) \equiv G'(1,t) - 1, \label{A}
\end{equation}
as given in equation~(4) of the main text.

To calculate the variance of the avalanche shape, we again use the generating function $G(s,t)$ to write
\begin{equation}
\text{var}\left(n(t)\right) = \left< n^2\right> - \left< n\right>^2 = G^{\prime\prime}\left(1,t\right)  + G'\left(1,t\right) - \left( G'(1,t)\right)^2 \label{v1}.
\end{equation}
Differentiating Supplementary Equation~(\ref{e25}) with respect to $s$, then setting $s$ equal to 1 (and using the identities $P_{1 0}(T-t) \equiv F(0,T-t)$ and $P_{11}(T-t) = F'(0,T-t)$) yields
\begin{equation}
G''(1,t) = \frac{F'(0,T-t) F(0,T-t)}{F'(0,T)} \left\{2 F''\left(F(0,T-t),t\right) + F'''\left( F(0,T-t),t\right)F(0,T-t)\right\}.
\end{equation}
Using Supplementary Equation~(\ref{2}) on the first term, and Supplementary Equation~(\ref{e29}) and differentiation of Supplementary Equation~(\ref{2}) on the terms in braces leads (after some algebra) to the result
\begin{align}
\text{var}\left(n(t)\right)  =& \frac{Q(T-t)}{\left[f(Q(T-t)-Q(T-t)\right]^2}\times \nonumber\\
    & \times \left\{ f(Q(T-t))\left[ f'(Q(T)-f'(Q(T-t)-Q(T-t) f''(Q(T-t)\right] \right.\nonumber\\
    & \quad \left. + Q(T-t)\left[f'(Q(T-t))^2-f'(Q(T)) f'(Q(T-t)) \right.\right. \nonumber\\
    &\quad  \left.\left. + f''(Q(T))\left[ f(Q(T))-Q(T)\right] +Q(T-t) f''\left( Q(T-t)\right)\right]\right\}. \label{vf}
\end{align}

\bb{
\section*{Supplementary Note 3: Exactly solvable  case}\label{app:exact} 
We consider binary fission, with \bb{$q_0=\frac{1+\mu}{2}$ and $q_2=\frac{1-\mu}{2}$} being the only non-zero offspring probabilities, where $\mu<1$ measures the deviation from criticality (the branching number is $\xi=1-\mu$). The generating function for the offspring distribution is
\begin{equation}
f(s) = \sum_k q_k\, s^k =  \frac{1+\mu}{2}+\frac{1-\mu}{2} s^2,
\end{equation}
and the ordinary differential equation of Supplementary Equation (\ref{5}) can be solved exactly for $Q(t)$ to yield
\begin{equation}
Q(t) = \frac{(1+\mu)\left(1-e^{-\mu t}\right)}{1+\mu-(1-\mu)e^{-\mu t}}. \label{Q1}
\end{equation}
Using these functions in Supplementary Equation~(\ref{4}) gives the exact avalanche shape
\begin{equation}
A(t) = G'(1,t)-1 = \frac{(1-\mu^2)\left(1-e^{-\mu t}\right)\left(1-e^{-\mu (T-t)}\right)}{\mu\left(1+\mu - (1-\mu)e^{-\mu T}\right)},
\end{equation}
which is clearly invariant under interchange of $t$ and $T-t$ at a fixed value of $T$. Hence the average avalanche shape is symmetric about the point $t=T/2$ for all values of $\mu$ (cf.~Fig.~2 of the main text).

In the critical limit $\mu\to 0$, Supplementary Equation (\ref{Q1}) simplifies to
\begin{equation}
Q(t) = \frac{t}{2+t}
\end{equation}
and the average avalanche shape is a parabolic function of $t$ with maximum at $t=T/2$:
\begin{equation}
A(t) =\frac{t}{2+T}(T-t) = \frac{T^2}{2+T}\frac{t}{T}\left(1-\frac{t}{T}\right),
\end{equation}
The corresponding variance of the avalanche shape is found from Supplementary Equation~(\ref{vf}) to be
\begin{equation}
\frac{T^4}{2(2+T)^2}\left(\frac{t}{T}\right)^2\left(1-\frac{t}{T}\right)^2,
\end{equation}
and so the coefficient of variation for this critical case is $\text{CV}=1/\sqrt{2}$, independent of $t$ and $T$.
}
\section*{Supplementary Note 4: Derivation of non-terminating avalanche shape}\label{app:NT}

We consider here the average shape of all avalanches that have not terminated by the observation time $T$. We use the notation introduced in Supplementary Note~2, but here we define (for a fixed observation time $T>0$) the avalanche path probability $\tilde{\pi}_n(t)$ to be the probability that $n$ particles are alive at time $t$ ($0\le t \le T$), conditioned on there being exactly one particle alive at time $0$ and {at least one} particle alive at time $T$:
\begin{equation}
\tilde{\pi}_n(t)=\text{Prob}\left\{ Z(t)=n | Z(0)=1 \text{ and } Z(T)>0\right\}.
\end{equation}
Using the Markov property, we have
\begin{equation}
\tilde{\pi}_n(t) \propto P_{1 n}(t) \left[ 1- P_{n 0}(T-t)\right] = P_{1 n}(t) \left[ 1- P_{1 0}(T-t)^n\right].
\end{equation}
The correctly normalized avalanche path probability distribution is therefore
\begin{align}
\tilde{\pi}_n(t) &= \frac{ P_{1 n}(t) \,\left[ 1- P_{1 0}(T-t)^n\right]}{\sum_n P_{1 n}(t)\,\left[ 1- P_{1 0}(T-t)^n\right]} \nonumber \\
&=  \frac{ P_{1 n}(t) \,\left[ 1- P_{1 0}(T-t)^n\right]}{1-Q(T)},
 \label{1b}
\end{align}
where we have used the Markov property of Supplementary Equation~(\ref{Markov}) in the denominator to write
\begin{equation}
\sum_m P_{1 m}(t) P_{1 0}(T-t)^m = F\left(P_{1 0}(T-t),t\right) = F\left(F(0,T-t),t\right) = F(0,T)=Q(T).
\end{equation}

The average non-terminating avalanche shape is then given by
\begin{align}
A_{NT}(t) & = \sum_n n \tilde{\pi}_n(t) \nonumber  \\
&=\sum_n \frac{n\, P_{1 n}(t) \,\left[ 1- Q(T-t)^n\right]}{1-Q(T)} \nonumber \\
&= \frac{F'(1,t)-Q(T-t)F'\left(Q(T-t),t\right)}{1-Q(T)}.
\end{align}
Using the relation given by Supplementary Equation~(\ref{2}) for the second term in the numerator we get
\begin{equation}
A_{NT}(t) = \frac{F'(1,t) - Q(T-t) \frac{f(Q(T))-Q(T)}{f(Q(T-t))-Q(T-t)}}{1-Q(T)} \label{ANT1},
\end{equation}
so it remains only to find an expression for $F'(1,t)$.

Recall that $F'(1,t)=\sum_n n P_{1 n}(t)$, which is the expected number of particles alive at time $t$. This quantity therefore measures the average number of events observed at time $t$ across an ensemble comprised of {all} avalanches (with no conditioning on avalanche duration); as noted in the main text, it provides a simple measure of whether an observed ensemble of avalanches is from a subcritical, critical, or supercritical system. Differentiating the backward Kolmogorov equation (the second of Supplementary Equations~(\ref{KB})) with respect to $s$ and setting $s$ equal to $1$ gives the ordinary differential equation satisfied by $F'(1,t)$:
\begin{equation}
\frac{d}{d t}F'(1,t) = \left[ f'(1)-1\right] F'(1,t),
\end{equation}
which has solution
\begin{equation}
F'(1,t) = \exp\left[(\xi-1)t\right] \label{numevents}
\end{equation}
(recall $\xi=f'(1)$ is the branching number). Therefore, critical systems (those with $\xi=1$) have a constant expected number of (unconditioned) events per unit time, while in sub- (respectively, super-) critical systems, the number of events decays (resp., grows) exponentially with the age $t$ of the avalanches.

Substituting from Supplementary Equation~(\ref{numevents}) into Supplementary Equation~(\ref{ANT1}) gives a formula for the average non-terminating avalanche shape in terms of the extinction fraction $Q(t)$; the asymptotic analysis of this shape is given in Supplementary Note~6.

\section*{Supplementary Note 5: Discrete-time branching processes} \label{app:discrete}
Here we consider discrete-time (Galton-Watson) branching processes instead of the continuous-time process used in Supplementary Notes~2--4. The arguments leading to Supplementary Equation~(\ref{B}) remain valid in the discrete-time case, but we cannot now use the Kolmogorov equations to further simplify Supplementary Equation~(\ref{B}). However, the Galton-Watson process has the governing equation \cite{AthreyaNeybook}
\begin{equation}
F(s,t+1)=f\left(F(s,t)\right), \label{d1}
\end{equation}
from which we can derive the relations
\begin{equation}
F'(s,t+1)=f'\left(F(s,t)\right) F'(s,t) \label{d2}
\end{equation}
and
\begin{equation}
F''(s,t+1) = f'\left( F(s,t)\right) F''(s,t) + f''\left(F(s,t)\right) \left[F'(s,t)\right]^2 \label{d3},
\end{equation}
with initial conditions
\begin{equation}
F(s,0)=s, \quad F'(s,0)=1, \quad F''(s,0)=0.
\end{equation}
For any chosen values of the variable $s$ and duration $T$, we can iterate Supplementary Equations~(\ref{d1})--(\ref{d3}) to find the values of $F(s,t)$, $F'(s,t)$ and $F''(s,t)$ at any (integer) time value $t$ between $0$ and $T$; evaluating at the appropriate value of $s$ then enables us to calculate the average avalanche shape $A(t)=G'(1,t)-1$ from Supplementary Equation~(\ref{B}).

\begin{figure}
\centering
\epsfig{figure=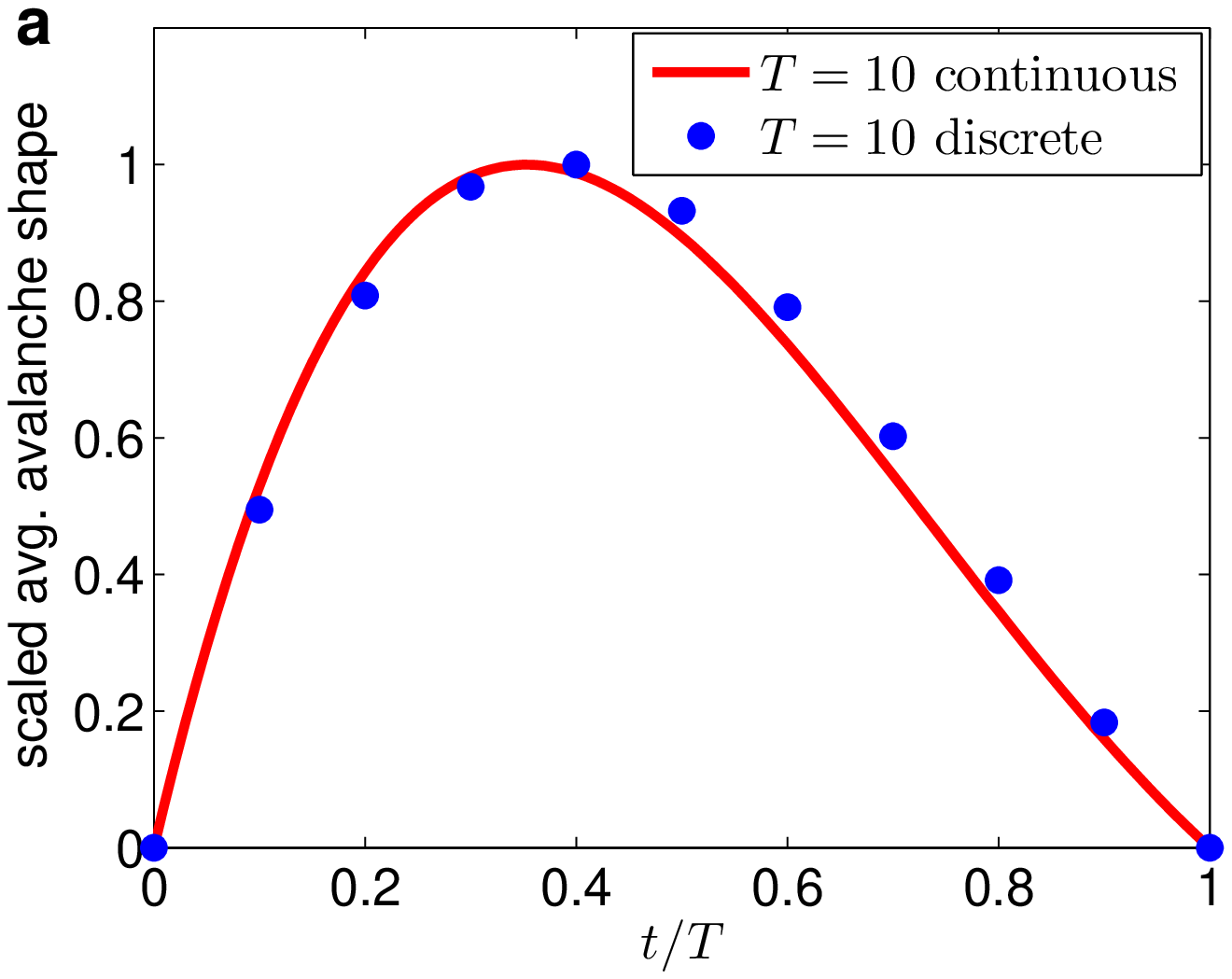,width=6.0 cm} 
\epsfig{figure=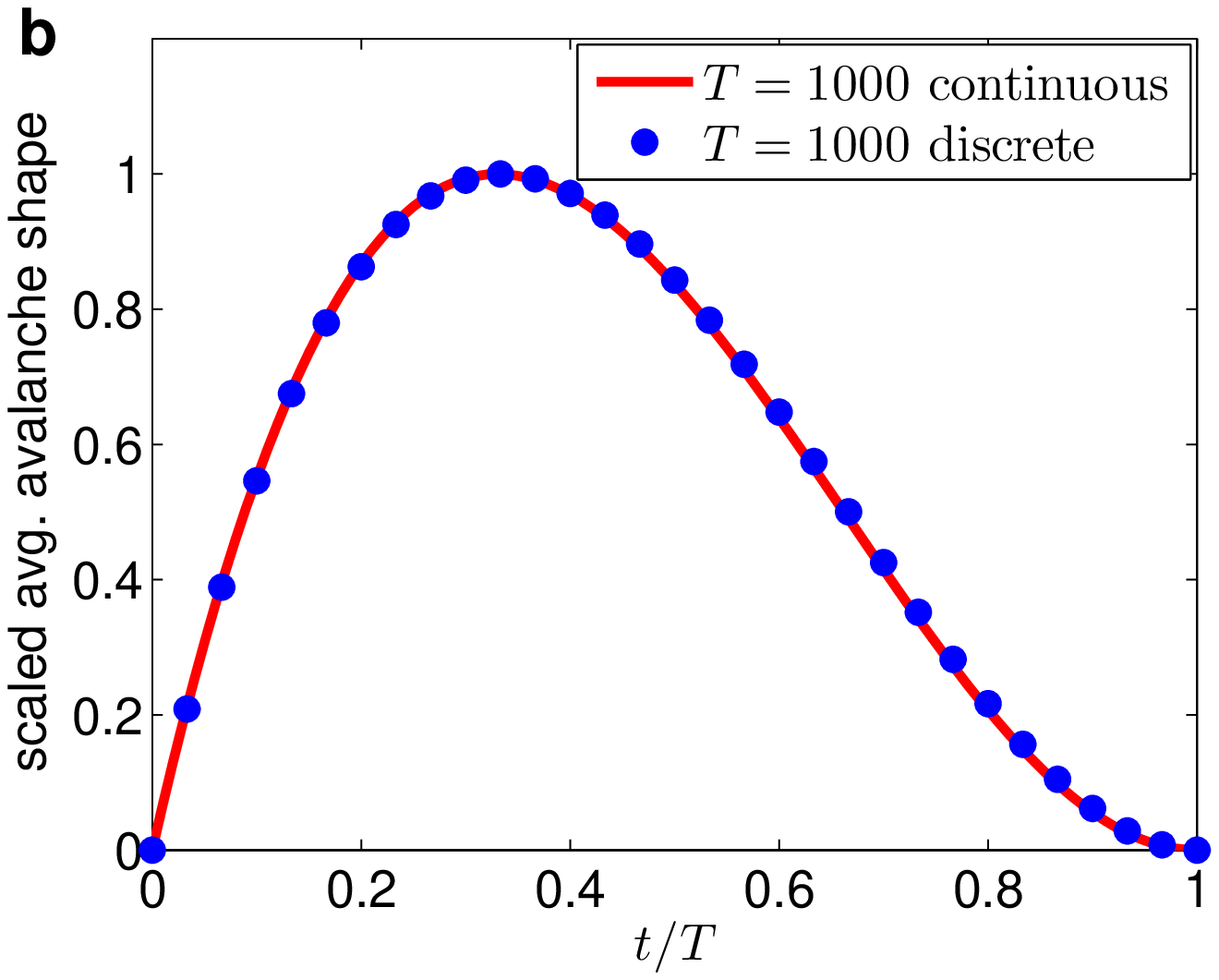,width=6.0 cm}
\caption{{\bf $\mathbf{\left|\right.}$ Comparing discrete-time and continuous-time results. }Average avalanche shapes at criticality ($\xi=1$) from the continuous-time approach of equation~(4) of the main text (curves) and the discrete-time approach of Supplementary Equations~(\ref{d1})--(\ref{d3}) (symbols). The offspring distribution $q_k$ is power-law, with exponent $\gamma=2.5$. The results of both approaches are very similar even for relatively short-duration avalanches ($T=10$ in panel (a)), and are almost indistinguishable for long-lived avalanches ($T=1000$ in panel (b)).}\label{figdiscrete}
\end{figure}

Supplementary Figure~\ref{figdiscrete} demonstrates that the discrete-time  and continuous-time (from equation~(4) of the main text) formulations give very similar results for the average avalanche shapes; accordingly, we concentrate in the main text on the continuous-time case, for which analytical insights (e.g., asymptotic behaviour) are easier to obtain.

\section*{Supplementary Note 6: Large-$T$ asymptotics at criticality}\label{app:asymptotics}
To consider the asymptotic behaviour of the avalanche shape for a general offspring distribution, we begin with the large-time asymptotics of $Q(t)$. Supplementary Equation (\ref{5}) can be rewritten as
\begin{equation}
\int \frac{d Q}{f(Q)-Q} = \int dt \label{int},
\end{equation}
and the large-$t$ asymptotic behaviour corresponds to $Q$ values near 1. Expanding $f(Q)$ near $Q=1$, and assuming that $q_k\sim D k^{-\gamma}$ as $k\to \infty$ with $2<\gamma<3$ yields \cite{Wilfbook}
\begin{equation}
f(Q)\sim 1- f'(1) (1-Q) + D \Gamma(1-\gamma) (1-Q)^{\gamma-1} \quad\text{ as } Q\to1^- \label{PL}
\end{equation}
{where $\Gamma$ is the Gamma function.}
If the offspring distribution has a finite second moment then Supplementary Equation~(\ref{PL}) is replaced by
\begin{equation}
f(Q)\sim 1- f'(1) (1-Q) + \frac{f''(1)}{2} (1-Q)^{2} \quad\text{ as } Q\to1^- \label{finite2mom}.
\end{equation}
In the critical case $f'(1)=1$ and substituting from Supplementary Equation~(\ref{PL}) into Supplementary Equation~(\ref{int}) gives
\begin{equation}
\int \frac{dQ}{D \Gamma(1-\gamma)(1-Q)^{\gamma-1} }= \int dt \nonumber\\
\end{equation}
which integrates to
\begin{equation}
1-Q \sim C_1 t^{-\frac{1}{\gamma-2}} \quad \text{ as }t\to\infty \label{asy},
\end{equation}
where $C_1=\left[D (\gamma-2)\Gamma(1-\gamma)\right]^{-\frac{1}{\gamma-2}}$. (If $f''(1)$ is finite, then Supplementary Equation (\ref{asy}) still holds with $\gamma$ replaced by 3 and $C_1$ replaced by $2/f''(1)$.)

Differentiating Supplementary Equation (\ref{PL}) (or Supplementary Equation (\ref{finite2mom})) and substituting for $Q(t)$ from Supplementary Equation (\ref{asy}) leads to \begin{equation}
f'(Q(t)) \sim 1- C_2 \, t^{-1} \quad\text{ as } t\to \infty \label{asy2}
\end{equation}
and hence Supplementary Equation (\ref{4}) yields the asymptotic avalanche shape, in the limit $T\to\infty$ and $T-t\to\infty$, as
\begin{equation}
A(t)\sim C_3 \,T^\frac{1}{\gamma-2} \frac{t}{T}\left(1-\frac{t}{T}\right)^\frac{1}{\gamma-2},\label{asyshape}
\end{equation}
where $C_2$ and $C_3=(\gamma-1)/C_1$ are independent of $t$ and $T$, and this formula holds also for finite $f''(1)$ if $\gamma$ is replaced by 3. We note that the asymptotic avalanche shape is parabolic for $\gamma=3$ (i.e., if $f''(1)$ is finite), but its peak is at $t/T=(\gamma-2)/(\gamma-1)< 1/2$ for $\gamma$ values between 2 and 3.

Similarly, the asymptotic behaviour of the average non-terminating avalanche shape at criticality is determined from Supplementary Equation~(\ref{ANT1}) to be (in the limit $T\to \infty$ and $T-t\to \infty$):
\begin{equation}
A_{NT}(t)\sim \left\{ \begin{array}{cl}C_4 \,t\left(2-\frac{t}{T}\right)  & \text{ if } f''(1)\text{ is finite,}\\ \\
                  C_4 \, T^\frac{1}{\gamma-2}\left(1-\left(1-\frac{t}{T}\right)^\frac{\gamma-1}{\gamma-2}\right) & \text{ if } q_k \sim k^{-\gamma}\text{ as }k\to \infty, \text{ with }2<\gamma<3,
                    \end{array}
                    \right.\label{NTasy}
\end{equation}
where $C_4$ is independent of $t$ and $T$.

The asymptotic behaviour of the variance of the avalanche shape is found by analyzing Supplementary Equation~(\ref{vf}) in the same way, yielding
\begin{equation}
\text{var}\left(n(t)\right)\sim T^\frac{2}{\gamma-2}\frac{\gamma-1}{C_1^2}\frac{t}{T}\left(1-\frac{t}{T}\right)^\frac{2}{\gamma-2}\left(3-\gamma+\frac{t}{T}(\gamma-2)\right)
\end{equation}
(with $\gamma=3$ again recovering the finite-$f''(1)$ case). Using this formula and the asymptotics of the mean avalanche shape from Supplementary Equation~(\ref{asyshape}) gives the result for the coefficient of variation of the avalanche shape  in equation~(10) of the main text.

\section*{\bb{Supplementary Note 7: Effect of number of avalanches}}\label{app:numA}
\bb{In this Note we examine how the results of the numerical simulations are affected by the number $n_A$ of avalanches that are initiated in each simulation (and subsequently filtered by their duration $T$) in order to calculate the average avalanche shape and the other characteristic temporal shapes. Specifically, we focus on the neuronal dynamics example of Figure~6 of the main text. As described in the Methods section, the results in each panel of Fig.~6 are obtained by repeating 24 replica experiments, each of which calculates the temporal profiles of $n_A=10^7$ avalanches. The average avalanche shape and the other temporal shapes of interest are calculated in each experiment, and the plots in Fig.~6 show the average of these results over the 24 replica experiments, with error bars denoting the standard deviations of the measure over the set of 24 replicas.

In Supplementary Figure~\ref{fignA1} we perform similar experiments on the same networks as in Fig.~6, but we use an order of magnitude fewer avalanches in each experiment when calculating the temporal profiles, i.e., $n_A=10^6$. Similarly, in Supplementary Figure~\ref{fignA2} we reduce the number of avalanches simulated by another factor of 10, to $n_A=10^5$, and in Supplementary Figure~\ref{fignA3}, the value of $n_A$ is decimated again, to $n_A=10^4$. Comparing these figures to the original Figure~6, it is clear that using fewer avalanches leads to increased fluctuations from experiment to experiment, with consequent increases in error bars. Nevertheless, the main qualitative features that we highlight, such as the left skew of the average avalanche shape in the left column panels (corresponding to power-law networks), remain detectable even at lower values of $n_A$. Recall also that the set of avalanches that have not terminated by time $T$ is typically much larger than the set of avalanches that terminate precisely at time $T$, so we remark that the new temporal characteristics we suggest in the main text, such as the average non-terminating avalanche shape, may be more suitable for experimentalists testing for criticality than the traditional average avalanche shape. The left column panels of Supplementary Figure~\ref{fignA3} give some evidence in support of this claim: in this case, using a relatively small data set, we find no avalanches that terminate precisely at $T=40$ and so the black curve is absent in panel (a) of Supplementary Figure~\ref{fignA3}. However, the corresponding curve for the average non-terminating avalanche shape remains well-defined at $T=40$, see panel (c). Thus, the use of the new characteristic temporal shape, and its comparison with the analytical predictions in Supplementary Notes 4 and 6---especially the asymptotic result of Supplementary Equation (\ref{NTasy})---can prove useful to experimentalists seeking to quantify whether a system is in a critical state.}
\begin{figure}
\centering
\epsfig{figure=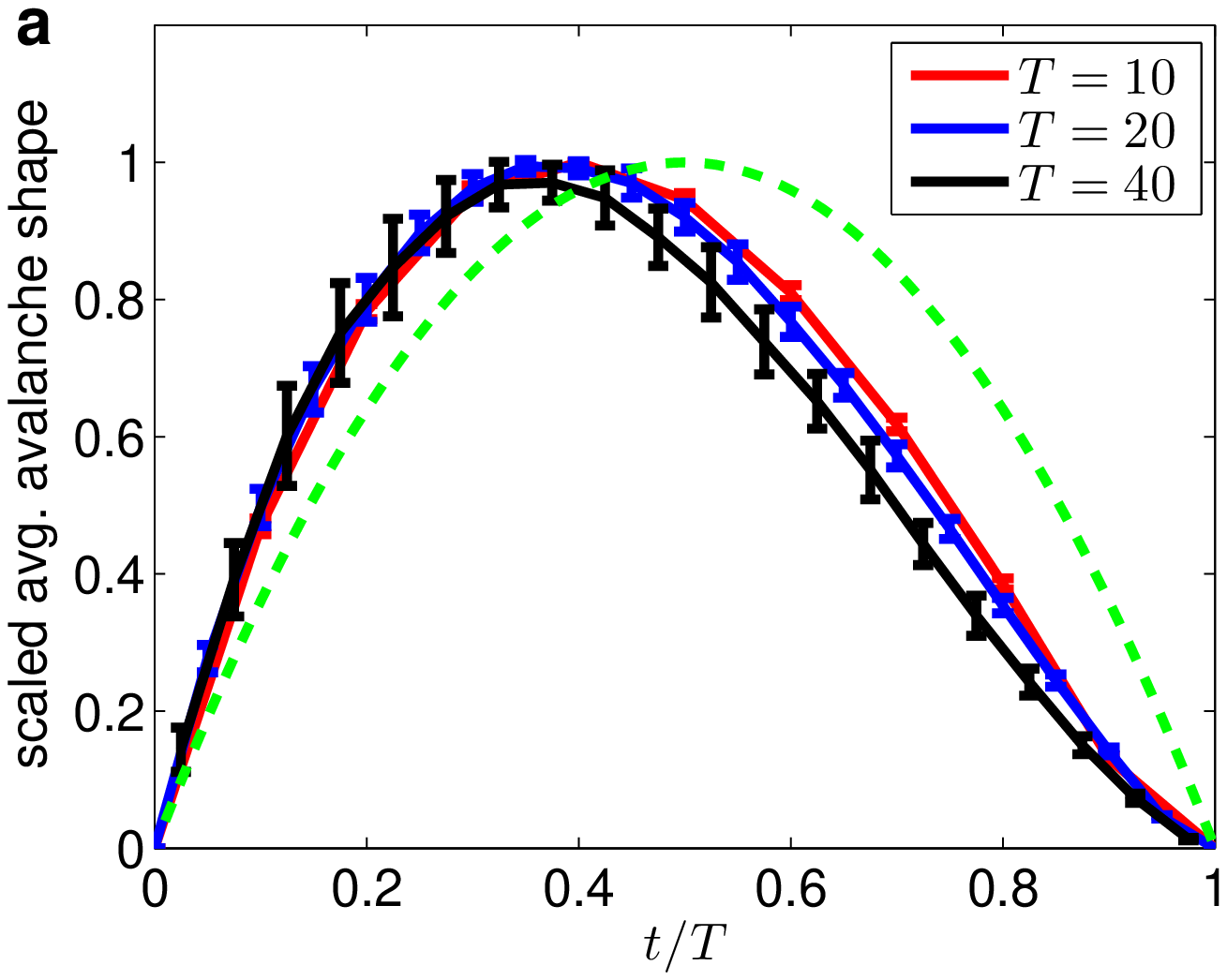,width=6.0 cm} 
\epsfig{figure=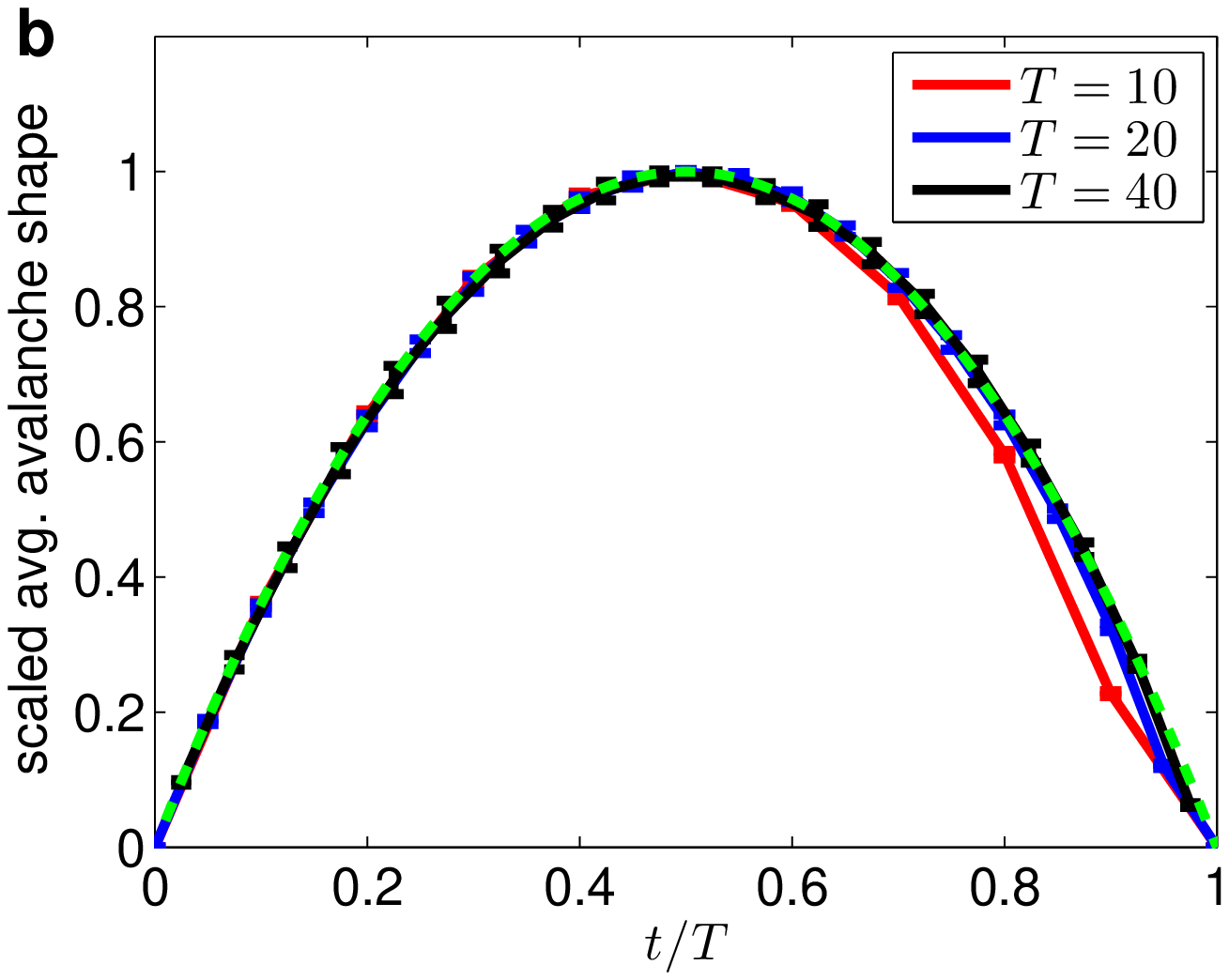,width=6.0 cm}
\epsfig{figure=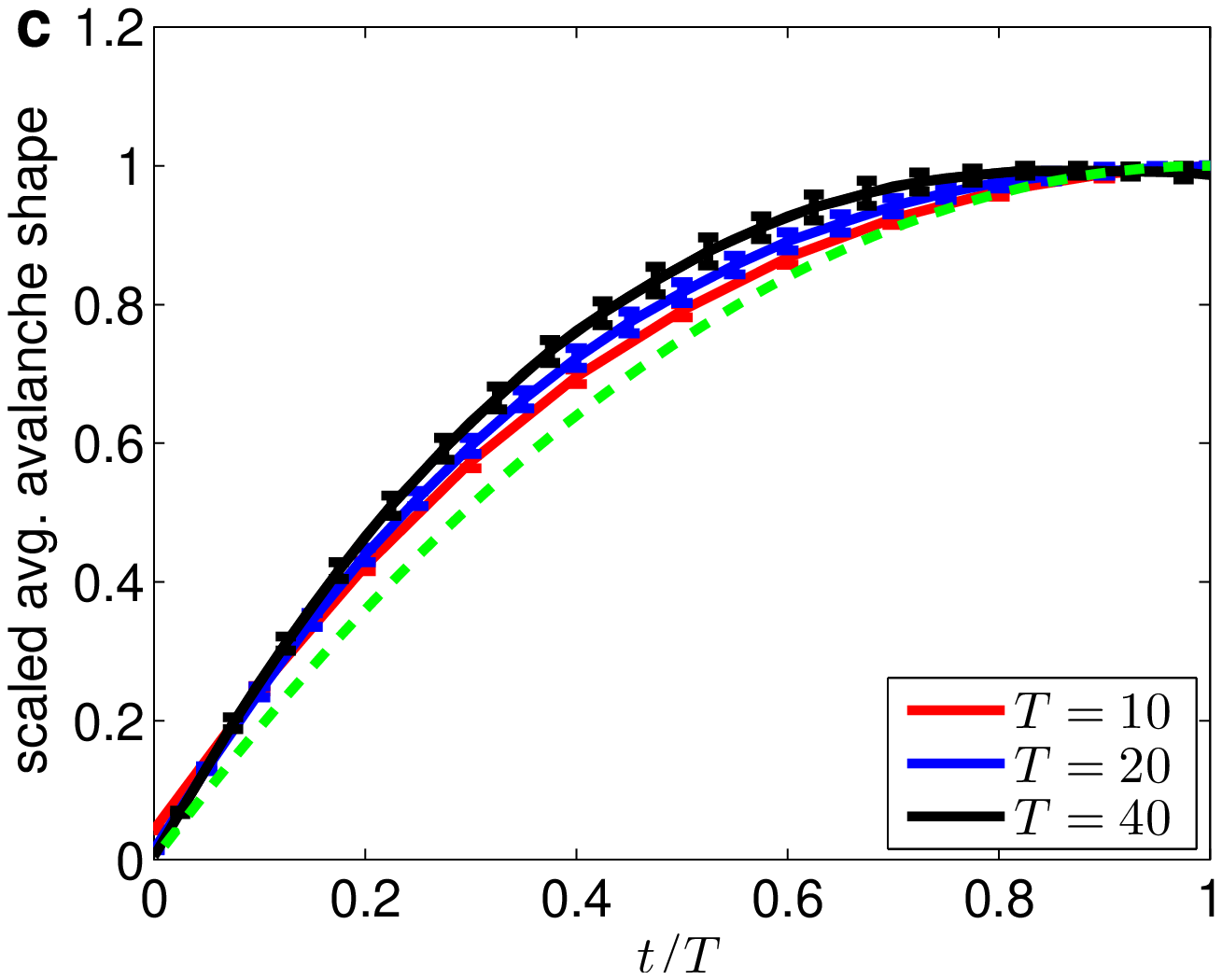,width=6.0 cm}
\epsfig{figure=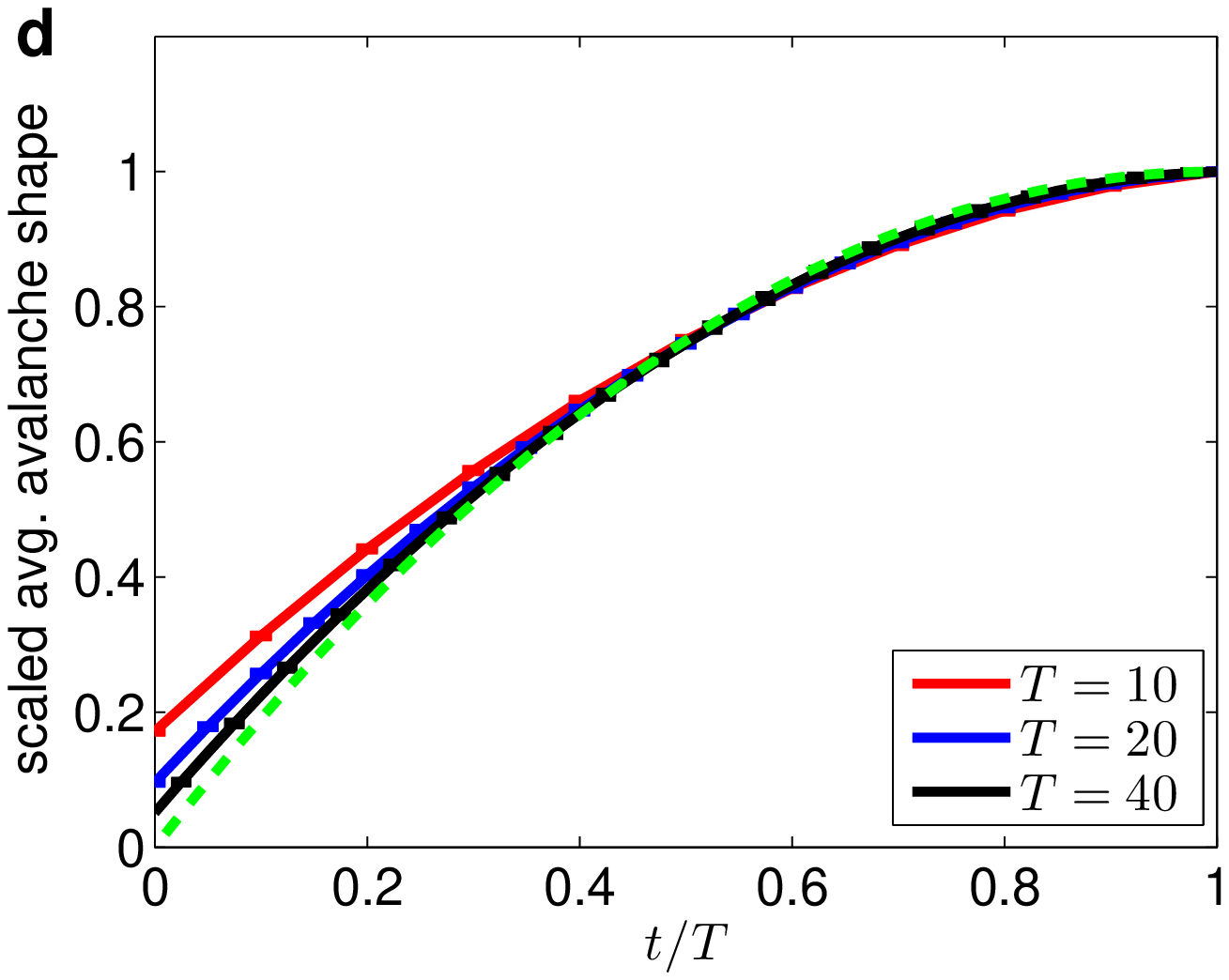,width=6.0 cm}
\epsfig{figure=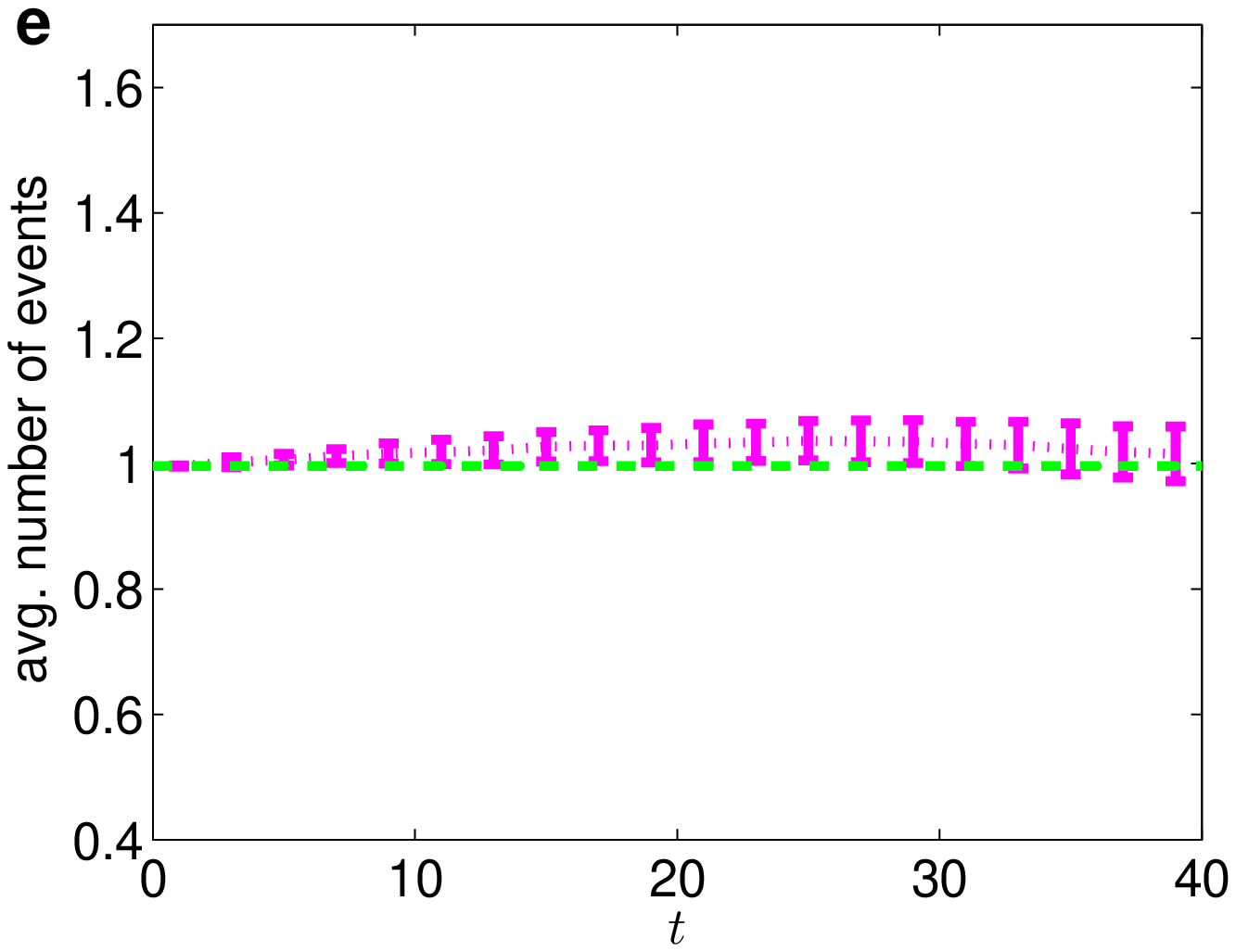,width=6.0 cm}
\epsfig{figure=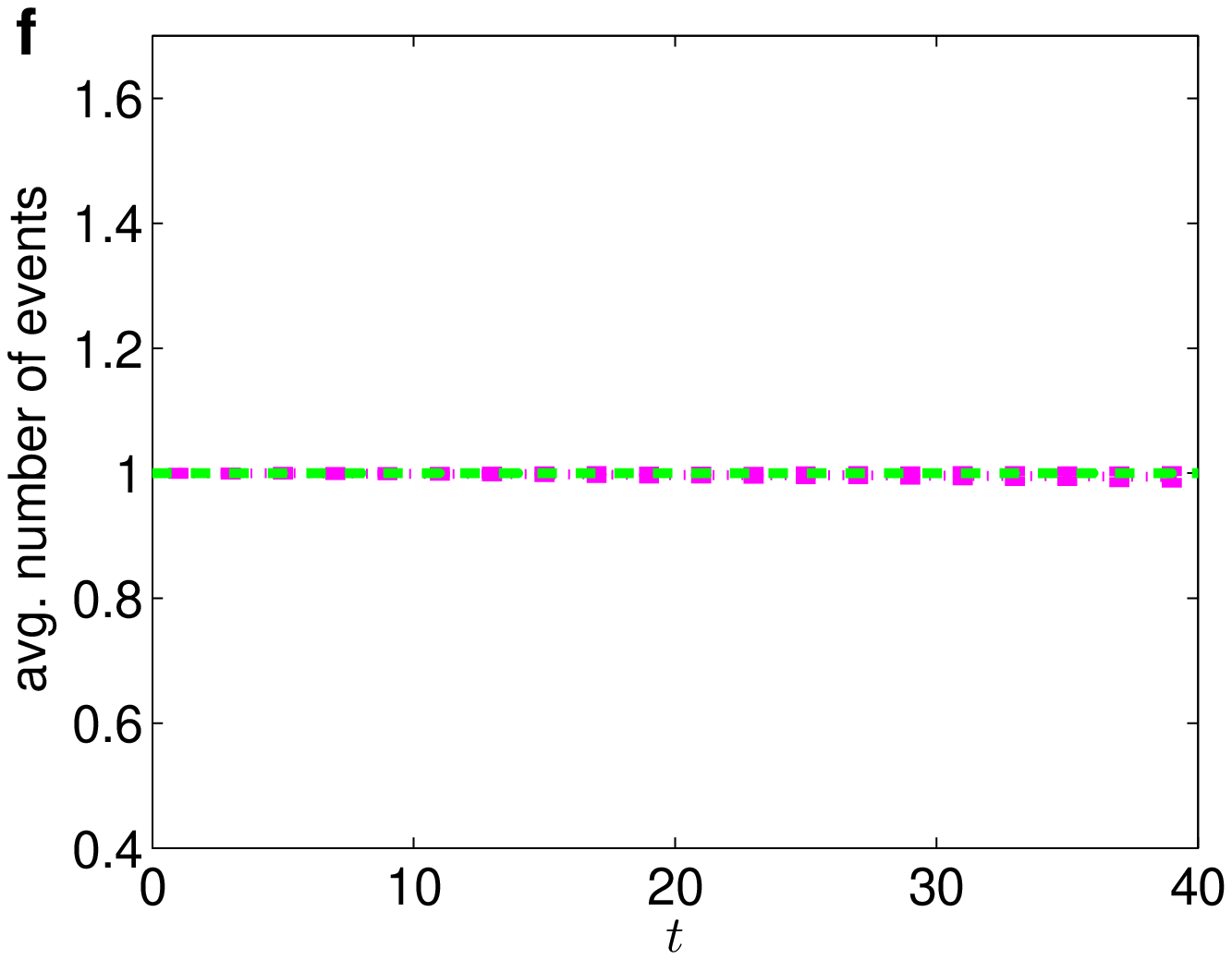,width=6.0 cm}
\caption{{{\bf $\mathbf{\left|\right.}$ Effect of number of avalanches: large ensemble. }} \bb{As Figure~6 of the main text, but using $n_A=10^6$ avalanches in each experiment.}
}\label{fignA1}
\end{figure}
\begin{figure}
\centering
\epsfig{figure=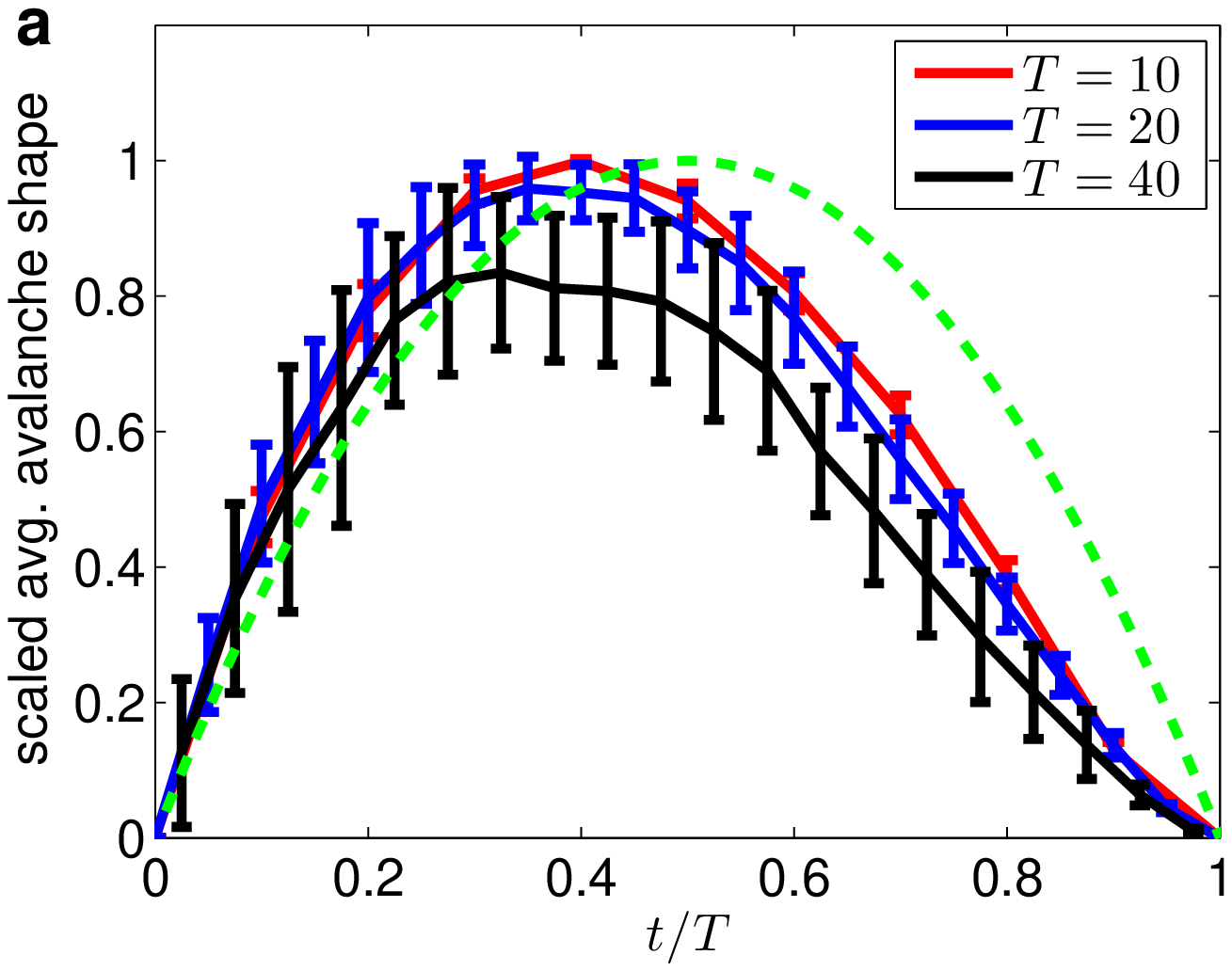,width=6.0 cm} 
\epsfig{figure=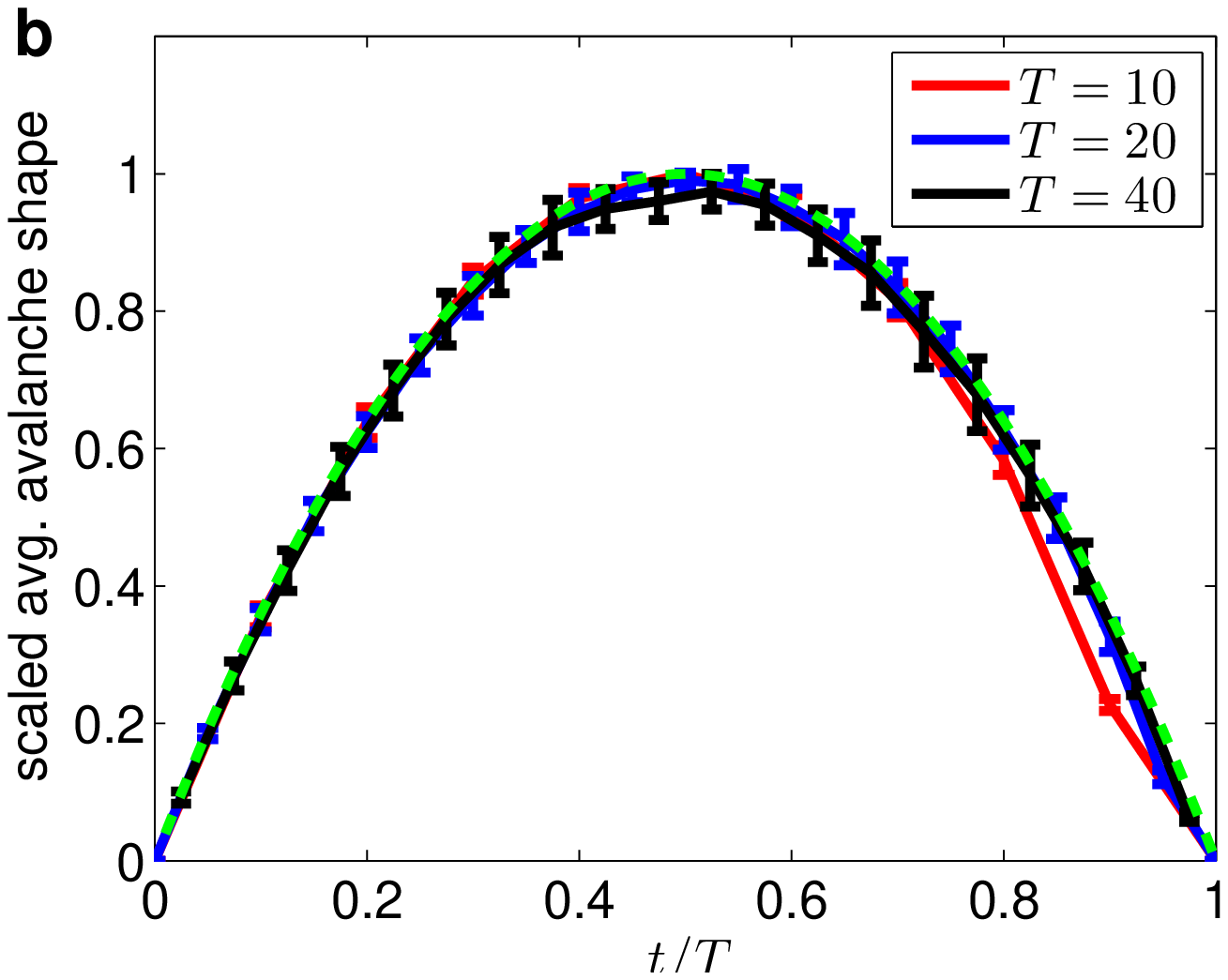,width=6.0 cm}
\epsfig{figure=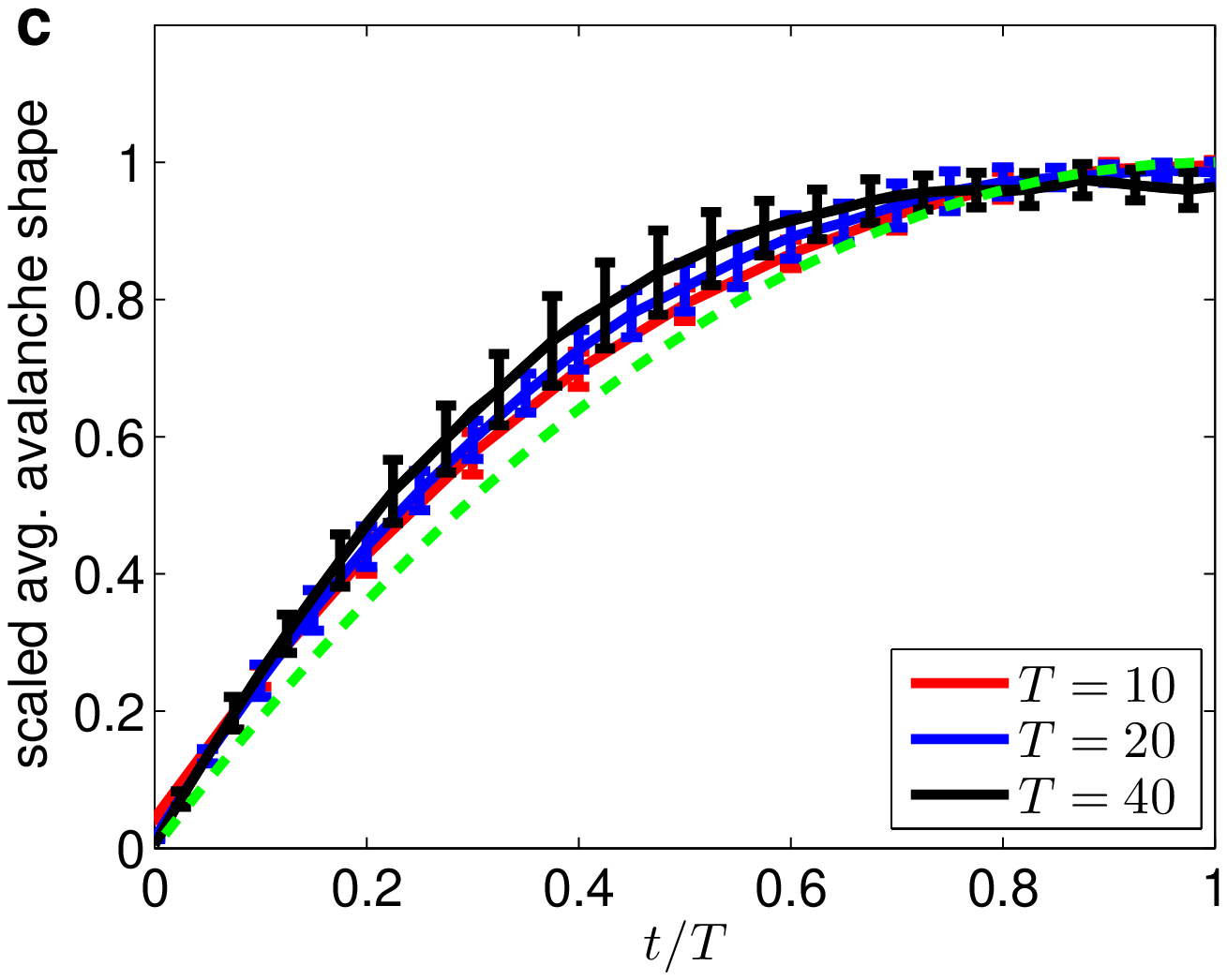,width=6.0 cm}
\epsfig{figure=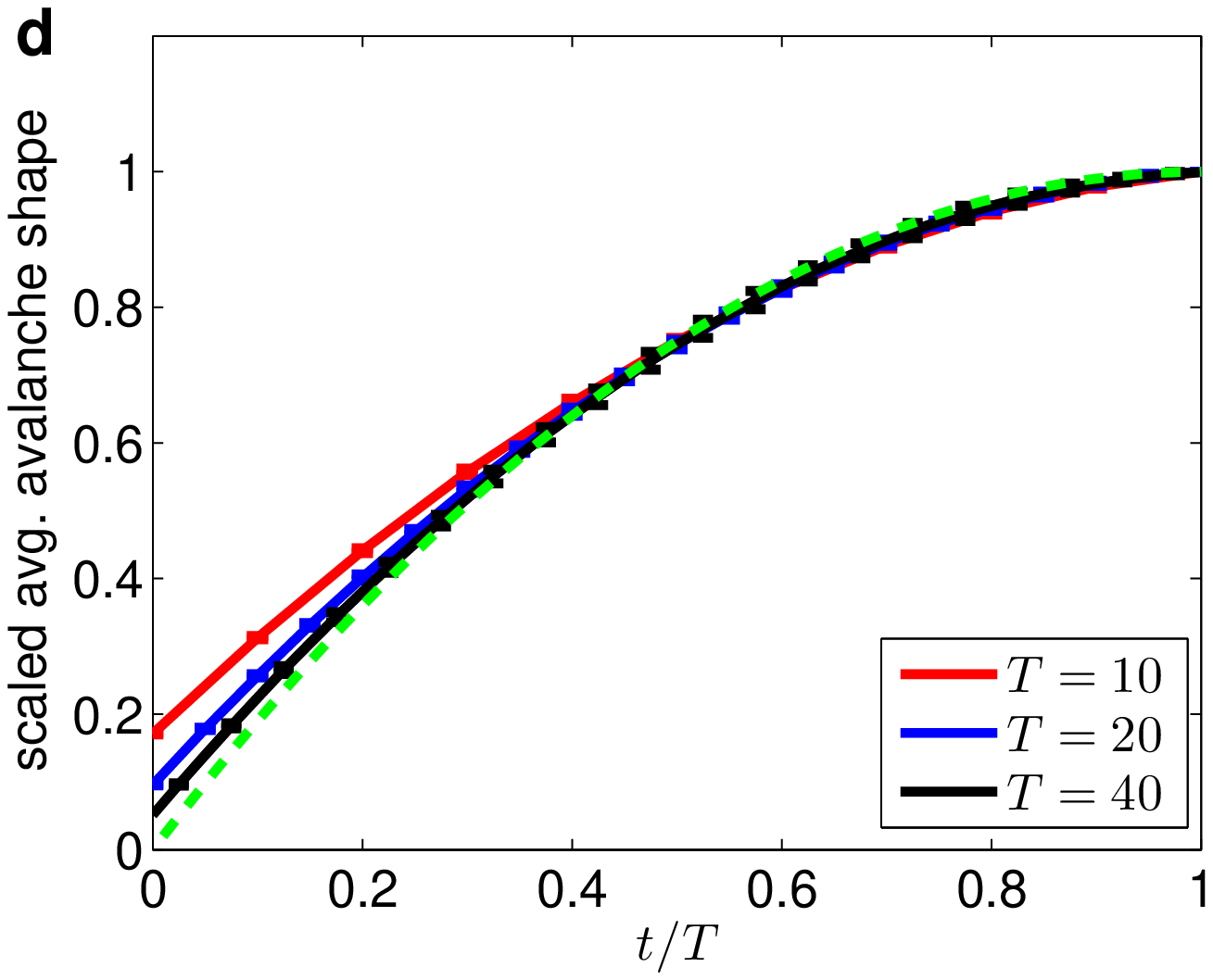,width=6.0 cm}
\epsfig{figure=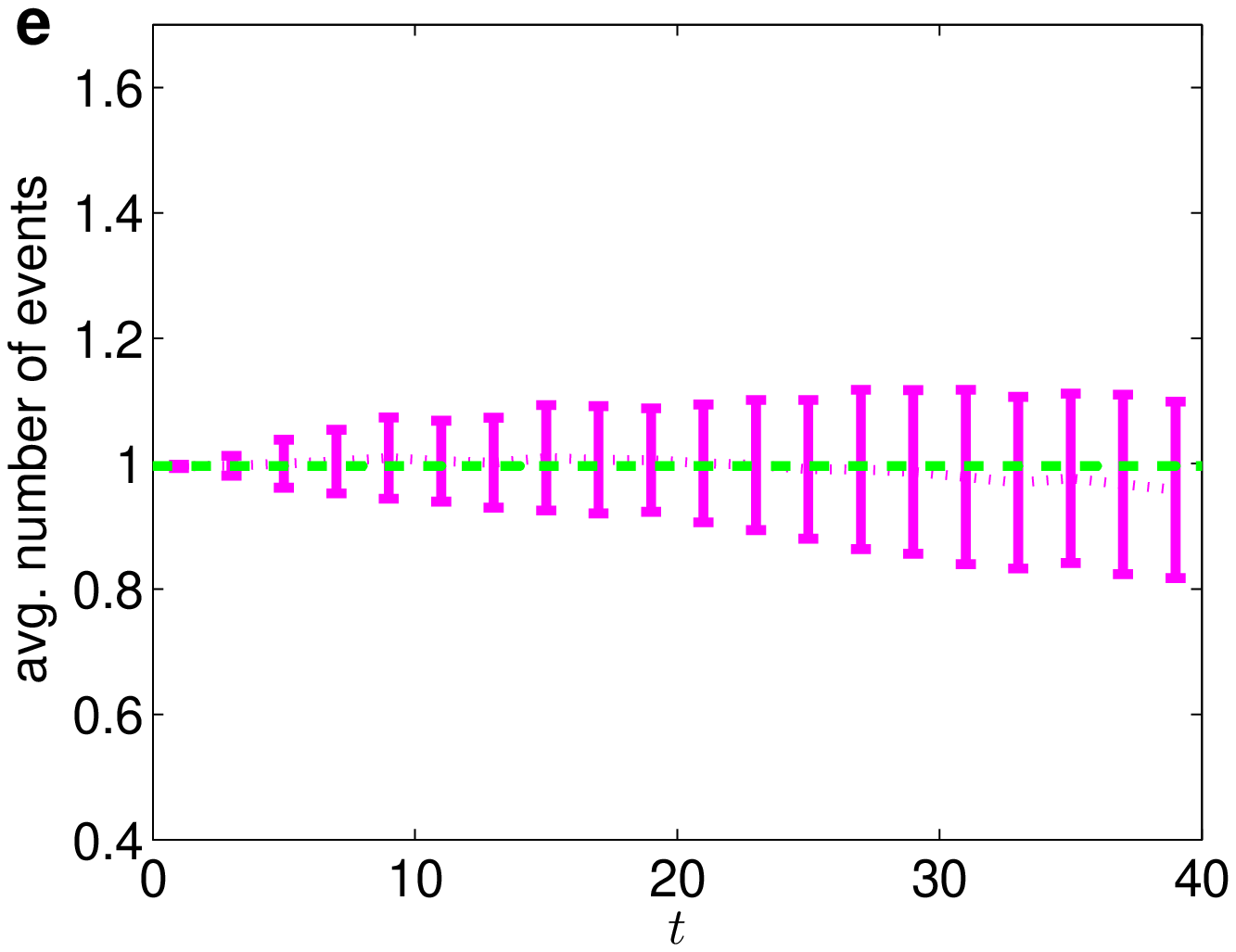,width=6.0 cm}
\epsfig{figure=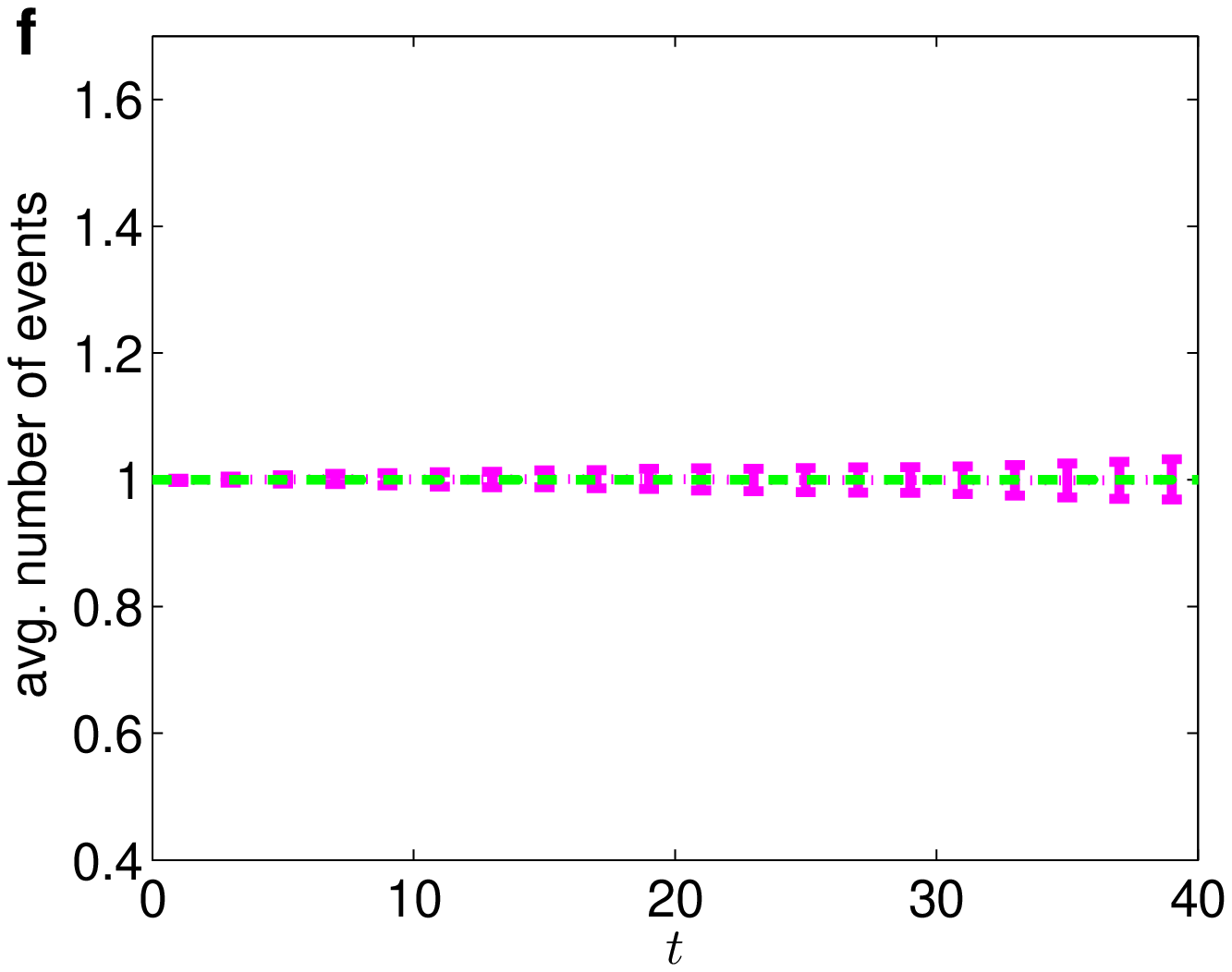,width=6.0 cm}
\caption{{\bf $\mathbf{\left|\right.}$ Effect of number of avalanches: medium ensemble. }\bb{As Figure~6 of the main text, but using $n_A=10^5$ avalanches in each experiment.}
}\label{fignA2}
\end{figure}
\begin{figure}
\centering
\epsfig{figure=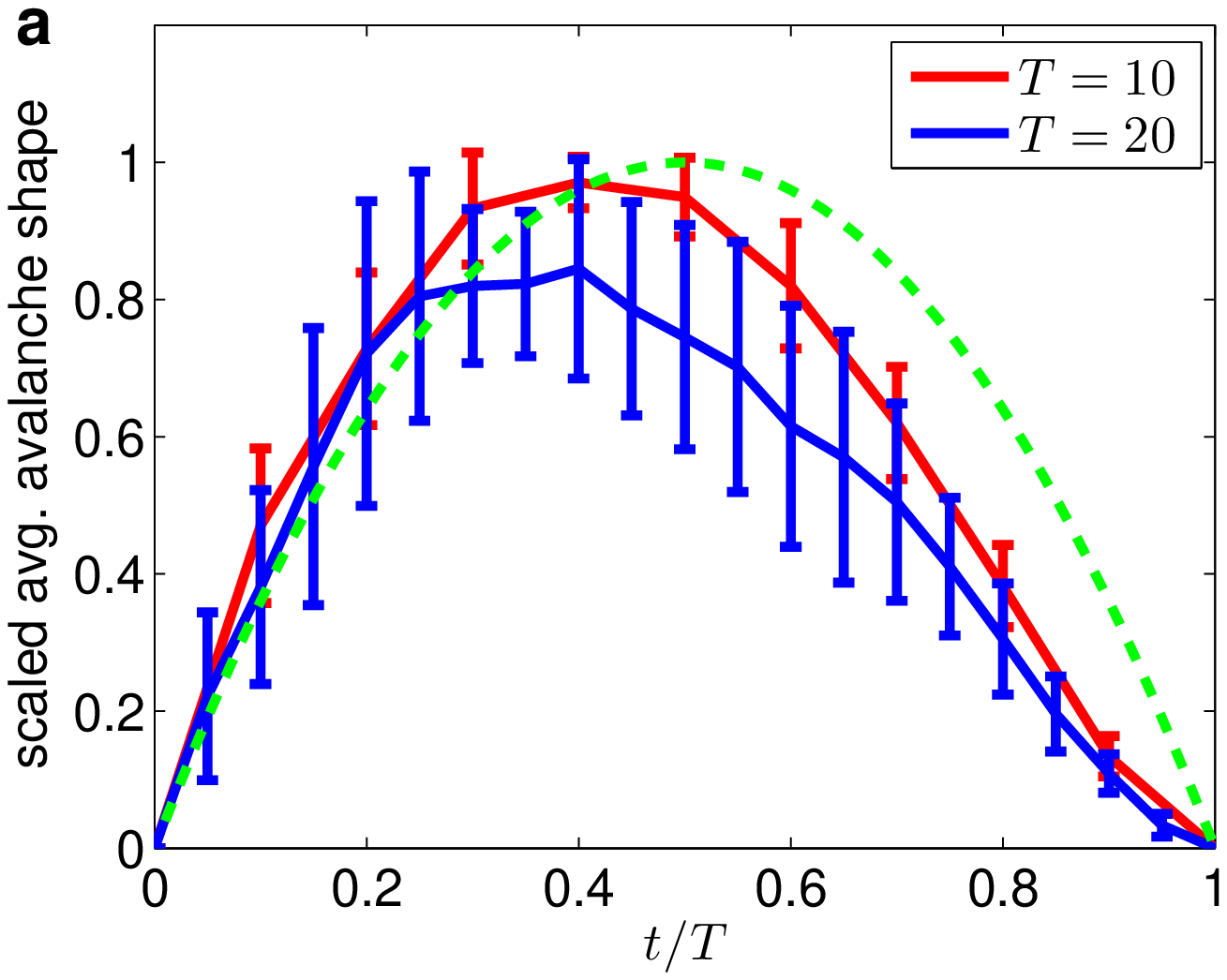,width=6.0 cm} 
\epsfig{figure=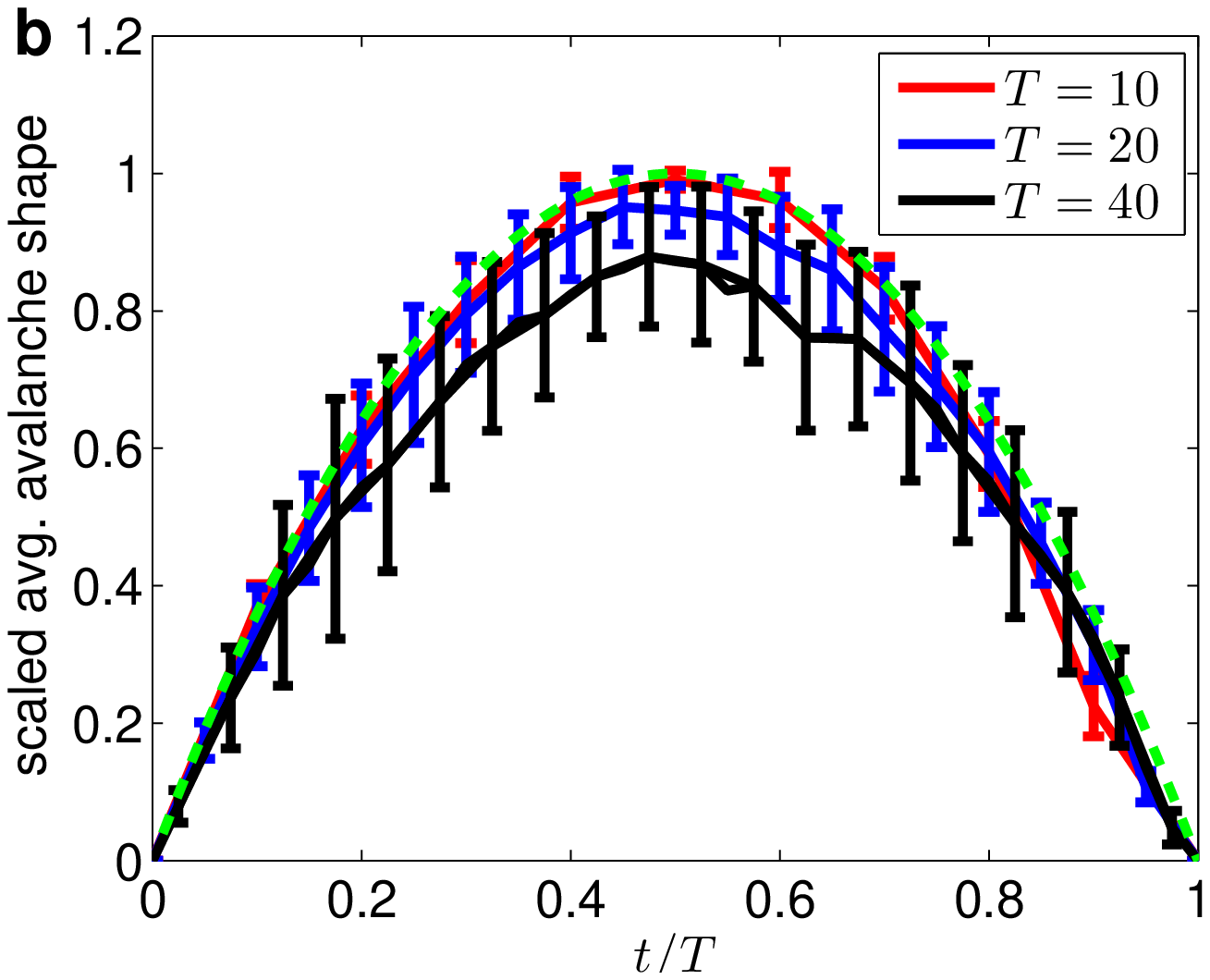,width=6.0 cm}
\epsfig{figure=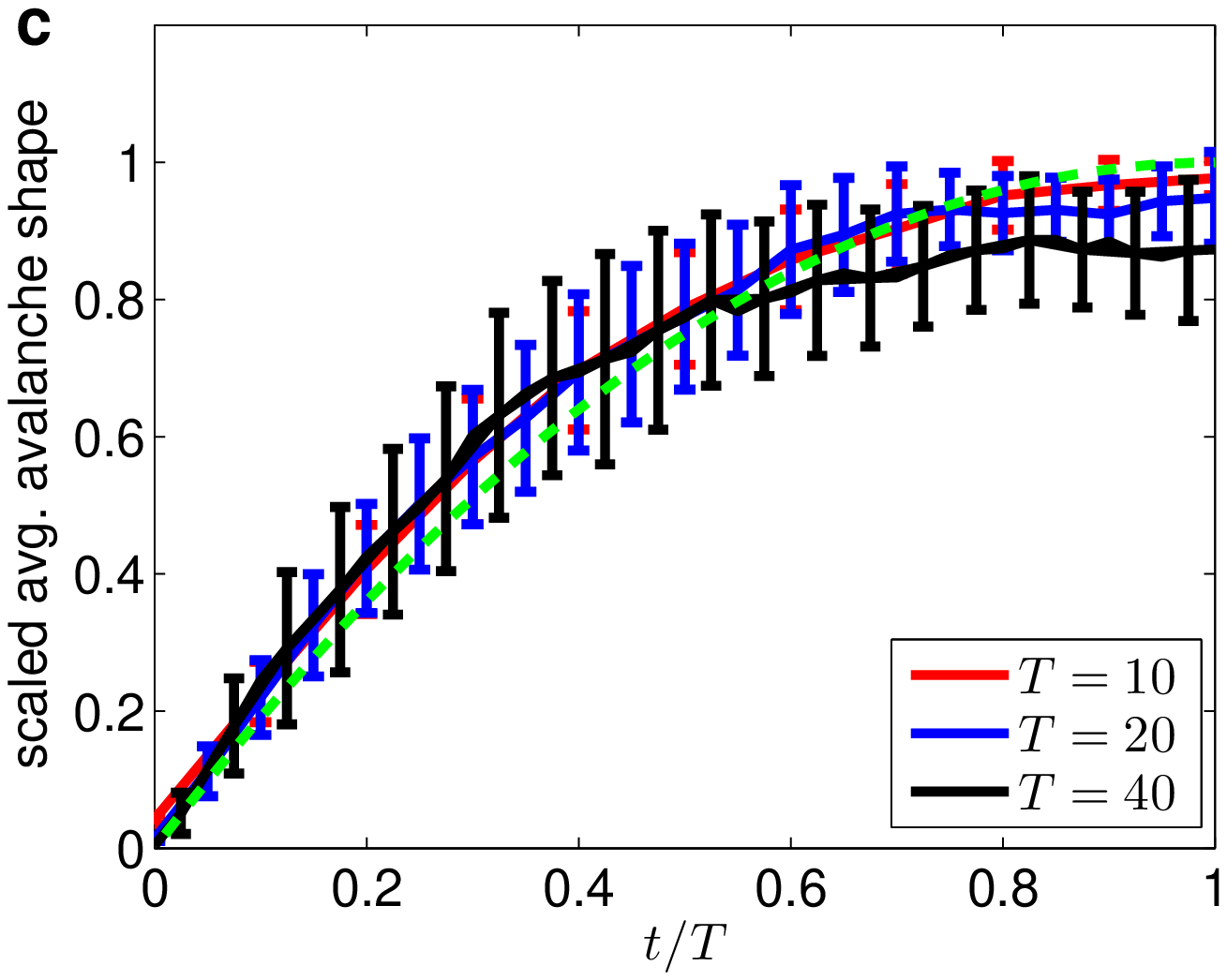,width=6.0 cm}
\epsfig{figure=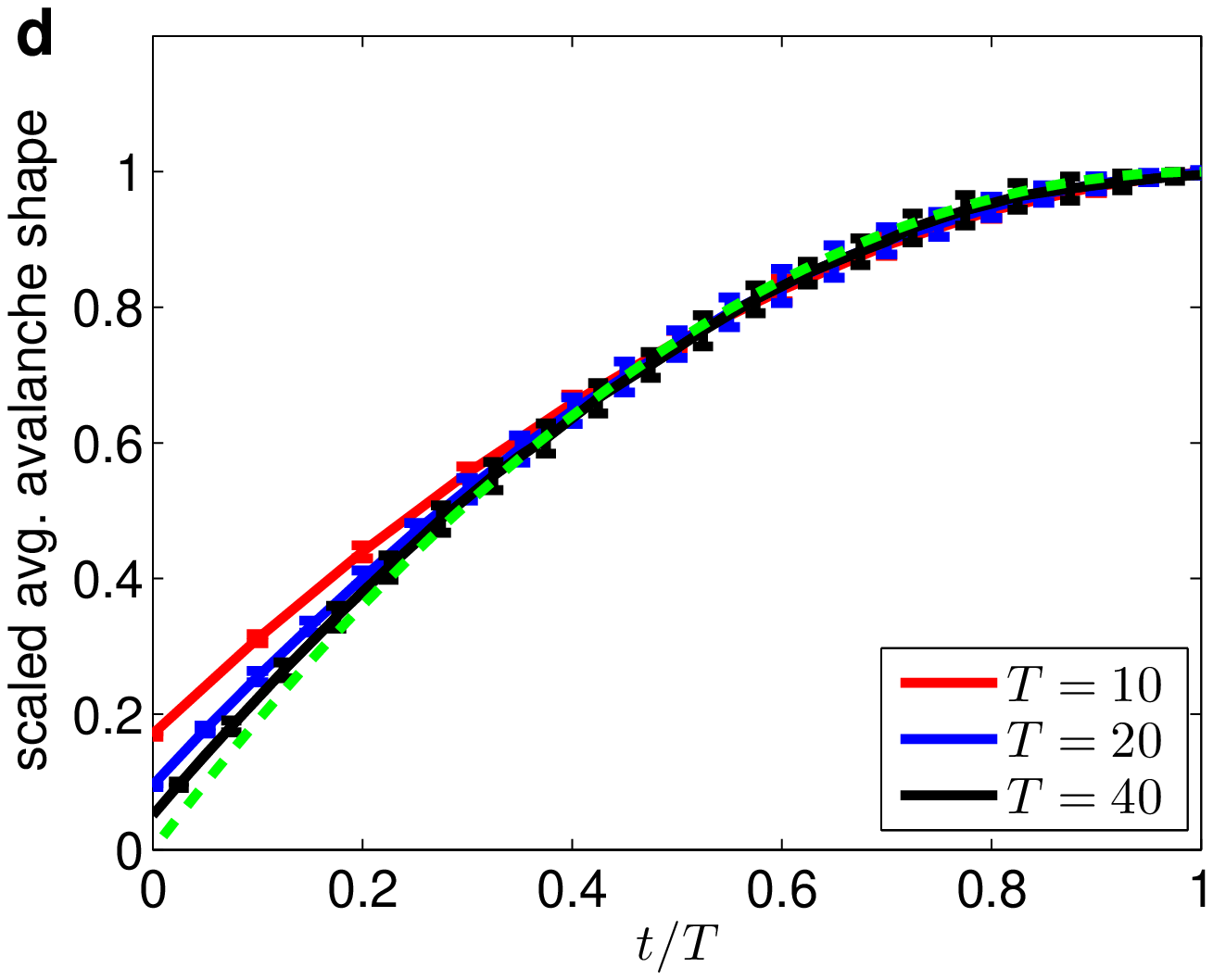,width=6.0 cm}
\epsfig{figure=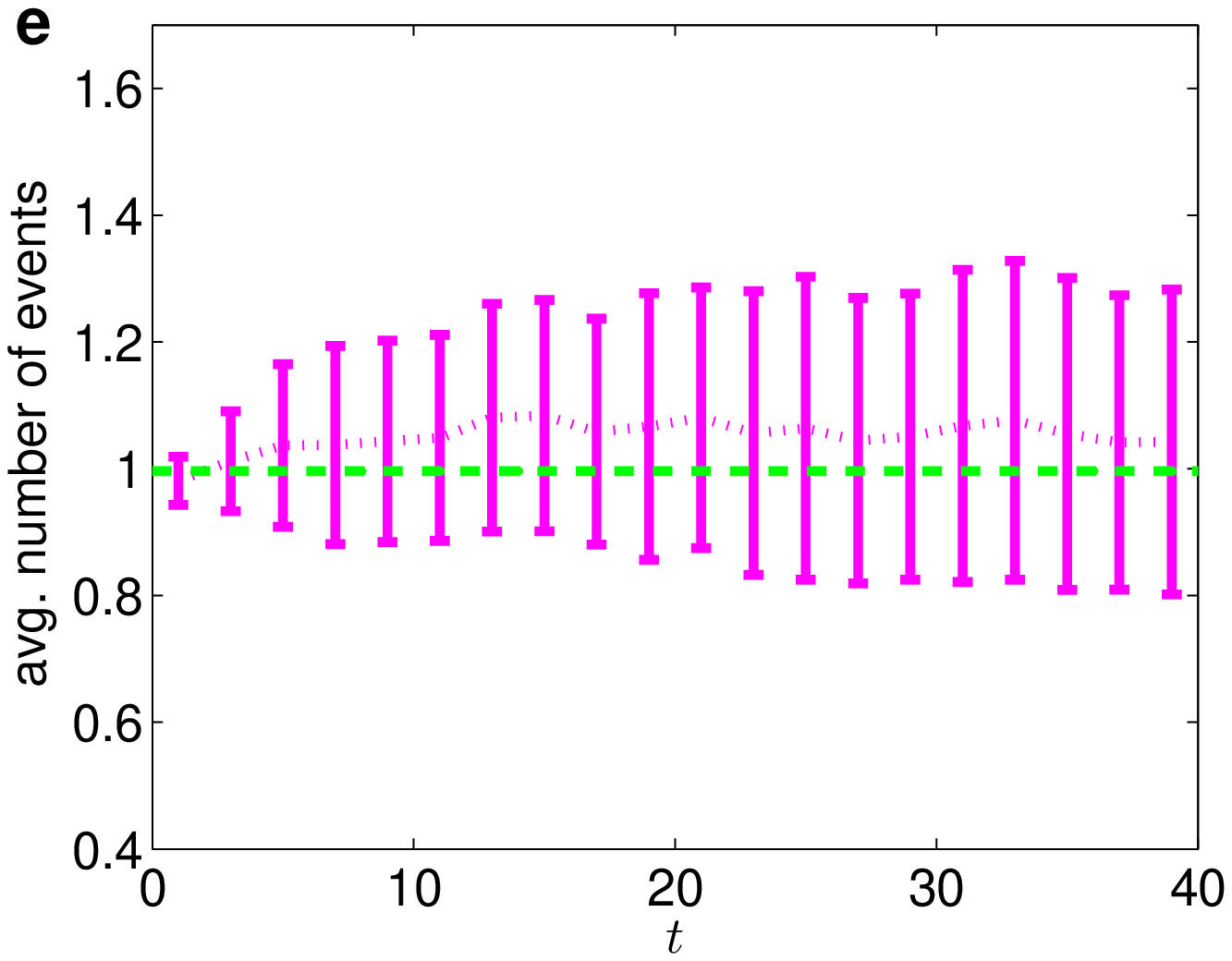,width=6.0 cm}
\epsfig{figure=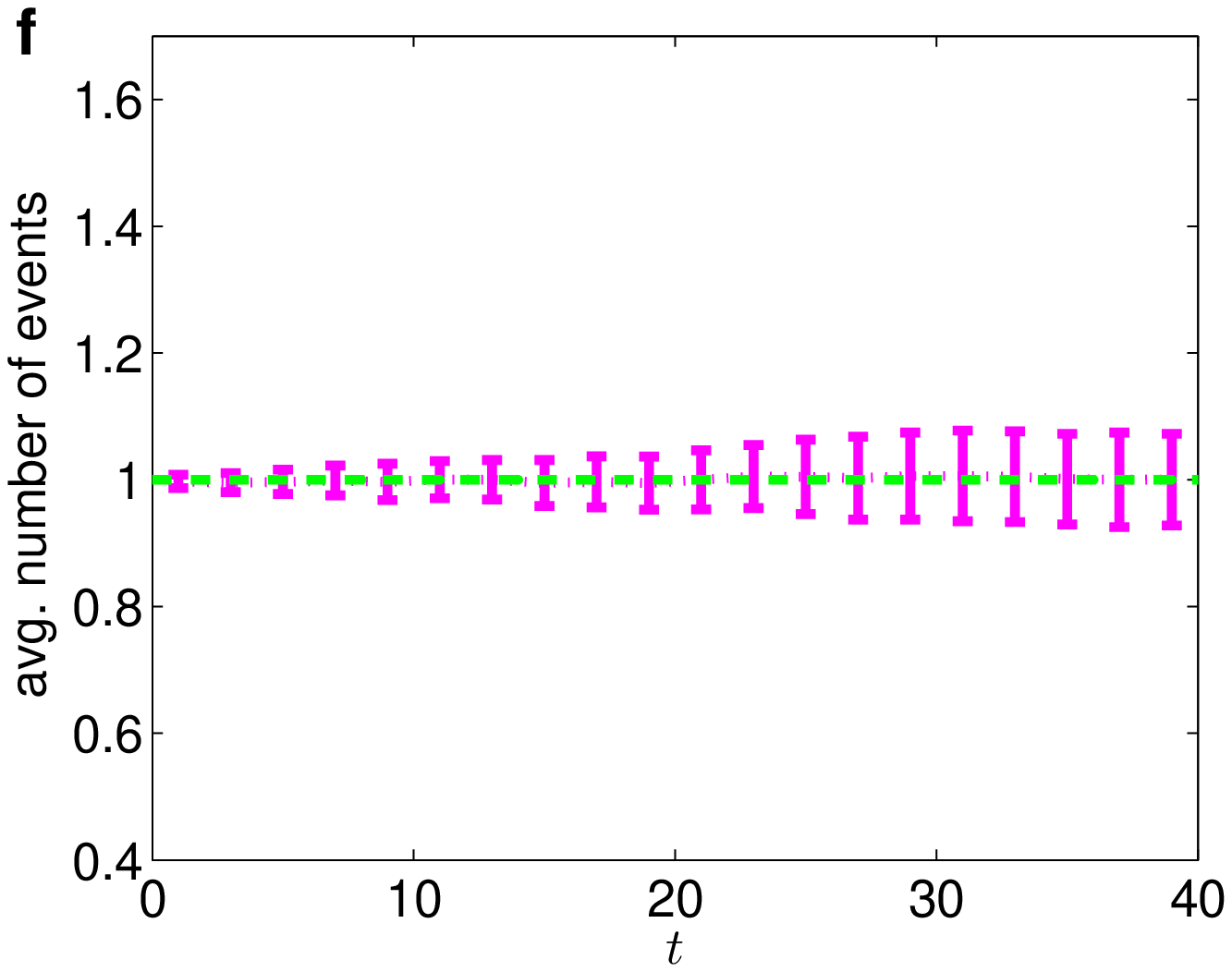,width=6.0 cm}
\caption{{\bf $\mathbf{\left|\right.}$ Effect of number of avalanches: small ensemble. }\bb{As Figure~6 of the main text, but using $n_A=10^4$ avalanches in each experiment.}
}\label{fignA3}
\end{figure}
\clearpage

\section*{\bb{Supplementary Note 8: Further information on Figure 8}}\label{app:realworld}
\bb{Figure 8 of the main text shows results of simulations of the meme propagation model of \cite{GleesonPRL14} on the empirical Twitter network of \cite{SNAPdata,SNAPref}. Despite the fact that the empirical network has many structural features that invalidate the assumptions of our mathematical derivation (e.g., reciprocal links, clustering, degree correlations), the results of Fig.~8 appear to obey the qualitative predictions of our theory, so demonstrating that the theory's usefulness extends beyond the regime of strict assumptions required for its derivation. However, a possible counterargument to this claim could point to the structural complexity of the empirical network and hypothesize that mesoscopic or macroscopic structural features (e.g., clustering, communities) may play a more important role in determining the temporal profiles of the cascade than the degree distribution does. In this Note we therefore explore (and refute) this hypothesis, using suitably rewired versions of the empirical network \cite{Karsai11}.

For the panels in the right-hand column of Fig. 8 (panels (b), (d) and (f)), we use a directed Erd\H{o}s-R\'{e}nyi network with the same mean degree as the empirical network but with all other structural characteristics randomised. We create this network by simply taking each directed link of the original network and reassigning both its end nodes to be randomly-chosen nodes. This rewired network is used in Fig.~8 to show that symmetric avalanche shapes are obtained when the degree distribution does not have fat tails, as expected from the theory.

In Supplementary Figure~\ref{figSNAPconfigmodel} we show simulation results for a differently rewired version of the original empirical network. The rewiring algorithm used here (called ``$p_{jk}$-rewiring'' in \cite{GleesonPRL14}) preserves the degree distribution of the original network, but removes (or drastically reduces) clustering, reciprocal links, and any macro-scale structure (similar to the rewiring algorithm used in \cite{Karsai11}). The algorithm begins by severing all links of the original network, but with each node retaining its number $j$ of ``in-stubs'' and $k$ of ``out-stubs'' that represent the endpoints of the deleted original edges. We then randomly select one out-stub and one in-stub from the entire set of stubs and create a new edge joining the selected nodes. By repeating this procedure, each time selecting from the set of unused out-stubs and in-stubs, we create a network with precisely the same $(j,k)$ distribution as the original network, but with randomised meso- and macro-scopic structure: for example, the fraction of reciprocal links is reduced from 48$\%$ in the original network to 1$\%$ in the rewired network. The panels of Supplementary Figure~\ref{figSNAPconfigmodel} show the results of $n_A=1.4\times 10^6$ avalanches over  $n_R=6$ replicas on the rewired network, and should be compared with panels (a), (c) and (e) of Fig. 8, which correspond to the original empirical network. The asymmetric avalanche shapes remain clearly visible in the rewired case of Supplementary Figure~\ref{figSNAPconfigmodel}, which is evidence that the fundamental driver of the avalanche shape asymmetry is indeed the degree distribution (as examined in our theory), while meso- and macro-scopic structural effects are of lesser importance.}
\begin{figure}
\centering
\epsfig{figure=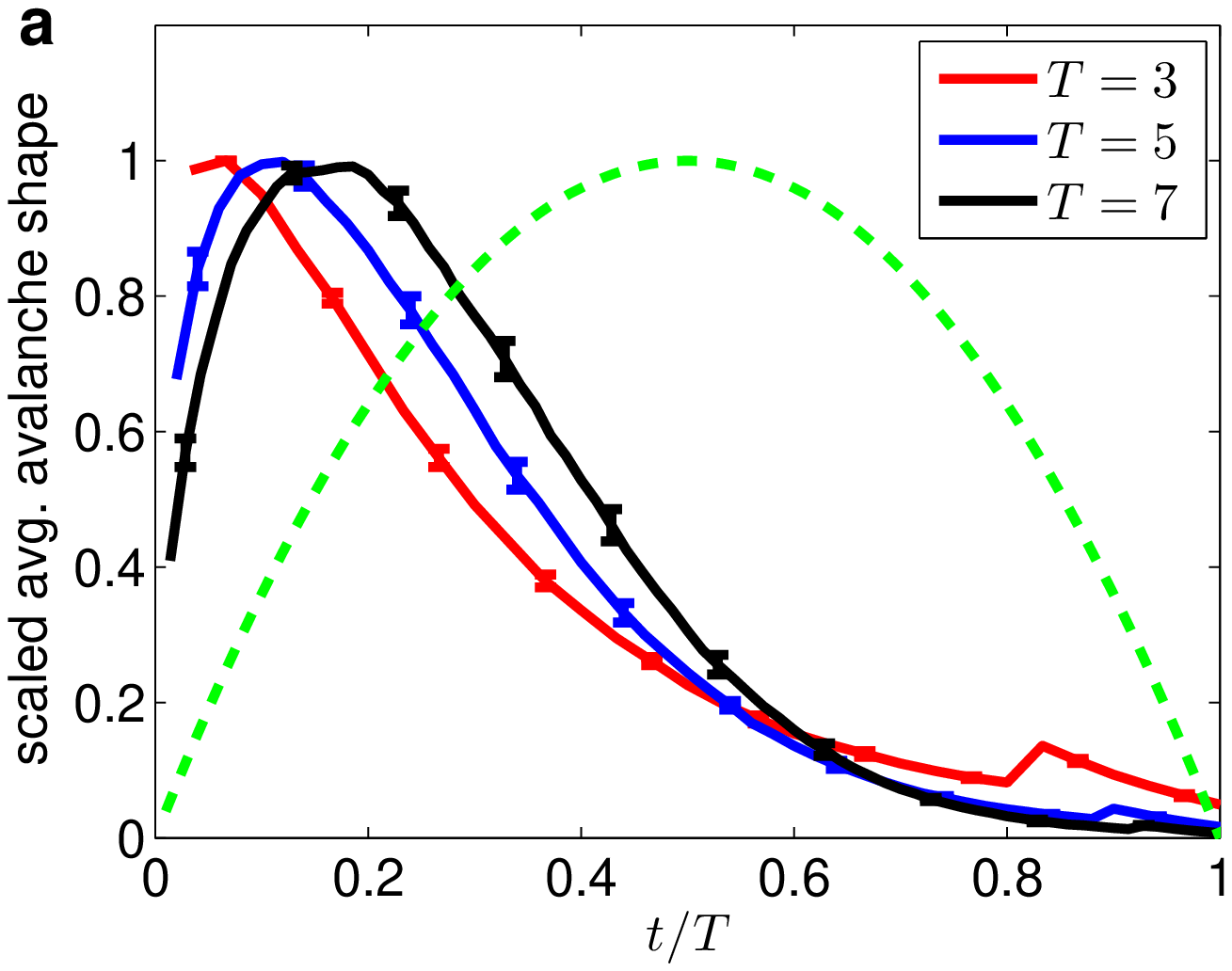,width=6.0 cm}\\ 
\epsfig{figure=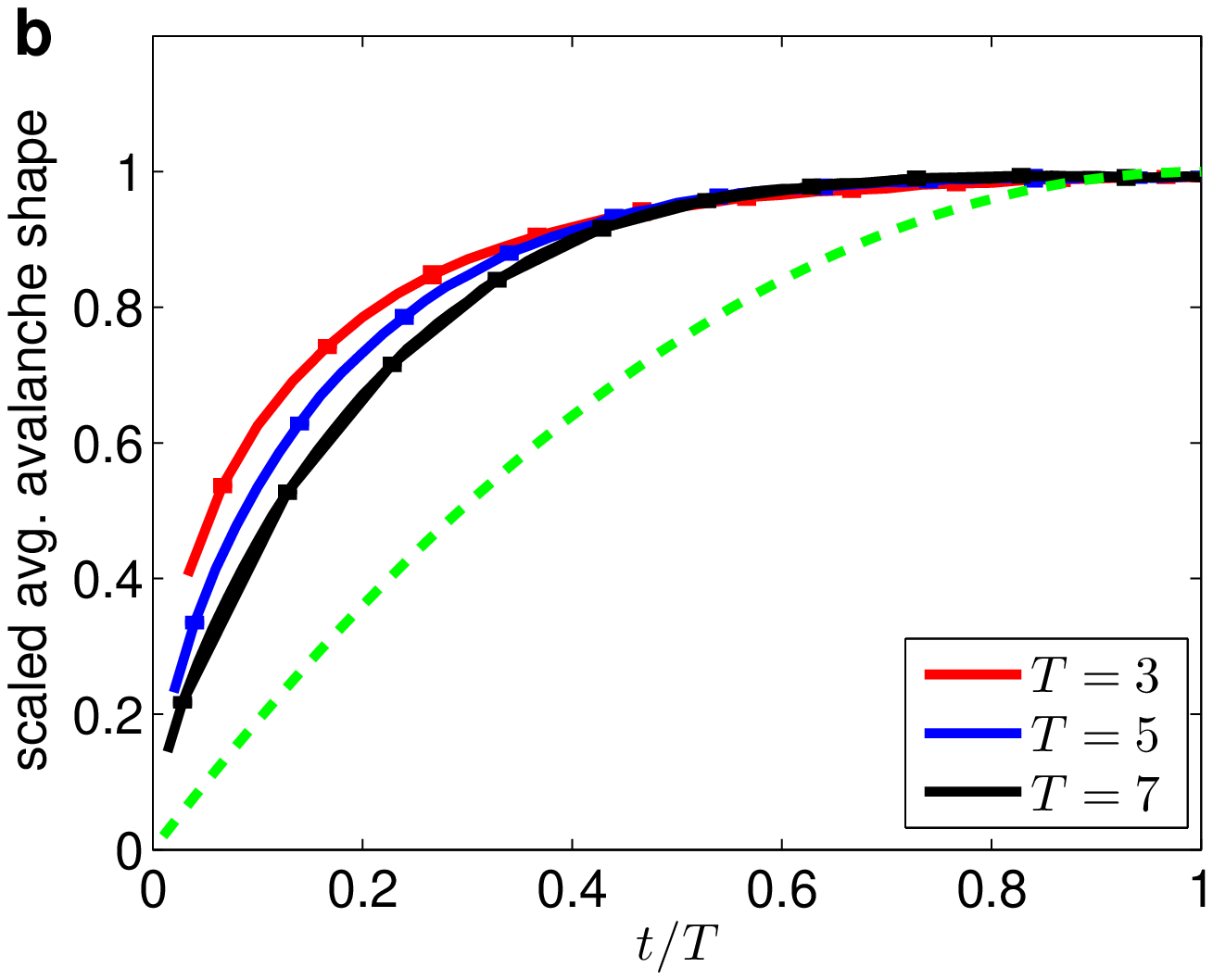,width=6.0 cm}\\
\epsfig{figure=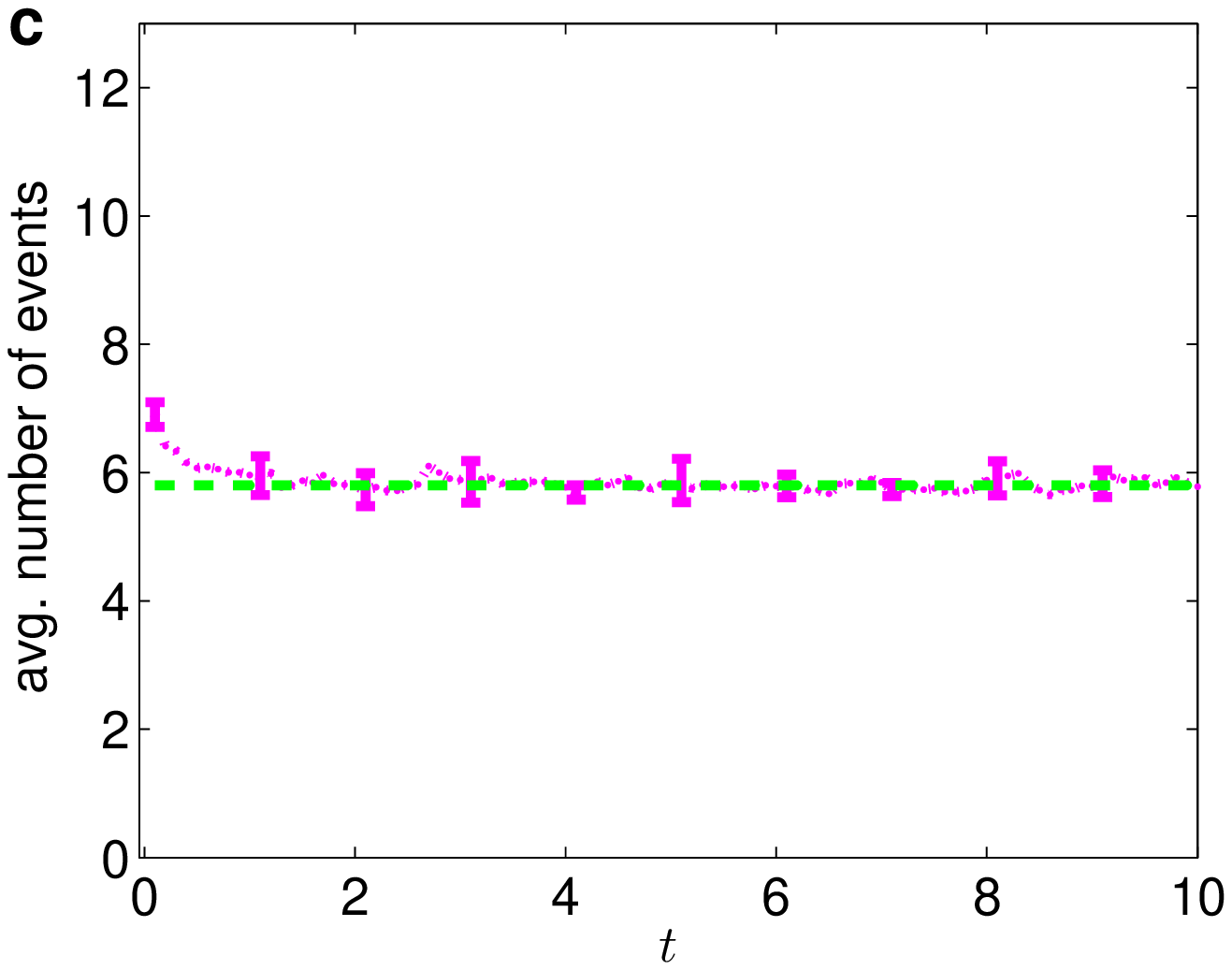,width=6.0 cm}
\caption{{\bf $\mathbf{\left|\right.}$ Cascades on rewired Twitter network.} \bb{As Figure 8 of the main text, but for simulations on a $p_{j k}$-rewired version of the empirical Twitter network, which retains the degree distribution of the original network but otherwise randomizes the structure. Note that the qualitative features seen in the left column of Fig.~8 remain apparent here, which is evidence that these features are principally controlled by the degree distribution of the network (as in our theory), rather than being caused by meso- or macro-scopic structure in the empirical network. See Methods (in main text) for definition of error bars. }}\label{figSNAPconfigmodel}
\end{figure}


\section*{\bb{Supplementary Note 9: Example of non-Markovian dynamics}}\label{app:nonMarkovian}
\bb{
The theoretical derivation of all our analytical results assume Markovian (Poissonian) temporal dynamics. As we note in the Discussion section, extending the theory to non-Markovian cascades is a significant challenge that lies beyond the scope of the current work. However, as a preliminary test of whether our results are robust to violations of the Markov assumption, in this Note we describe the results of numerical simulations of a non-Markovian extension of the meme-propagation model of \cite{GleesonPRL14}.

The original model of \cite{GleesonPRL14} is described in terms of discrete (but small) time steps of $\Delta t= 1/N$, where $N$ is the number of nodes in the network. For large $N$ (we use $N=10^5$ nodes), the discrete-time dynamics is a good approximation to Markovian continuous-time updating, where each node waits an exponentially-distributed time $\tau_n$ after its last tweet before it tweets again. The time scales of the model are chosen so that the mean inter-event time (average of $\tau_n$)  is 1.

\begin{figure}
\centering
\epsfig{figure=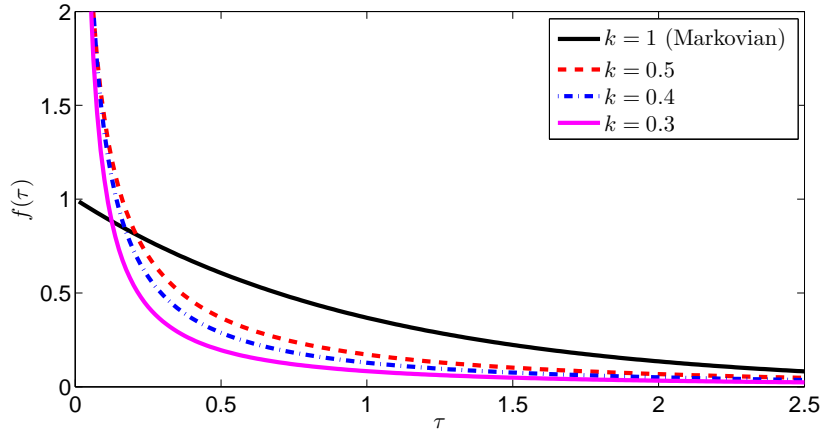,width=12.5 cm}
\caption{{\bf $\mathbf{\left|\right.}$ Distributions of inter-event times.} \bb{The unit-mean Weibull probability density function $f(\tau)=\frac{k}{\lambda}\left(\frac{\tau}{\lambda}\right)^{k-1}e^{-\left(\tau/\lambda\right)^k}$ with shape parameter $k$ and scale parameter $\lambda=1/\Gamma\left(1+1/k\right)$, for the values of $k$ used in the numerical simulations of Supplementary Figures~\ref{fignM1} to \ref{fignM3}.
}
}\label{figcomparepdfs}
\end{figure}
For the non-Markovian generalization of the model, simulations are run in continuous time and each user becomes active (sends a tweet) at times that are separated by intervals $\tau_n$ drawn from a unit-mean Weibull distribution with shape parameter $k$. To ensure the dynamics are stationary from the beginning, the time to the first tweeting of each node is drawn from the corresponding residual waiting time distribution \cite{Jo14}. For $k=1$, the inter-event time distribution is exponential, and we recover the results of the original Markovian model shown in Figure~5 of the main text. In Supplementary Figures~\ref{fignM1} through \ref{fignM3} we successively decrease the parameter $k$ to investigate the impact of increasingly non-Markovian dynamics, where short inter-event times become more probable \cite{Gleeson16}, see Supplementary Figure~\ref{figcomparepdfs}. In each case we use $n_A=1.9\times 10^5$ avalanches in $n_R=6$ replicas; the innovation parameter is $\mu=0$. Supplementary Figure~\ref{fignM1} shows results for $k=0.5$, which are qualitatively similar to the $k=1$ Markovian case of Fig.~5, except for some early-time effects (compare, in particular, the panels (e) and (f) in Fig. 5 and in Supplementary Figure~\ref{fignM1}). Our main result, that the symmetry of the avalanche shape functions depends on whether or not the offspring distribution
 has a power-law tail, certainly seems robust to the non-Markovian effects (compare panels (a) and (b)). We therefore continue towards increasingly non-Markovian dynamics in Supplementary Figures~\ref{fignM2} and \ref{fignM3}, where the Weibull shape parameter is further reduced to, respectively, $k=0.4$ and $k=0.3$. The early-time effect is more extreme in these cases, which also impacts upon the early-time avalanche shapes. Nevertheless, the qualitative results do seem to be robust for at least moderately non-Markovian dynamics, indicating that the domain of possible applicability of our theoretical results is not limited solely to Markovian dynamics. Of course, a rigorous analysis of non-Markovian effects requires an extensive theoretical analysis; we hope the present paper will form a basis for such future work.
}

\begin{figure}
\centering
\epsfig{figure=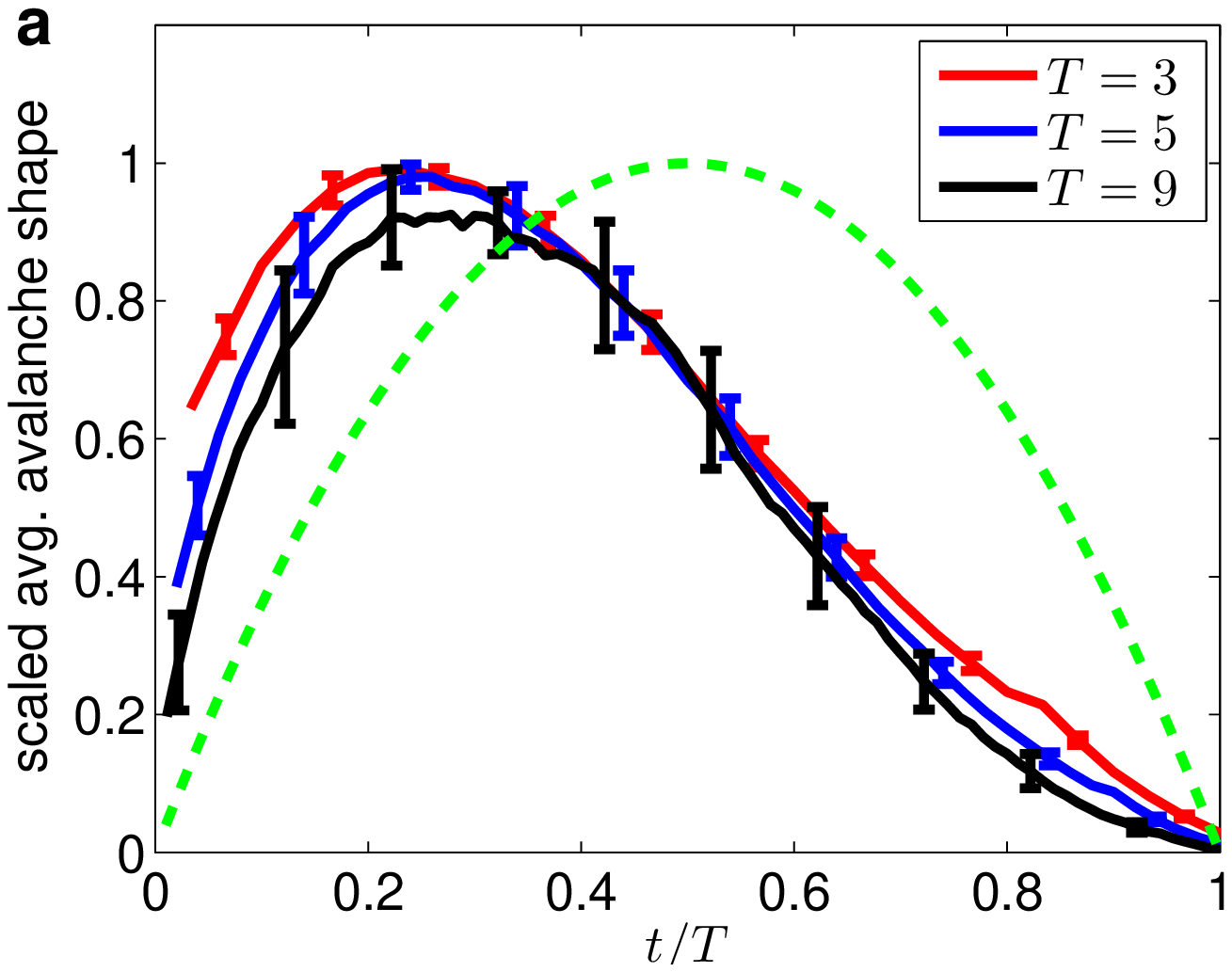,width=6.0 cm} 
\epsfig{figure=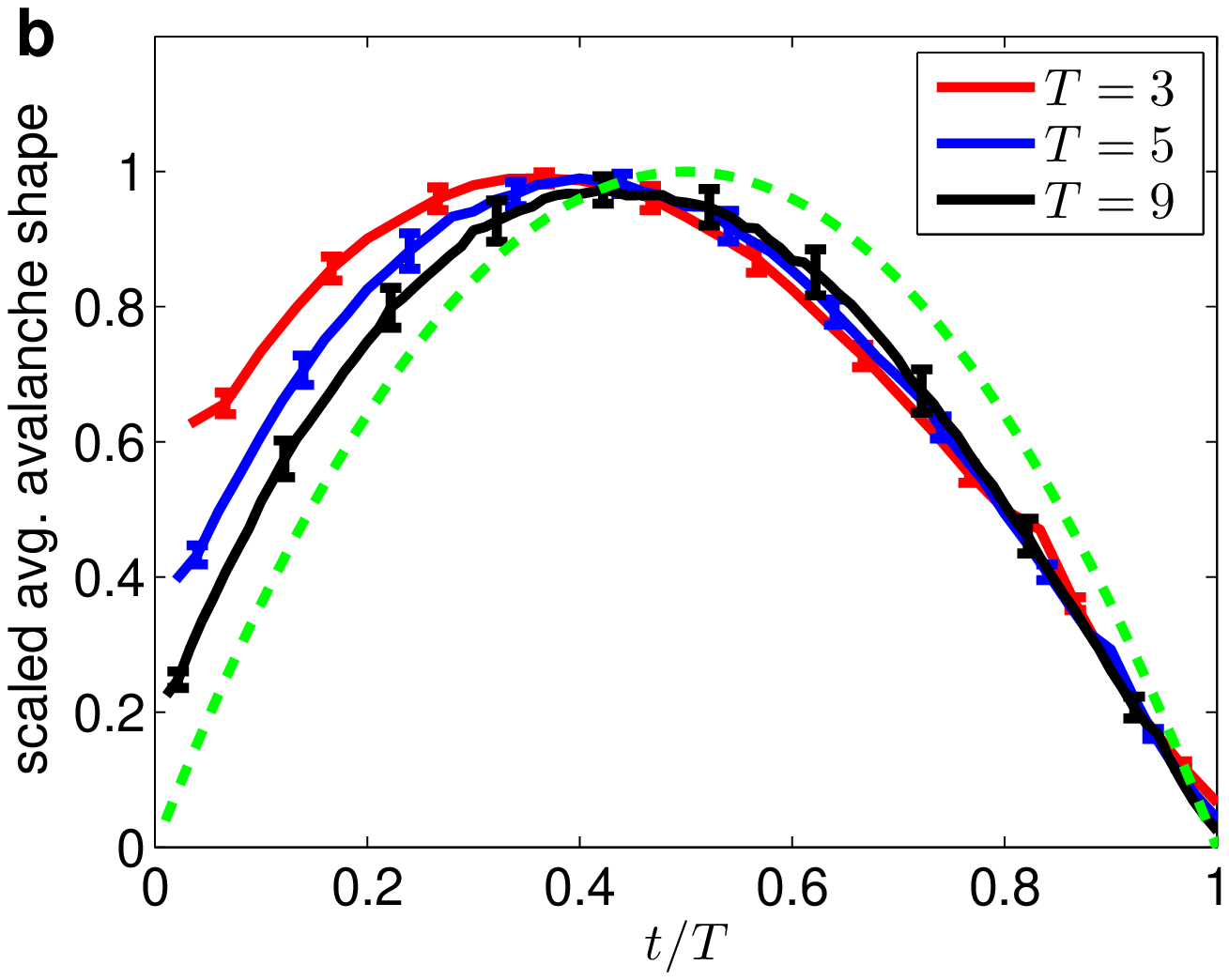,width=6.0 cm}
\epsfig{figure=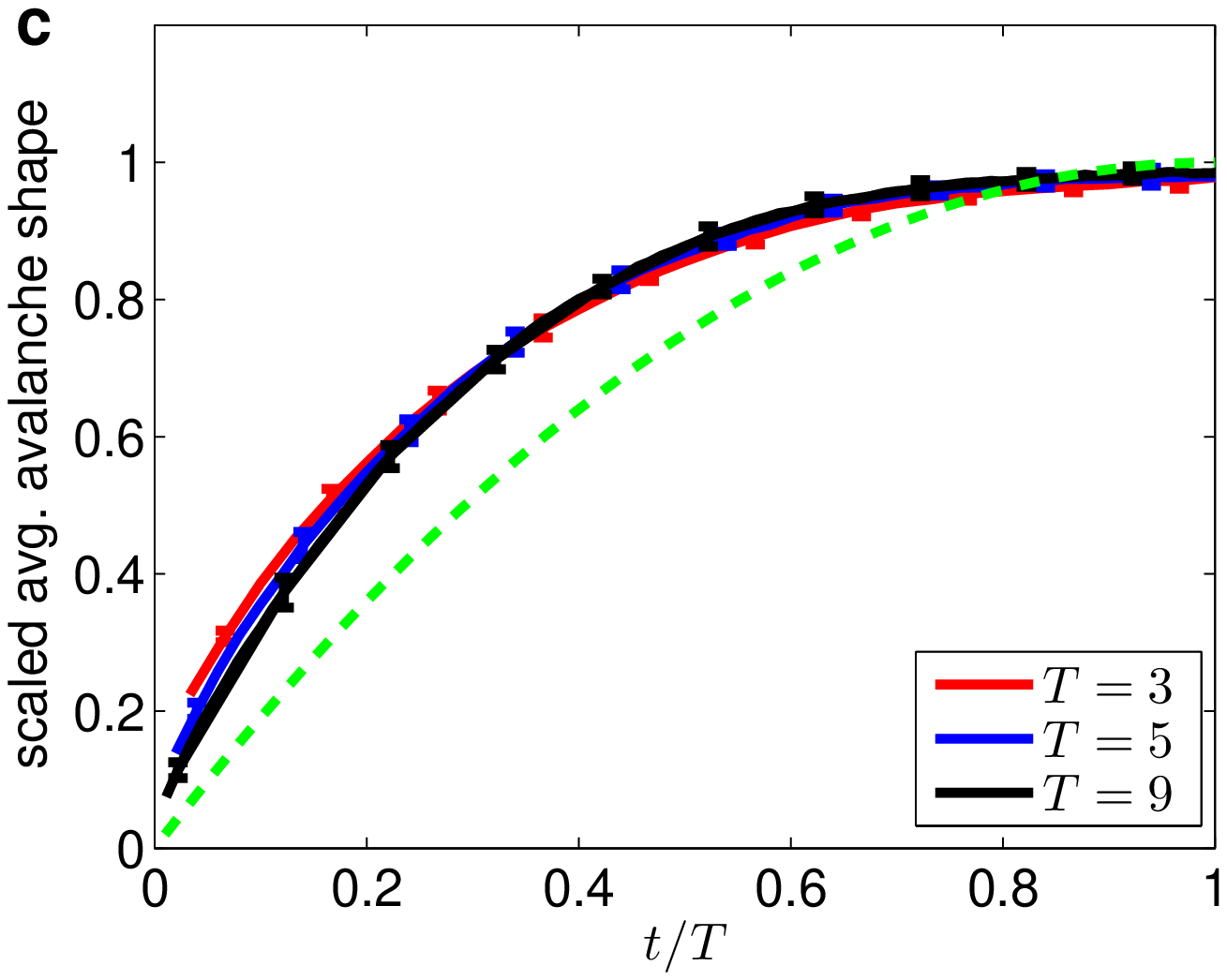,width=6.0 cm}
\epsfig{figure=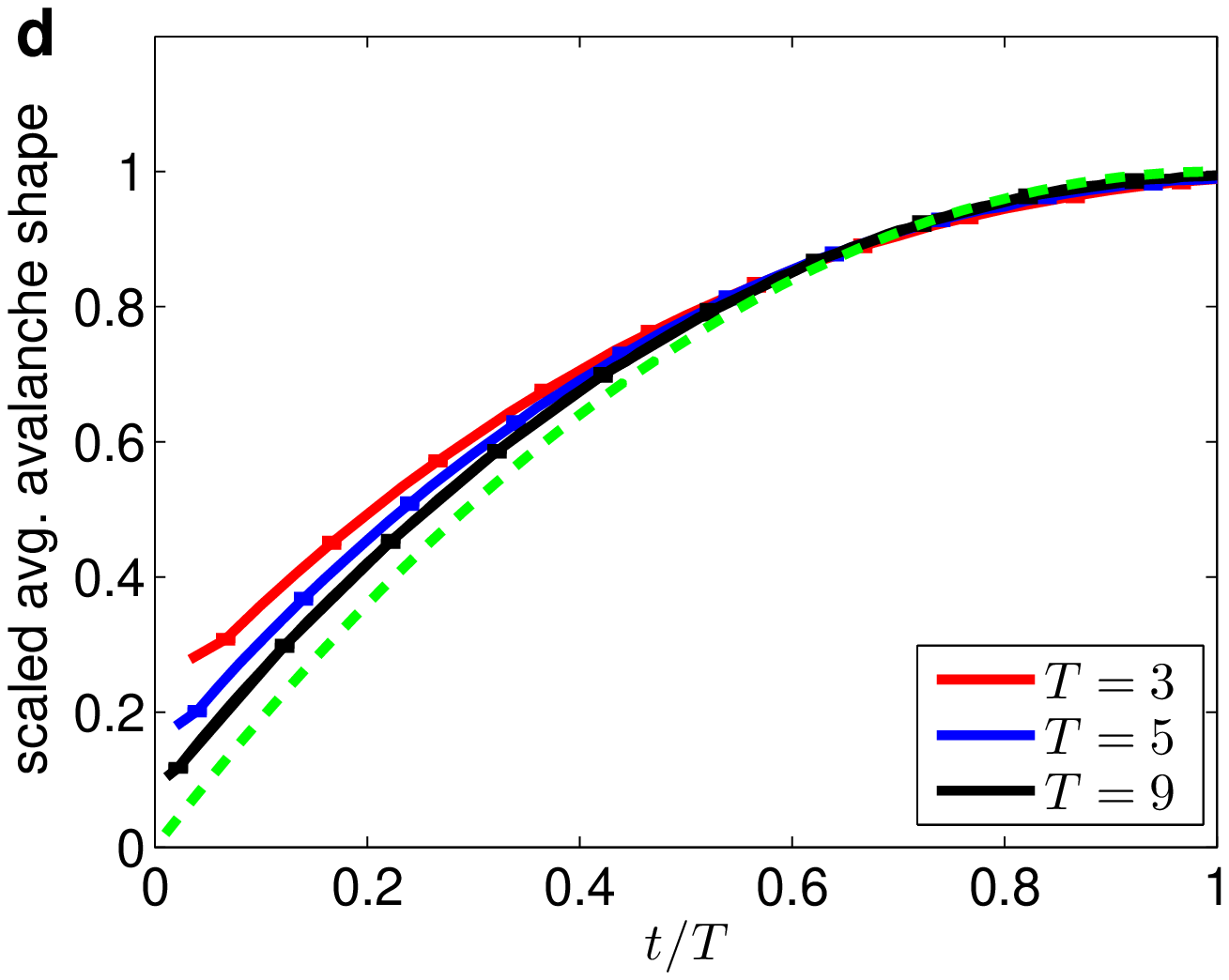,width=6.0 cm}
\epsfig{figure=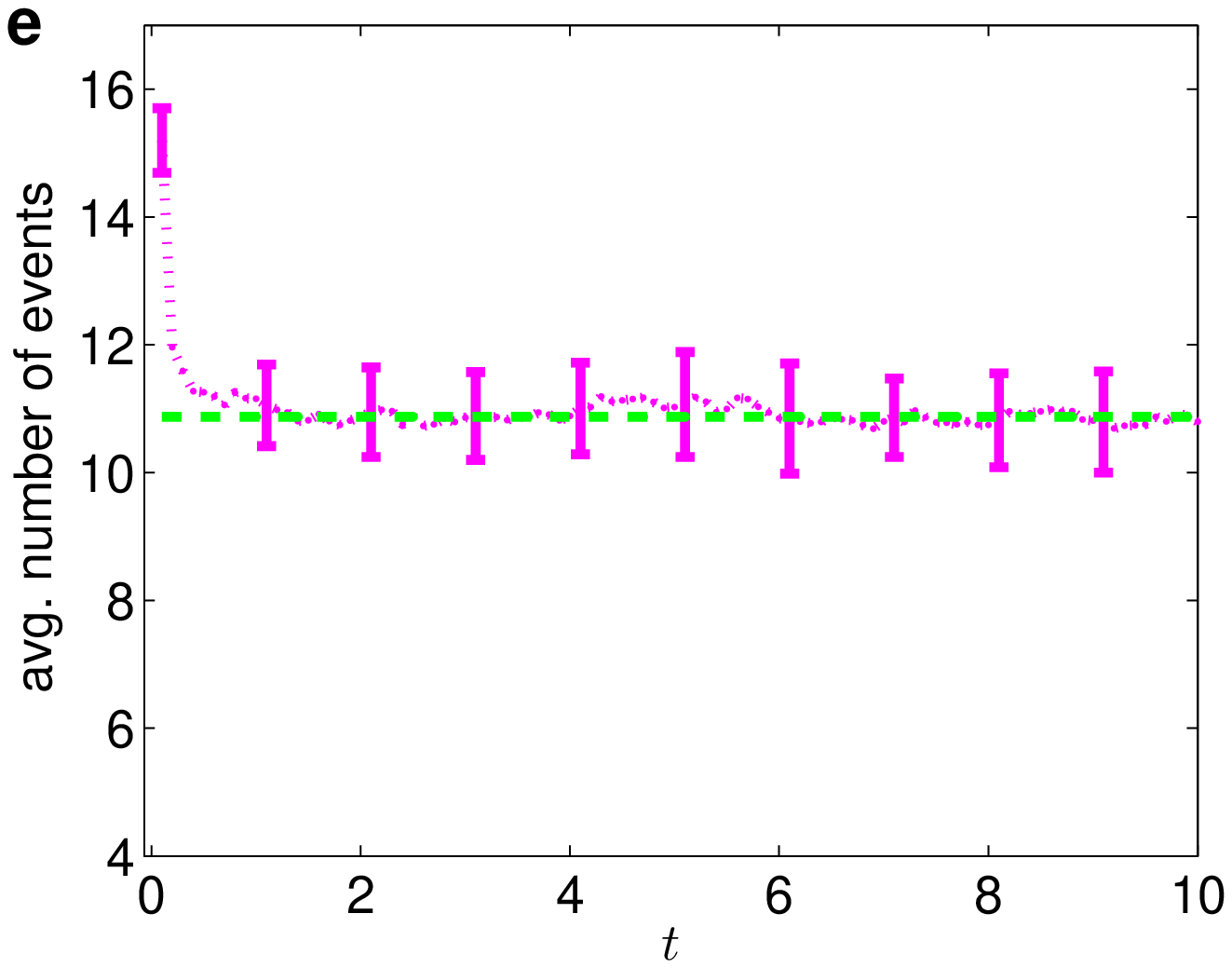,width=6.0 cm}
\epsfig{figure=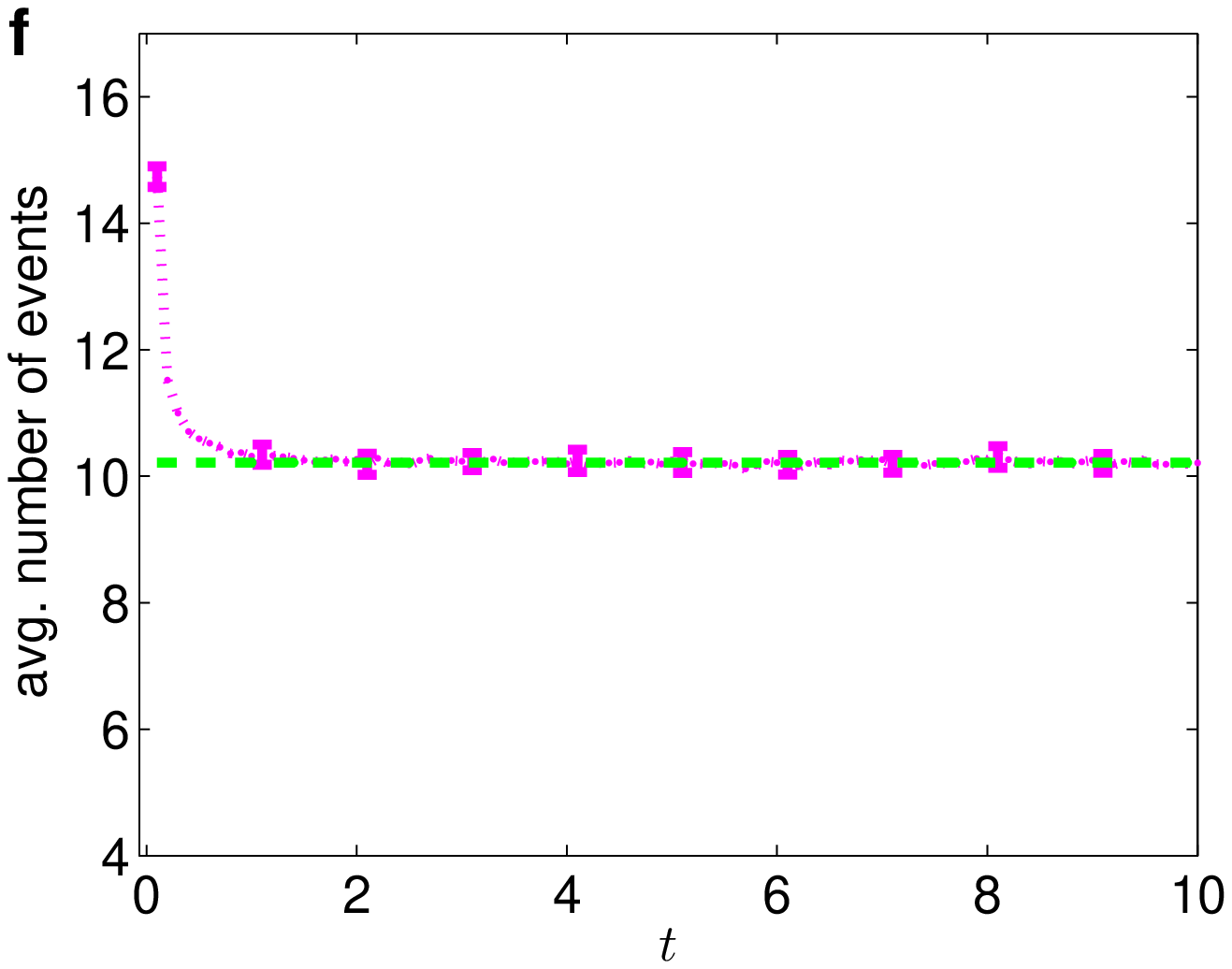,width=6.0 cm}
\caption{{\bf $\mathbf{\left|\right.}$ Non-Markovian dyanamics, $k=0.5$. }\bb{Non-Markovian generalization of the meme propagation model; compare to the Markovian case in Fig.~5 of the main text. Here the shape parameter of the Weibull distribution of inter-event times is $k=0.5$. Note that panels (a) through (d) are quite similar to the corresponding panels of Fig.~5; panels (e) and (f) clearly show the non-Markovian effects on the  early time evolution of the avalanches, but the long-time behaviour is again similar to that of Fig.~5. See Methods (in main text) for definition of error bars.}
}\label{fignM1}
\end{figure}

\begin{figure}
\centering
\epsfig{figure=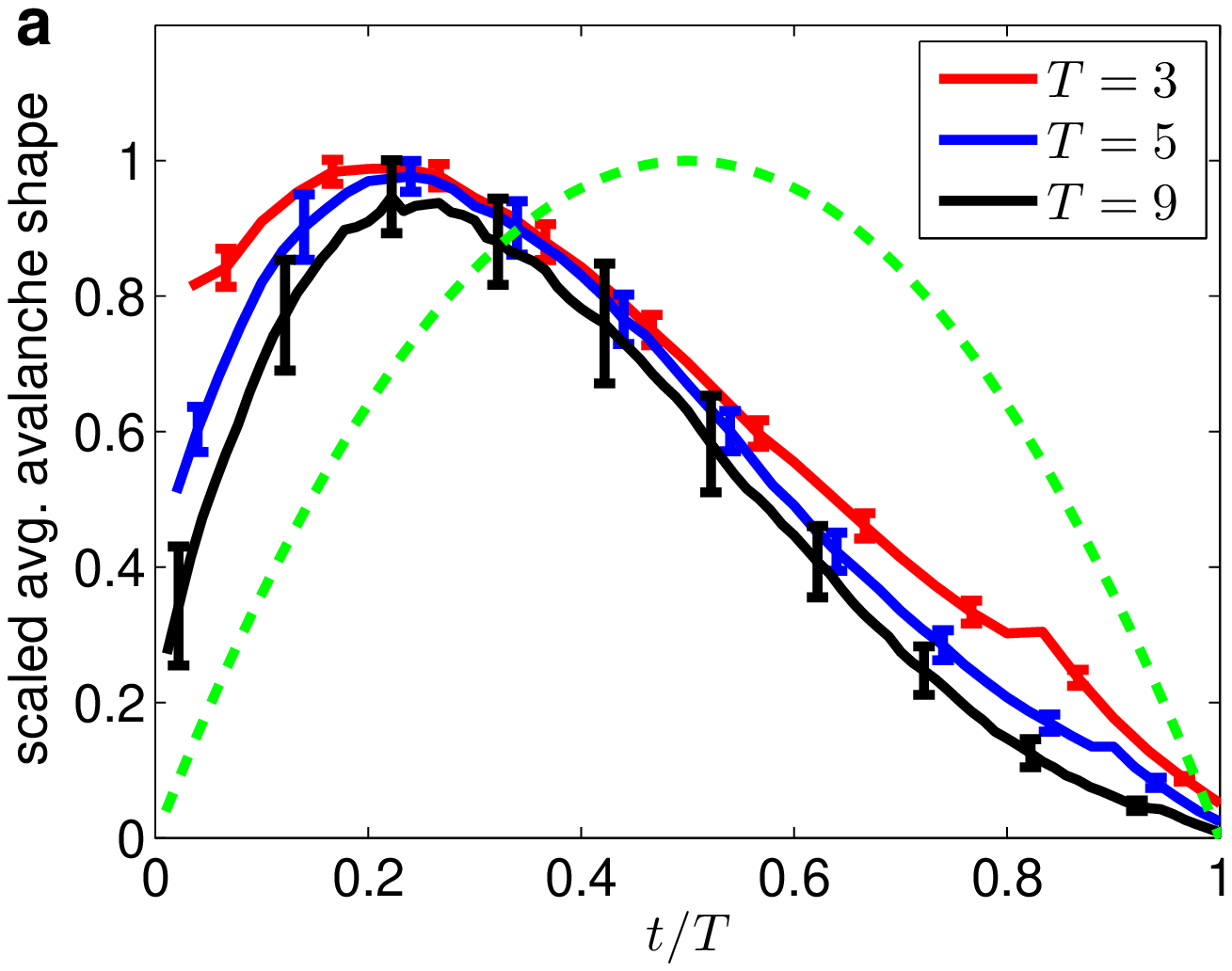,width=6.0 cm} 
\epsfig{figure=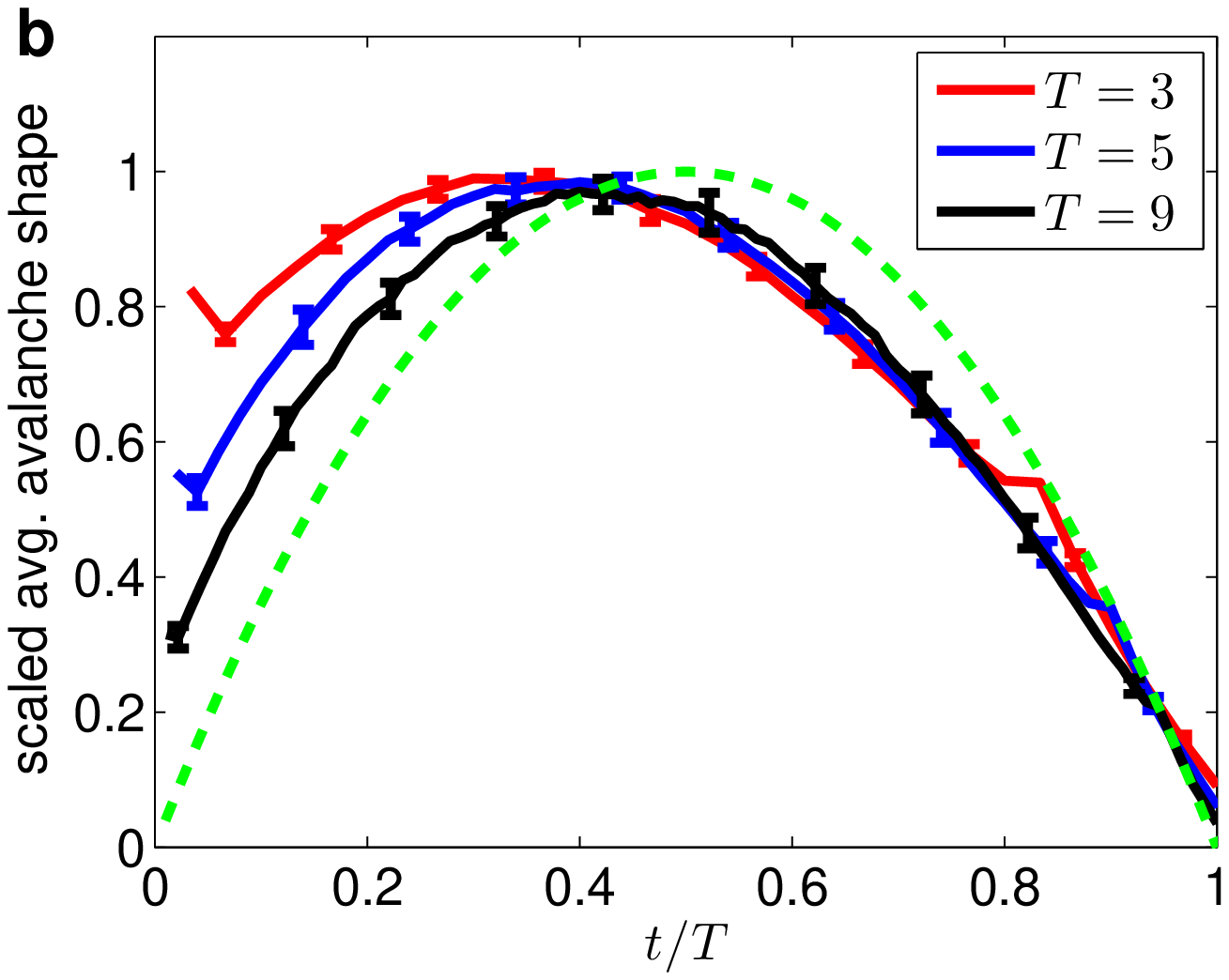,width=6.0 cm}
\epsfig{figure=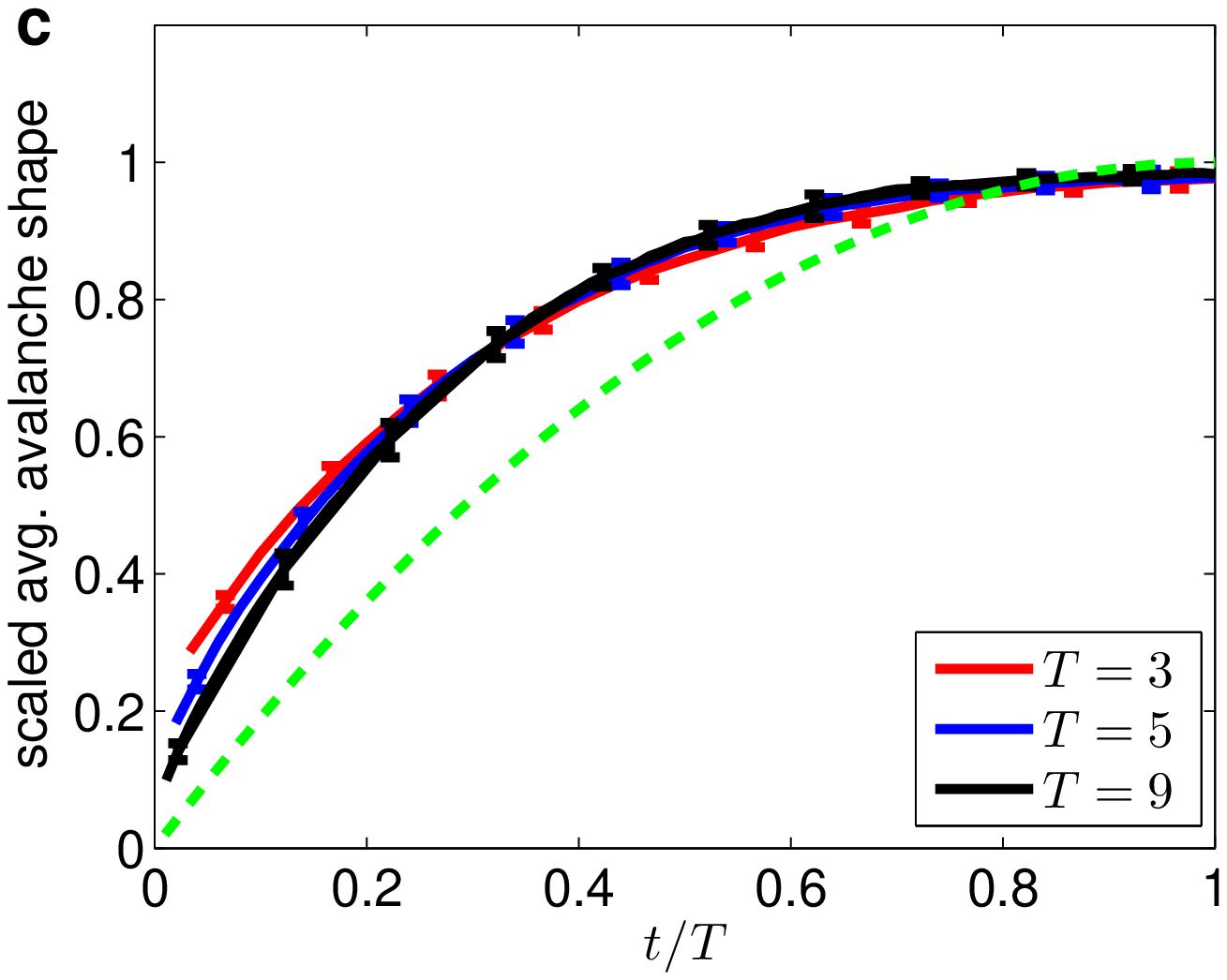,width=6.0 cm}
\epsfig{figure=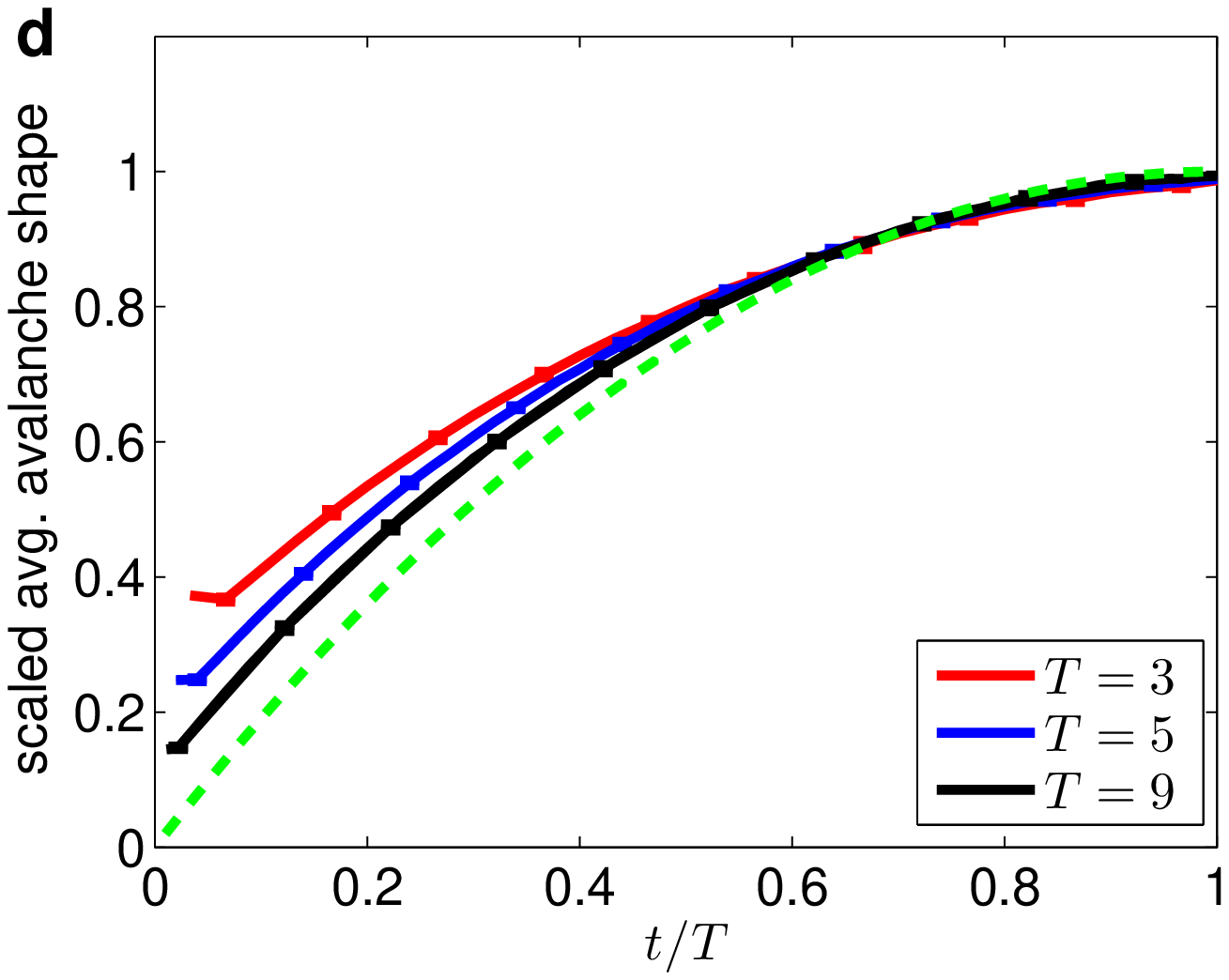,width=6.0 cm}
\epsfig{figure=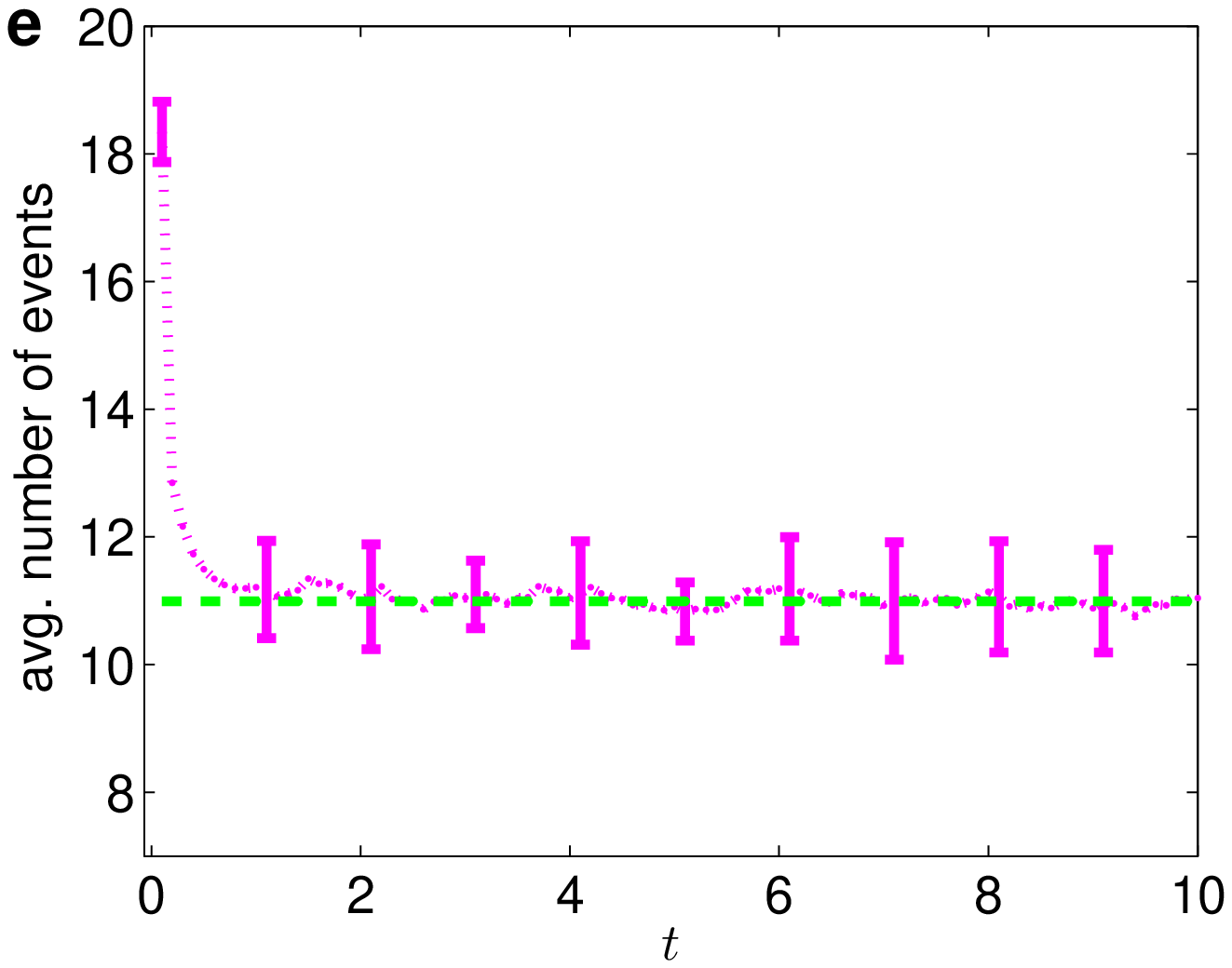,width=6.0 cm}
\epsfig{figure=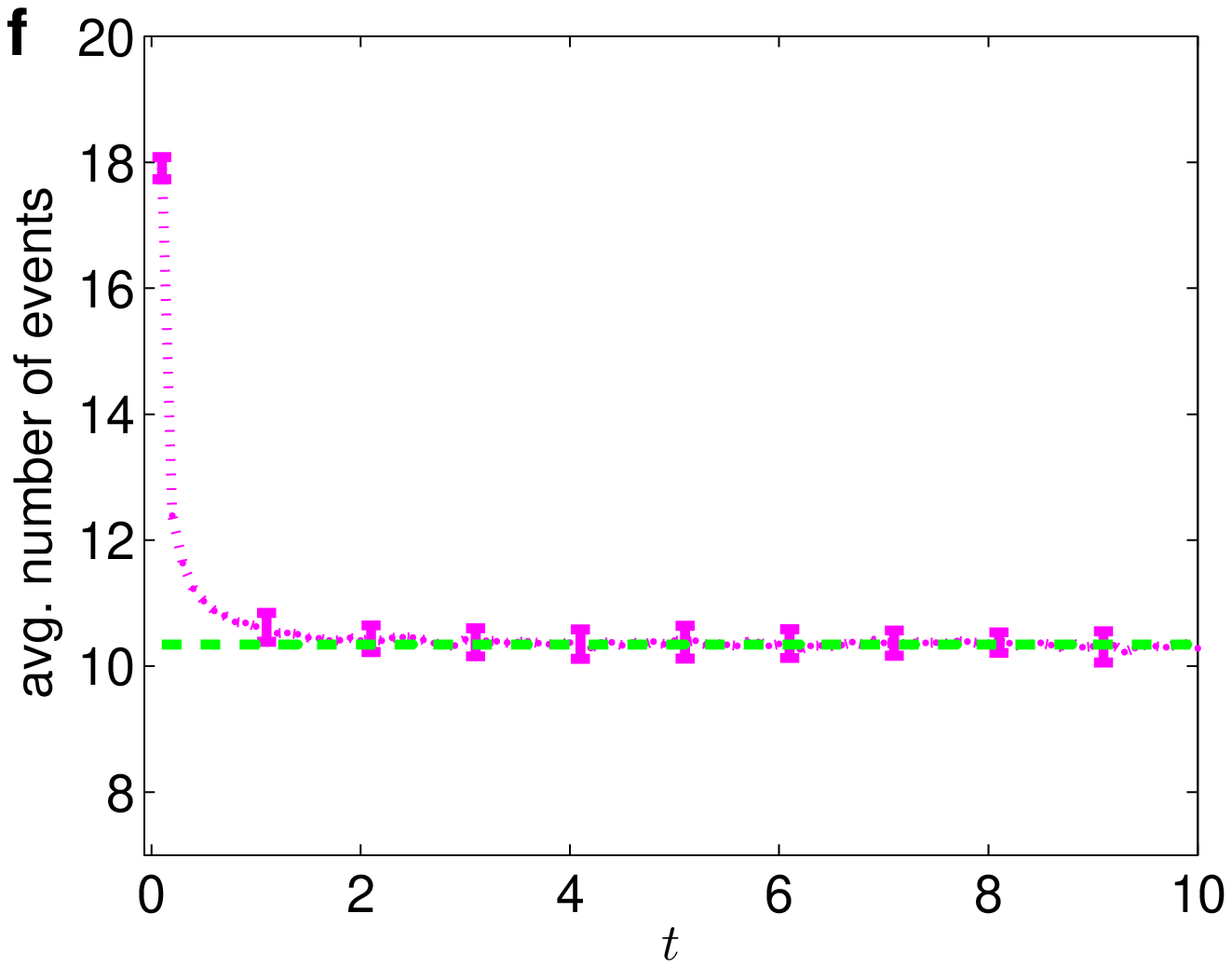,width=6.0 cm}
\caption{{\bf $\mathbf{\left|\right.}$ Non-Markovian dyanamics, $k=0.4$. }\bb{As Supplementary Figure~\ref{fignM1}, but for Weibull shape parameter $k=0.4$. Note the increased vertical scale in panels (e) and (f).}
}\label{fignM2}
\end{figure}

\begin{figure}
\centering
\epsfig{figure=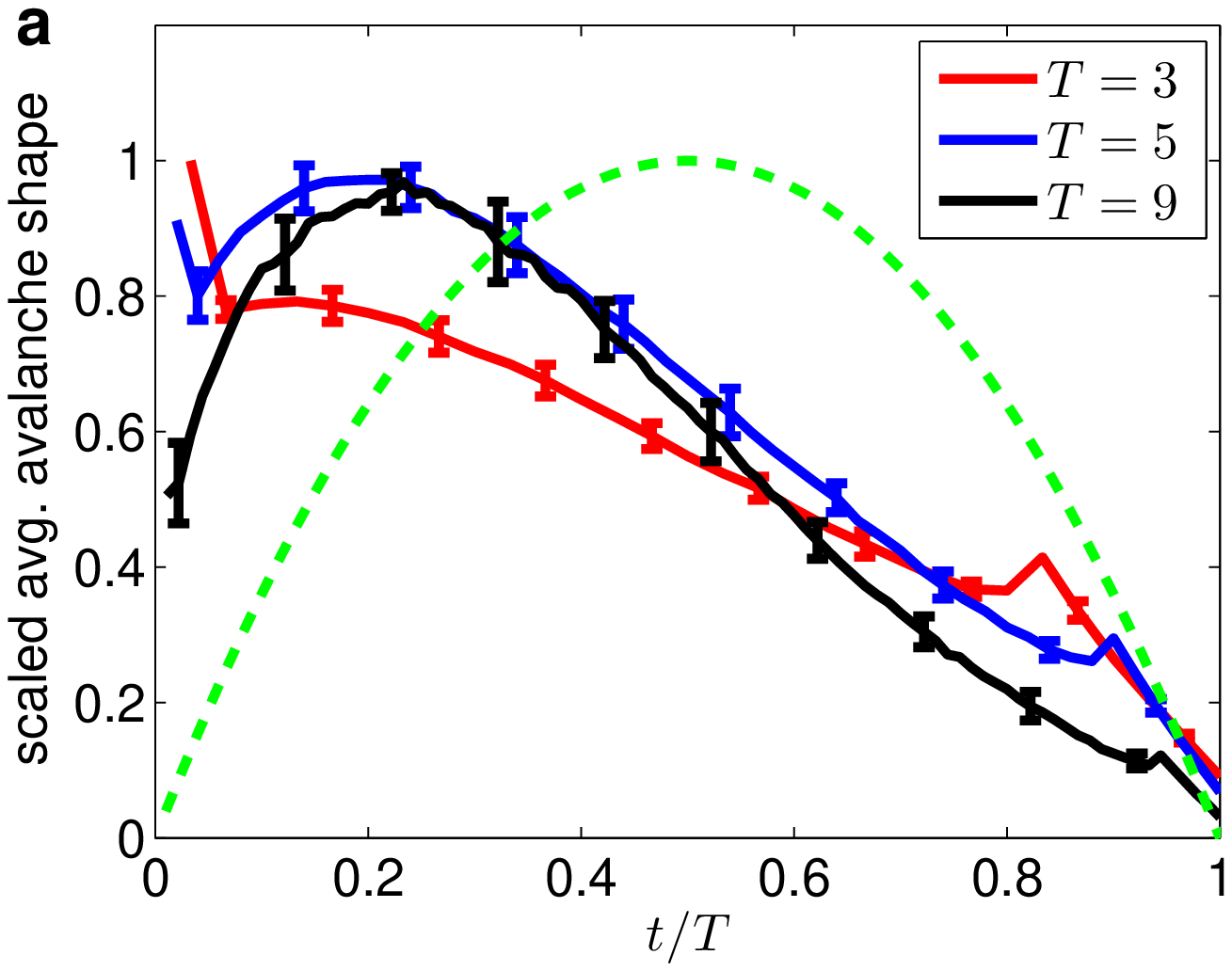,width=6.0 cm} 
\epsfig{figure=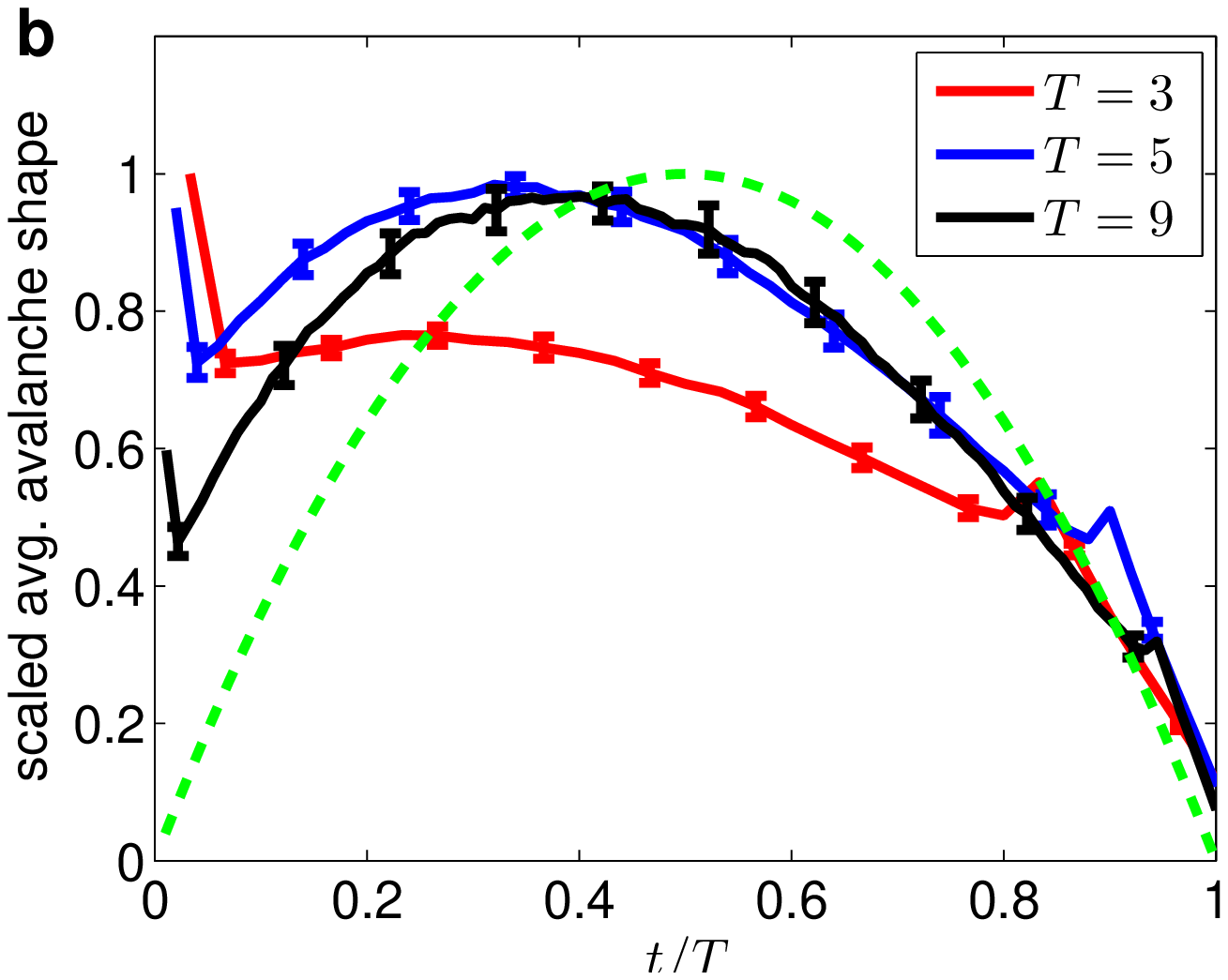,width=6.0 cm}
\epsfig{figure=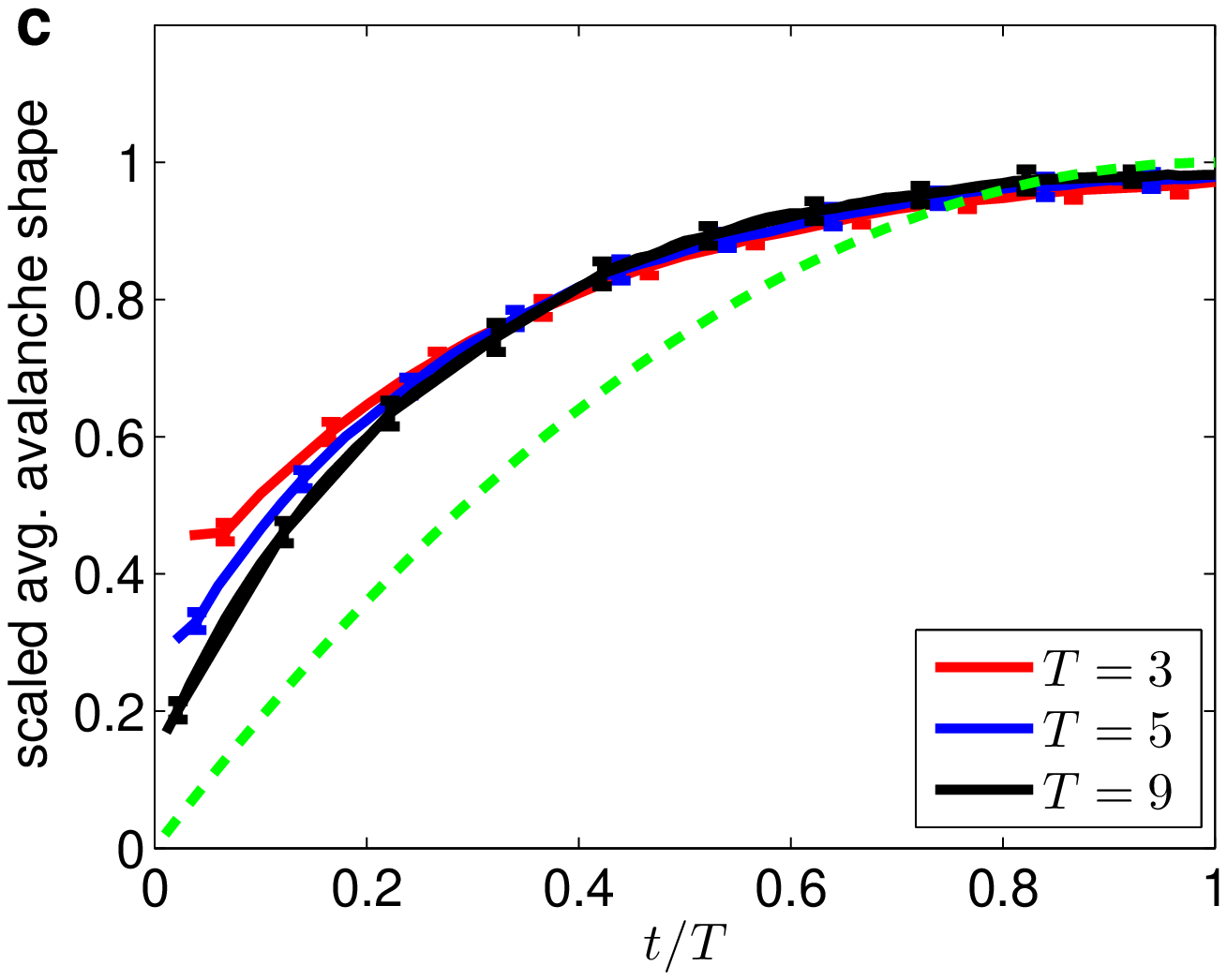,width=6.0 cm}
\epsfig{figure=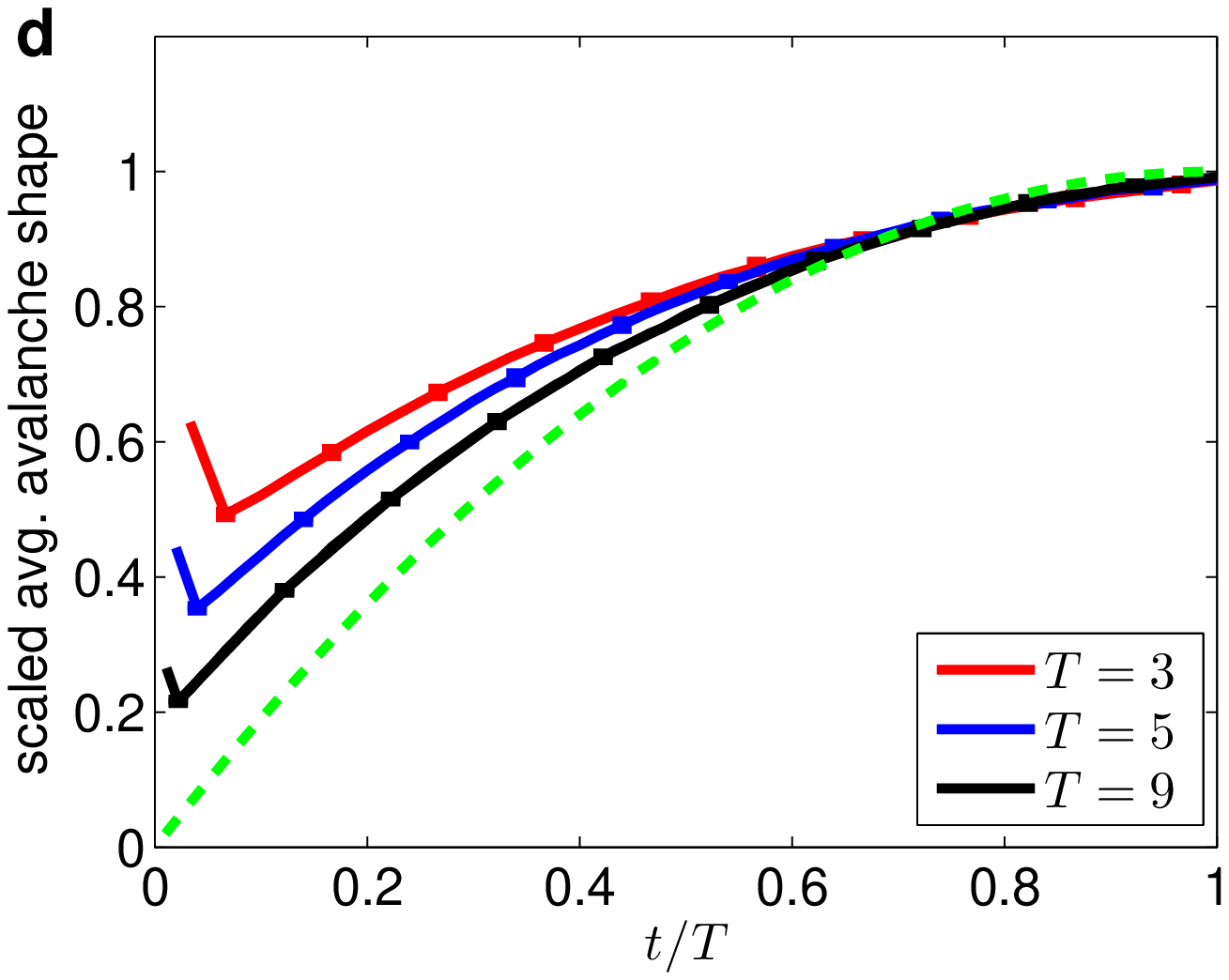,width=6.0 cm}
\epsfig{figure=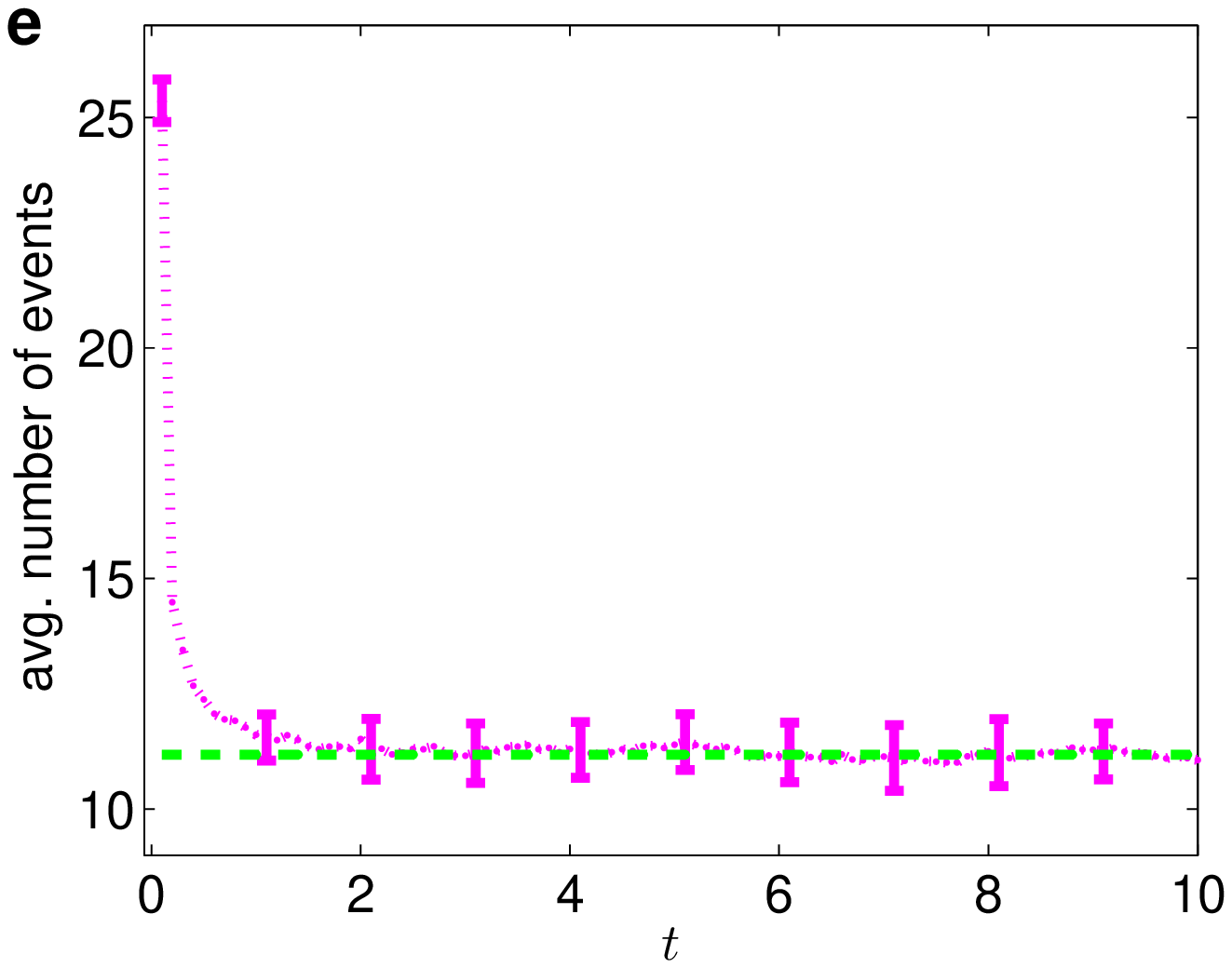,width=6.0 cm}
\epsfig{figure=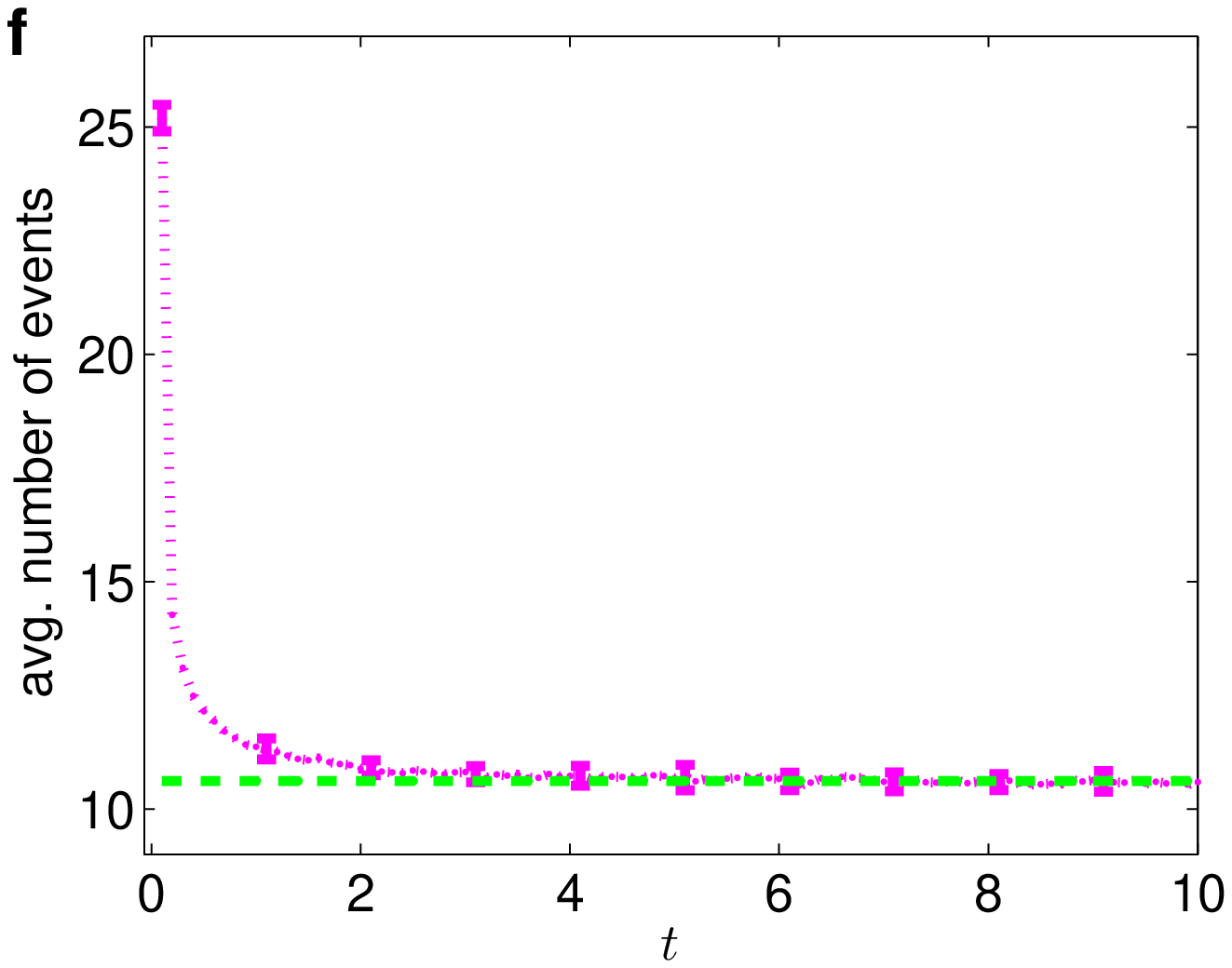,width=6.0 cm}
\caption{{\bf $\mathbf{\left|\right.}$ Non-Markovian dyanamics, $k=0.3$. } \bb{As Supplementary Figure~\ref{fignM1}, but for Weibull shape parameter $k=0.3$. Note the increased vertical scale in panels (e) and (f).}
}\label{fignM3}
\end{figure}


\end{document}